\documentclass[%
 aip,
rsi,
 amsmath,amssymb,
 reprint,%
]{revtex4-1}

\usepackage{graphicx}% Include figure files
\usepackage{dcolumn}% Align table columns on decimal point
\usepackage{bm}% bold math
\usepackage{xcolor}
\usepackage{siunitx}

\usepackage[utf8]{inputenc}
\usepackage[T1]{fontenc}
\usepackage{mathptmx}

\begin{document}

\preprint{AIP/123-QED}

\title[BASE-STEP]{BASE-STEP: A transportable antiproton reservoir for fundamental interaction studies}
% Force line breaks with \\

\author{C. Smorra}
\email{chsmorra@uni-mainz.de}
\affiliation{Institut für Physik, Johannes Gutenberg-Universit\"at, Mainz, Germany}
\affiliation{RIKEN, Fundamental Symmetries Laboratory, Wako, Japan}
 
\author{F. Abbass}
\affiliation{Institut für Physik, Johannes Gutenberg-Universit\"at, Mainz, Germany}

\author{M. Bohman}
\affiliation{RIKEN, Fundamental Symmetries Laboratory, Wako, Japan}
\affiliation{Max-Planck-Institut für Kernphysik, Heidelberg, Germany}

\author{Y. Dutheil}
\affiliation{CERN, Geneva, Switzerland}

%\author{S. Gavranovic}
%\affiliation{Institut für Physik, Johannes Gutenberg-Universit\"at, Staudingerweg 7, D-55128 Mainz, Germany}

\author{A. Hobl}
\affiliation{Bilfinger Noell GmbH, W\"urzburg, Germany}

%\author{R. Moller}
%\affiliation{Institut für Physik, Johannes Gutenberg-Universit\"at, Staudingerweg 7, D-55128 Mainz, Germany}

\author{D. Popper}
\affiliation{Institut für Physik, Johannes Gutenberg-Universit\"at, Mainz, Germany}

%\author{F. Rohland}
%\affiliation{Institut für Physik, Johannes Gutenberg-Universit\"at, Staudingerweg 7, D-55128 Mainz, Germany}

\author{B. Arndt}
\affiliation{RIKEN, Fundamental Symmetries Laboratory, Wako, Japan}
\affiliation{Max-Planck-Institut für Kernphysik, Heidelberg, Germany}
\affiliation{GSI Helmholtzzentrum für Schwerionenforschung GmbH, Darmstadt, Germany}

\author{B. B. Bauer}
\affiliation{Institut für Physik, Johannes Gutenberg-Universit\"at, Mainz, Germany}
\affiliation{RIKEN, Fundamental Symmetries Laboratory, Wako, Japan}

\author{J. A. Devlin}
\affiliation{RIKEN, Fundamental Symmetries Laboratory, Wako, Japan}
\affiliation{CERN, Geneva, Switzerland}

\author{S. Erlewein}
\affiliation{RIKEN, Fundamental Symmetries Laboratory, Wako, Japan}
\affiliation{Max-Planck-Institut für Kernphysik, Heidelberg, Germany}
\affiliation{CERN, Geneva, Switzerland}

\author{M. Fleck}
\affiliation{RIKEN, Fundamental Symmetries Laboratory, Wako, Japan}

%\author{P. Geissler}
%\affiliation{RIKEN, Fundamental Symmetries Laboratory, Wako, Japan}
%\affiliation{CERN, Geneva, Switzerland}

\author{J. I. J{\"a}ger}
\affiliation{RIKEN, Fundamental Symmetries Laboratory, Wako, Japan}
\affiliation{Max-Planck-Institut für Kernphysik, Heidelberg, Germany}
\affiliation{CERN, Geneva, Switzerland}

\author{B. M. Latacz}
\affiliation{RIKEN, Fundamental Symmetries Laboratory, Wako, Japan}
%\affiliation{Max-Planck-Institut für Kernphysik, Heidelberg, Germany}
\affiliation{CERN, Geneva, Switzerland}

\author{P. Micke}
\affiliation{Max-Planck-Institut für Kernphysik, Heidelberg, Germany}
\affiliation{CERN, Geneva, Switzerland}

\author{M. Schiffelholz}
\affiliation{Institut f{\"u}r Quantenoptik, Leibniz Universität Hannover, Hannover, Germany}
\affiliation{Physikalisch-Technische Bundesanstalt, Braunschweig, Germany}

\author{G. Umbrazunas}
\affiliation{RIKEN, Fundamental Symmetries Laboratory, Wako, Japan}
\affiliation{Eidgen{\"o}ssisch Technische Hochschule Z{\"u}rich, Z{\"u}rich, Switzerland}

\author{M. Wiesinger}
\affiliation{Max-Planck-Institut für Kernphysik, Heidelberg, Germany}

\author{C. Will}
\affiliation{Max-Planck-Institut für Kernphysik, Heidelberg, Germany}

\author{E. Wursten}
\affiliation{RIKEN, Fundamental Symmetries Laboratory, Wako, Japan}
\affiliation{CERN, Geneva, Switzerland}

\author{H. Yildiz}
\affiliation{Institut für Physik, Johannes Gutenberg-Universit\"at, Mainz, Germany}

%\author{M. J. Borchert}
%\affiliation{RIKEN, Fundamental Symmetries Laboratory, 2-1 Hirosawa, Wako, Saitama 351-0198, Japan}
%\affiliation{Institut f{\"u}r Quantenoptik, Leibniz Universität Hannover, D-30167 Hannover, Germany}
%\affiliation{Physikalisch-Technische Bundesanstalt, D-38116 Braunschweig, Germany}

%\author{V. Grunhofer}
%\affiliation{Institut für Physik, Johannes Gutenberg-Universit\"at, Staudingerweg 7, D-55128 Mainz, Germany}

\author{K. Blaum}
\affiliation{Max-Planck-Institut für Kernphysik, Heidelberg, Germany}

\author{Y. Matsuda}
\affiliation{Graduate School of Arts and Sciences, University of Tokyo, Tokyo, Japan}

\author{A. Mooser}
\affiliation{Max-Planck-Institut für Kernphysik, Heidelberg, Germany}

\author{C. Ospelkaus}
\affiliation{Institut f{\"u}r Quantenoptik, Leibniz Universität Hannover, Hannover, Germany}
\affiliation{Physikalisch-Technische Bundesanstalt, Braunschweig, Germany}

\author{W. Quint}
\affiliation{GSI Helmholtzzentrum für Schwerionenforschung GmbH, Darmstadt, Germany}

\author{A. Soter}
\affiliation{Eidgen{\"o}ssisch Technische Hochschule Z{\"u}rich, Z{\"u}rich, Switzerland}

\author{J. Walz}
\affiliation{Institut für Physik, Johannes Gutenberg-Universit\"at, Mainz, Germany}
\affiliation{Helmholtz-Institut Mainz, Mainz, Germany}

\author{Y. Yamazaki}
\affiliation{RIKEN, Fundamental Symmetries Laboratory, Wako, Japan}

\author{S. Ulmer}
\affiliation{RIKEN, Fundamental Symmetries Laboratory, Wako, Japan}
\affiliation{Heinrich Heine Universit\"at D\"usseldorf, D\"usseldorf, Germany}

\date{\today}% It is always \today, today,

\begin{abstract}
Currently, the only worldwide source of low-energy antiprotons is the AD/ELENA facility located at CERN. 
To date, all precision measurements on single antiprotons have been conducted at this facility and provide stringent tests of the fundamental interactions and their symmetries. 
However, the magnetic field fluctuations from the facility operation limit the precision of upcoming measurements. 
To overcome this limitation, we have designed the transportable antiproton trap system BASE-STEP to relocate antiprotons to laboratories with a calm magnetic environment. We anticipate that the transportable antiproton trap will facilitate enhanced tests of CPT invariance with antiprotons, and provide new experimental possibilities of using transported antiprotons and other accelerator-produced exotic ions. 
We present here the technical design of the transportable trap system. This includes the transportable superconducting magnet, the cryogenic inlay consisting of the trap stack and the detection systems, and the differential pumping section to suppress the residual gas flow into the cryogenic trap chamber.
\end{abstract}

\maketitle

\section{Physics Motivation}
Our present understanding of the fundamental interactions and cosmology has left us with several unresolved issues. For example, the origin of the matter-antimatter asymmetry in our universe \cite{Din03}, the composition of dark matter, and its interaction with Standard Model particles have yet to be understood \cite{Ber05}. 
Recently, searches for new physics at low energies have improved their measurement precision resulting in an increased sensitivity to new interactions \cite{Saf18}. 
Prominent examples are the precision measurements of the electron magnetic moment \cite{Han08,Fan23} that are ultimately compared to independent measurements of the fine structure constant \cite{Par18,Mor20}, the muon $(g-2)$ measurement \cite{Abi21}, or the searches for the permanent electric dipole moments of the electron \cite{Cai17,And18} and the neutron \cite{Abe20}.
However, only a few low-energy precision tests have been conducted on antiparticle systems. 
In particular, all recent measurements with antiprotonic systems, including antiprotonic helium spectroscopy \cite{Hor16}, antihydrogen spectroscopy \cite{Ahm18, Ahm20, Char20}, and the high-precision measurements of the antiproton's fundamental properties \cite{Ulm15, Smo17, Bor22}, have been performed exclusively at a single facility - the Antiproton Decelerator (AD) at CERN \cite{Hor13} and its low-energy extension ELENA \cite{Bar18}.\\

The BASE collaboration (Baryon-Antibaryon Symmetry Experiment) has performed the most precise tests of the combined charge, parity and time-reversal (CPT) symmetry with trapped protons and antiprotons to date\cite{Ulm15,Smo17,Sch17,Bor22}. 
These include improved limits on a possible difference of the charge-to-mass ratios of the proton and the antiproton, which, under certain assumptions, also act as a test of the weak equivalence principle for antiprotons \cite{Hug91}. 
BASE has also provided an improved limit on the potential difference of the proton and antiproton magnetic moments at the parts-per-billion level. 
This test is sensitive to Lorentz- and CPT-violating physics\cite{Blu98, Din19}, and it provides the first direct limits on the coupling of antiprotons to axion-like dark matter \cite{Smo19}. 
Cyclotron frequency ratio measurements have confirmed the Standard Model predictions to 16 parts-per-trillion (ppt) relative precision and constrain perturbations of the trapped antiproton's level splittings by new interactions with an energy resolution of $2\times10^{-27}\,$GeV \cite{Bor22}. 
However, the lack of understanding of the baryon asymmetry requires further symmetry tests, and while measurements of CP violation in mesons \cite{Bed20} and studies of flavor composition of the sea quarks \cite{Dov21} are still ongoing, it seems highly valuable to also increase the precision of low-energy symmetry tests in the baryon sector. 
Experimentally though, the precision of proton/antiproton comparisons based on measuring the in-trap cyclotron frequency $\nu_c =  q B/(2\pi m)$ and Larmor (spin-precession) frequency $\nu_L = (g/2)\, q B/(2\pi m)$ is currently limited by magnetic field fluctuations in the accelerator hall of the AD. 
Here, $B$ is the magnetic field strength, and $q$, $m$, and $g$ are the charge, mass, and $g$-factor of the trapped particle. 
The most obvious sign that magnetic field fluctuations are limiting our measurements is that the measured cyclotron frequency fluctuations are five times lower during accelerator shut down than during normal AD/ELENA operation \cite{Ulmer2019a}. \\

BASE-STEP (STEP: Symmetry Tests in Experiments with Portable antiprotons) provides a foundation for the next generation of precision antiproton measurements in the BASE collaboration by developing transportable traps to bring antiprotons out of the magnetic field fluctuations of the AD accelerator hall and into dedicated low-noise laboratories. 
The relevance of transportable traps for precision measurements had been pointed out as early as in the 1990s \cite{Tse93, Deh95} as one of the most compelling use cases for Penning traps. 
In fact, transportable ion traps have a storied history. 
During the delivery of a new superconducting magnet, electrons were transported more than 5000 km across North America in a closed cryogenic vacuum chamber \cite{Tse93}. 
More recently, the sensitivity of atomic clocks to gravitational and relativistic time-dilation effects with systematic uncertainties at or below the level of 10$^{-18}$  have renewed the interest in transportable traps for atomic clocks \cite{Hun16, Bot19, Bre18, Hua20}. 
Transportable Sr lattice clocks, for example, have recently been used for pioneering measurements in relativistic geodesy \cite{Gro18}, and have been operated at two locations with a height difference of \SI{450}{\meter} to test general relativity \cite{Tak20}. 
In contrast to these experiments, however, antiprotons cannot be produced in the measurement device at the desired location, and the technical implementation of accommodating the injection of antiprotons, their storage and transportation are considerably more involved. 
Uniquely though, transporting antiprotons to radioactive ion beam facilities could also enable novel studies of nuclear structure using the antiproton annihilation signature \cite{Wad04}. 
The implementation of this concept is currently pursued by the PUMA collaboration \cite{Nak19,PUMA2022} which aims to use antiprotons to study radioactive isotopes produced at the ISOLDE facility of CERN. 
In the context of the BASE precision matter-antimatter comparison measurements, dedicated off-site experiments will allow us to take advantage of the lower magnetic-field noise environment, and fully profit from the developments of new laser-based sympathetic-cooling methods \cite{Boh17, Boh21, Wil22}, and quantum-logic inspired state readout \cite{Nie19, Mie21}. 
Thereby, antiproton precision experiments will be able to also benefit from the latest technology developed to improve the precision records of the most precise Penning-trap measurements, which currently operate only with matter systems \cite{Rau20, Mey18}. 
We also note that further dark matter search studies are possible with a distributed network of precision experiments as in magnetometer and atomic-clock based dark-matter searches \cite{Afa21, Wis18}, but using antiparticle-based clocks instead. 
Further, measurements in an underground location with a transportable trap would also enable a sensitive search for millicharged dark-matter particles \cite{Bud22}. \\

Here, we present the BASE-STEP antiproton trap system - a compact transportable cryogenic Penning-trap apparatus for the transport of antiprotons produced and trapped at the AD/ELENA facility, and for the subsequent transfer of antiprotons into an apparatus for precision experiments. We review the magnetic field limitations of precision measurements in the AD/ELENA facility and present the technical design of the transportable trap apparatus with particular focus on the transportable magnet, the trap system and the differential pumping system. \\  

\section{Antiproton precision measurements and magnetic-field limitations}
Tests of CPT invariance with single trapped protons and antiprotons are performed in cryogenic multi Penning-trap systems that measure frequency ratios in a strong magnetic field. Details of the basic measurement concepts have been reported in detail~\cite{Smo15,Smo20}.
Charge-to-mass ratio measurements of trapped particles in Penning traps are based on determining the cyclotron frequency $\nu_c$ by measuring the motional frequencies in the trap and using the invariance theorem \cite{Bro82}. Further, magnetic moment measurements require in addition the determination of the Larmor frequency $\nu_L$ by the application of the continuous Stern-Gerlach effect \cite{Deh86}. 
Both methods rely on non-destructive image-current measurements of the trapped particle's motional frequencies \cite{Nag16}. 
In particular, the CPT invariance tests require so far the determination of the antiproton Larmor frequency, and the cyclotron frequencies of the antiproton and the negative hydrogen ion H$^-$ 
\begin{eqnarray}
\nu_{L,\overline{p}} = \frac{1}{2 \pi}\frac{g_{\overline{p}}}{2}\left(\frac{q}{m}\right)_{\overline{p}}\,B\\
\nu_{c,\overline{p}} = \frac{1}{2 \pi}\left(\frac{q}{m}\right)_{\overline{p}}\,B\\
\nu_{c,H^-} = \frac{1}{2 \pi}\frac{1}{R}\left(\frac{q}{m}\right)_p\,B
\end{eqnarray}
in the AD/ELENA facility. In addition, an off-line measurement of the proton magnetic moment is necessary \cite{Sch17}. 
In charge-to-mass ratio measurements, the negative hydrogen ion is used as a stand-in for the proton with no loss in experimental precision \cite{Smo15}, and $R=1.001\,089\,218\,753\,80(3)$ is the mass ratio of the negative hydrogen ion to the proton \cite{Bor22} including the polarization shift\cite{Tho04} in a magnetic field of $B=1.944864\,$T. 
The ratio of the proton and antiproton charge-to-mass ratios, and the magnetic moment of the antiproton are then given by:
\begin{eqnarray}
\frac{(q/m)_{\overline{p}}}{(q/m)_p} = -\frac{1}{R} \frac{\nu_{c,\overline{p}}}{\nu_{c,H^-}}  \\
\frac{\mu_{\overline{p}}}{\mu_N} = -\frac{g_{\overline{p}}}{2} \frac{m_p}{m_{\overline{p}}} = -\frac{\nu_{L,\overline{p}}}{\nu_{c,\overline{p}}} \frac{m_p}{m_{\overline{p}}},
\end{eqnarray}
respectively, where $g_{\overline{p}}$ is the antiproton $g$-factor, and $\mu_N$ the nuclear magneton. 
Here, we require the magnetic field strength to be constant during the measurement of the involved frequencies, so that it cancels in the frequency ratios. 
To fulfill this condition as ideally as possible, precision Penning traps use persistent-mode superconducting magnets that achieve in the best cases even relative fluctuations of a few times 10$^{-11}$ in frequency ratio measurements \cite{Van99,Kro22}. \\

\begin{figure}
    \centering
    \includegraphics[width=\linewidth]{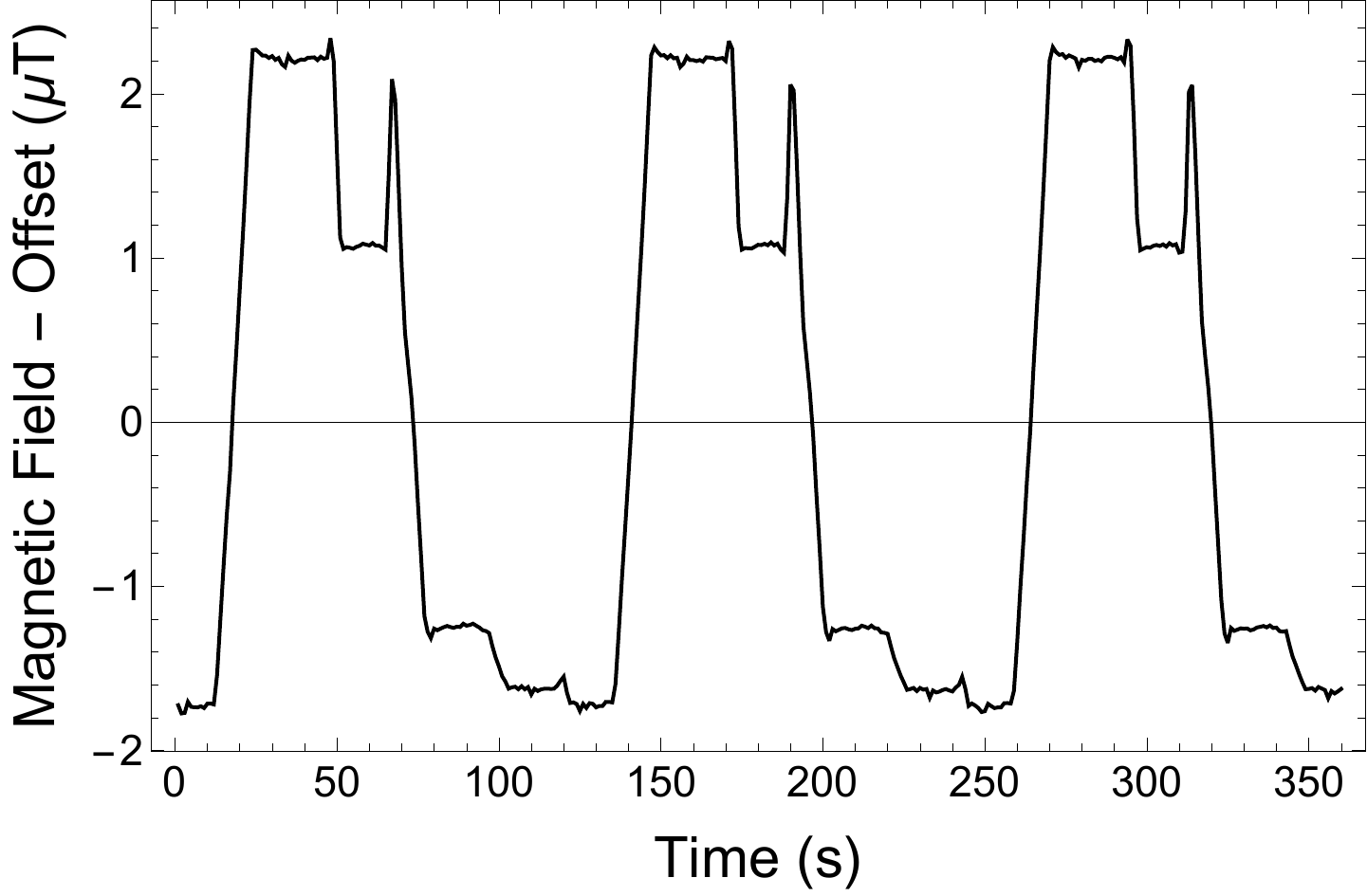}
    \caption{Measurement of the magnetic field fluctuations during the operation of the AD/ELENA facility.  The data was measured using a fluxgate sensor at three meters distance from the BASE trap system.}
    \label{fig:ADNoise}
\end{figure}

Compared to offline Penning trap measurements that deal with the intrinsic stability of the superconducting magnet, external magnetic-field fluctuations are an additional concern for measurements in the AD/ELENA complex. Here, the BASE experiment is exposed to the periodic magnetic-field ramps of the antiproton decelerator, see Fig.~\ref{fig:ADNoise}, which are measured with an FLC3-70 fluxgate magnetometer with a noise limitation of 0.12 nT/$\sqrt{\textrm{Hz}}$ about 3 meters from the trap center. 
These periodic ramps cause 4$\,\mu$T peak-to-peak shifts, and would cause relative frequency shifts on the 10$^{-6}$ level if they were present in the center of the trap.
To counteract these fluctuations, our trap system is operated with an advanced system of superconducting self-shielding solenoids \cite{Gab89, Dev19} to suppress the impact of the magnetic field ramps of the AD. 
In addition, helium pressure and temperature stabilization of the magnet reduce the fluctuations of the residual magnetization of materials near the trap and further improve the magnetic-field stability. \\

To characterize the impact of magnetic field fluctuations we measure a series of $2 N$ cyclotron frequencies and define the fluctuation of the resulting $N$ cyclotron frequency ratios from this series
\begin{eqnarray}
\sigma_r = \sqrt{\frac{1}{N}\sum_{i=1}^{N}\left(\frac{\nu_{c,2i}- R_{\mathrm{even/odd}} \nu_{c,2i-1}}{\nu_{c,2i}}\right)^2}
\end{eqnarray}
as figure of merit, where $R_{\mathrm{even/odd}}$ accounts for whether different particles were used for the odd and even measurements. It is $R_{\mathrm{even/odd}}=1$ for identical particles and otherwise it is given by the evaluated mean cyclotron-frequency ratio.
Individual cyclotron-frequency measurements $\nu_{c,i}$ are composed of the measurements of the three trap eigenfrequencies, the modified cyclotron frequency $\nu_+$, the axial frequency $\nu_z$, and the magnetron frequency $\nu_-$. The cyclotron frequency is obtained by application of the invariance theorem\cite{Bro82}: $\nu_c^2 = \nu_+^2 + \nu_z^2 + \nu_-^2$. 
In a series with $2 N$ measurements, we expect to reach a statistical uncertainty of $\delta_r = \sigma_r / \sqrt{N}$ in such a measurement sequence. 
Although one can in principle accumulate enough measurements $N$ to bring $\delta_r$ to the desired value, the quality of the measurement is better reflected by $\sigma_r$. 
If data needs to be accumulated over several months to reach the desired level of precision, it is challenging to observe and correct systematic shifts at the level of $\delta_r$.
If the measurement comprises more than $\sim$10$^3$ frequency ratios, it is already challenging to distinguish data points in the tails of the distribution from shifts due to external temporal perturbations that need to be excluded from the data and such measurements may suffer from even only a single undetected correlated quantity. 
Consequently, it is desirable to reduce $\sigma_r$ to improve antiproton precision measurements. \\

\begin{figure}
    \centering
    \includegraphics[width=\linewidth]{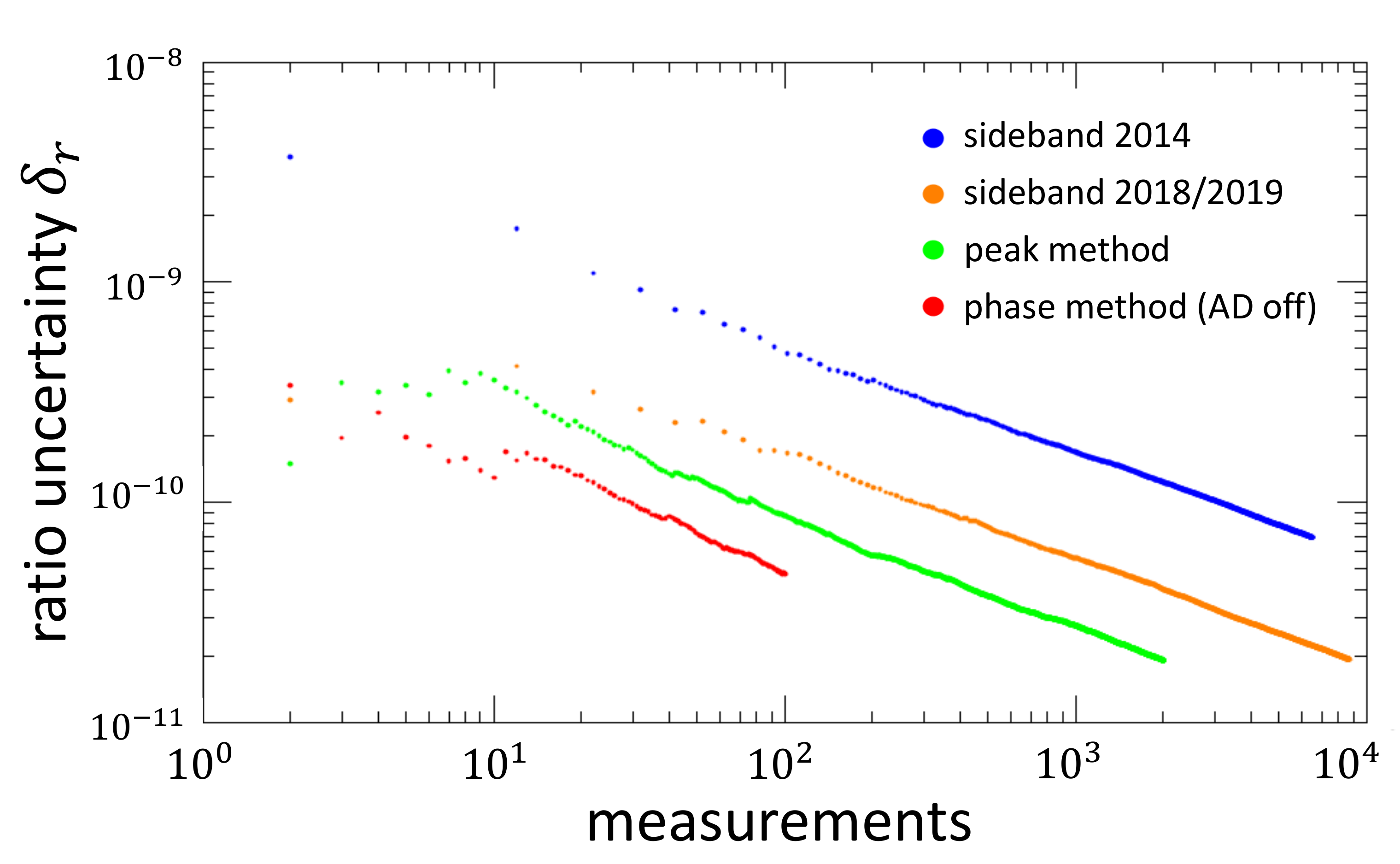}
    \caption{The statistical uncertainty of cyclotron frequency-ratio measurements in the BASE apparatus is shown for the first sideband-method measurement campaign \cite{Ulm15} (blue) and in an improved apparatus \cite{Bor22} (orange). We also show results from a peak campaign (green) in which we measure the excited cyclotron motion directly although with large systematic effects \cite{Bor22}. Finally, we show the stability using phase-sensitive techniques \cite{Stu11} that are possible while the AD/ELENA complex is not operating (red). 
    }
    \label{fig:RUnc}
\end{figure}

We summarize the observed frequency-ratio fluctuations in the BASE experiment in its accelerator-environment exposed magnetic field in Fig.~\ref{fig:RUnc}. 
The blue data points show the first antiproton charge-to-mass ratio measurement of BASE with $\sigma_r \sim 5.5$~parts per billion (ppb), recorded in 2014\cite{Ulm15}.
This measurement was conducted without a sophisticated environment stabilization and with a low magnetic shielding factor of only about 10. 
The data was acquired using the sideband method \cite{Cor90,Smo15}, which measures the cyclotron frequency by mode coupling of the axial and cyclotron modes with a quadrupolar rf-field with frequency $\nu_{rf} \approx \nu_+ - \nu_z$. 
This causes a periodic energy exchange between the two modes and enables the acquisition of the cyclotron frequency from recording the image-current spectrum on the axial detector.
The method is capable of averaging over the magnetic field fluctuations, and we synchronized the cyclotron-frequency measurements to the experiment cycle of the AD, so that each measurement is exposed to the same sequence of magnetic-field ramps in the AD cycle. 
Ultimately, we were able to compare the proton and antiproton charge-to-mass ratio with 69~parts-per-trillion (ppt) uncertainty under these conditions. \\

The orange points in Fig.~\ref{fig:RUnc} represent the results of sideband measurements conducted in an improved apparatus during 2018/2019 \cite{Bor22}. 
Here, a new superconducting magnet with a system of self-shielding superconducting solenoids was used and for certain types of external magnetic field changes shielding factors of up to 225 were reached \cite{Dev19}. 
Further, the temperature and helium pressure stabilization were improved and the vibrations of the cryogenic trap inlay were reduced. 
Under these conditions, BASE measured a ratio fluctuation of $\sigma_r \sim 1.7\,$ppb, and the best measurement sequence with the sideband method resulted in a $\delta_r = 19\,$ppt statistical and 23$\,$ppt systematic uncertainty \cite{Bor22}. 
Since $\nu_z \propto \sqrt{V_R}$, the sideband method is limited at this level of precision by the fact that all frequency information is obtained from axial frequency spectra. 
The stability of the trap voltage $V_R$, and the fit uncertainty of the axial and sideband spectra contribute to this limit and set the lowest achievable ratio fluctuation in BASE to $\sigma_r \gtrsim 1.67\,$ppb \cite{Bor22}. 
Further, an observed scaling of the measured cyclotron frequency as a function of the detuning of the axial frequency from the detection resonator frequency is the most significant contribution to the 23$\,$ppt systematic uncertainty mentioned above and becomes a significant limitation at this level of precision. 
We conclude that the sideband method is therefore not suited to further improve the antiproton charge-to-mass ratio measurement because $\sigma_r$ is limited at the 10$^{-9}$ level and improving the systematic effects of the frequency measurements via the axial dip spectra is also challenging. 
Nevertheless, it was possible to suppress the contributions of external magnetic-field noise to $\sigma_r$ for the sideband measurements to a level where they were not significant, and to reach an uncertainty on the level of 20$\,$ppt in the proton-to-antiproton charge-to-mass ratio comparison. However, significant improvements in upcoming $q/m$ comparisons require using direct cyclotron frequency measurements instead, such as cyclotron peak detection \cite{Gab99,Bor22}, or phase-sensitive methods\cite{Cor89,Stu11,BorTh}, and have been a subject of investigation as conclusion to sideband measurements. \\

The green points in Fig.~\ref{fig:RUnc} display the performance of the peak detection sequence that was part of the latest $q/m$ result \cite{Bor22}. 
Using an image-current detector at the cyclotron frequency \cite{Ulm13}, the power dissipated from an excited antiproton with $E_+ \sim 5\,$eV kinetic energy in the resonant detection circuit can be directly detected. 
The antiproton appears as a peak signal in the image-current FFT spectrum, and the peak frequency can be easily read out. 
Subsequent cyclotron-frequency measurements with this method show that $\sigma_r$ is a reduced by a factor 2 compared to the sideband method. 
However, the measured cyclotron frequencies are shifted, since excitation energy contributes about 1 ppb per 1 eV to the relativistic shift $E_+/m_{\overline{p}} c^2$, and other trap imperfections can cause additional systematic shifts. 
BASE has demonstrated that these uncertainties can be well controlled by employing a simultaneous axial-frequency measurement during the peak detection to determine $E_+$ from the shifted axial frequency \cite{Ket14}. 
This reduced the systematic uncertainty of this method and resulted in an H$^-$ ion-to-antiproton cyclotron-frequency ratio with $\delta_r = 18.5\,$ppt statistical and 13.5$\,$ppt systematic uncertainty \cite{Bor22}. \\

\begin{figure}
    \centering
    \includegraphics[width=\linewidth]{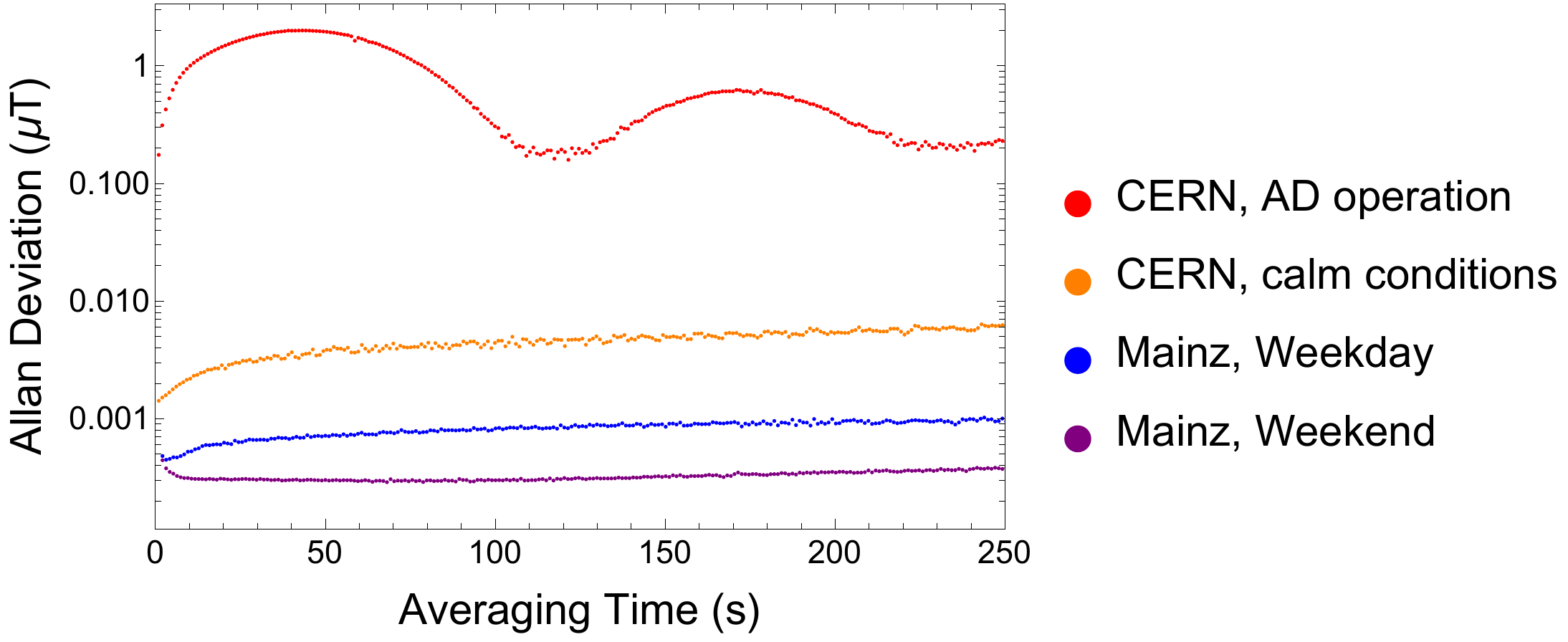}
    \caption{Measurements of the Allan deviation of the magnetic field noise as function of the averaging time.  We compare data recorded in the BASE experiment area at CERN with AD operation (red) and without (orange), and in the BASE-Mainz laboratory during weekdays (blue) and during the weekend (purple).}
    \label{fig:BfieldData}
\end{figure}

\begin{figure*}
    \centering
    \includegraphics[width=\linewidth]{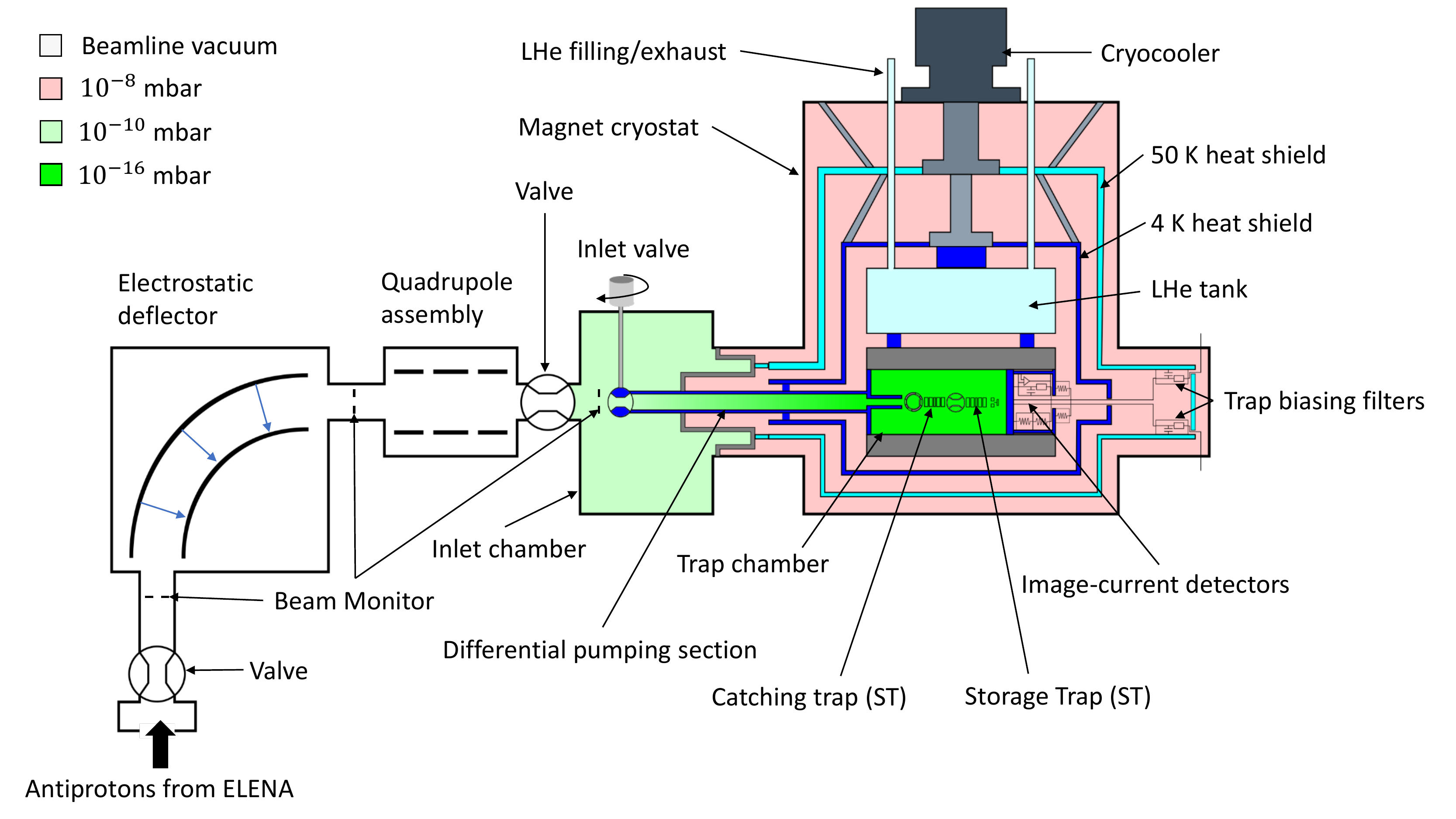}
    \caption{Schematic of the BASE-STEP apparatus for transporting antiprotons and its antiproton injection beamline connected to ELENA. 
    The colors represent different vacuum regions: the magnet cryostat isolation vacuum expected to be about $10^{-8}\,$mbar is shown in light red, the inlet chamber vacuum with a pressure of $10^{-10}\,$mbar in light green, and the trap chamber vacuum with local pressure reaching down to $10^{-16}\,$mbar in white. 
    The injection beamline is also expected to reach the $10^{-10}\,$mbar level. Further details are given in the text.}
    \label{fig:cutview}
\end{figure*}

Currently, the best technique for the cyclotron-frequency measurement is to use a phase-sensitive detection method \cite{Cor89,Stu11}. Here, an initial cyclotron phase is imprinted on the particle by an excitation drive, and the cyclotron frequency is determined by measuring the cyclotron phase as a function of the phase evolution time. 
Such measurements have been performed by BASE with protons during accelerator shutdown periods and are shown as red points in Fig.~\ref{fig:RUnc}.
The displayed measurement has $\sigma_r \approx 500\,$ppt, and the best measurement is a 10 hour long data set with fluctuations as low as $\sigma_r \approx 280\,$ppt in the BASE trap system \cite{BorTh}.
While in operation, the antiproton decelerator will cause phase slips in the phase-evolution time due the magnetic field ramps. 
As a result, the phase unwrapping procedure will be disturbed, and it will be a challenge to evaluate the phase-sensitive frequency measurements with high precision while the facility is in operation. 
Several Penning-trap experiments in dedicated precision laboratories have already demonstrated measurements at lower cyclotron-frequency ratio fluctuations with $\sigma_r < 100\,$ppt using phase-sensitive cyclotron frequency measurements \cite{Rau20,Sch20,Mey18,Hei17,Kro22}, and measured charge-to-mass ratios with statistical uncertainty below 10 ppt \cite{Mye13}. 
Since these laboratories are exposed to much lower magnetic field fluctuations, they are much more suited to determine magnetic-field dependent frequencies with high precision. \\

To highlight the effect of the magnetic-field fluctuations, we show as an example the comparison of the Allan deviation of the magnetic-field noise of representative data sets with 24 hours duration in Fig.~\ref{fig:BfieldData}.
The Allan deviation expresses the fluctuation of a continuously-measured quantity as a function of the averaging time, and is for example of fundamental importance in frequency metrology \cite{Ril95}.
The magnetic field data was recorded in the BASE experiment area inside CERN's antimatter facility and in the BASE-Mainz laboratory at the Institute for Physics in Mainz using the same FLC3-70 fluxgate magnetometer. 
The data recorded during the operation of the antiproton decelerator shows minima at the 100 nT level for averaging times that are integer multiples of the AD cycle time, see the red points in Fig.~\ref{fig:BfieldData}. 
At one half of the AD cycle time, the periodic perturbation of the AD magnet ramps causes the Allan deviation to reach up to 2$\,\mu$T.
During shutdown periods, the magnetic field noise is lower but other activities in the facility still contribute to a significant noise floor. 
The largest perturbation is due to the operation of the overhead crane that causes a jump of several $\mu$T when passing above the experiment zone. 
The orange data shows a measurement without crane motion, but with other user operations in the AD facility. 
In this particular measurement, the Allan deviation at 120 s is about 40 times lower than during AD operation.
The data from the BASE-Mainz laboratory are shown in blue for weekdays and in purple during the weekend. 
Here, the magnetic perturbations are mainly due to the motion of the elevators in the building which cause shifts of up to a few nT when passing the floor of the laboratory. Further contributions are due to displacment of magnetic materials while working in the laboratory. Both activities have reduced contributions during the weekend.
In both cases, the Allan deviation of the magnetic-field noise in the BASE-Mainz laboratory is below 1 nT for 120 s averaging time, and a factor 200 and 600 lower compared to AD operation in these particular measurements.
If a further reduction of magnetic noise is required, one could even conceive of measurements in a magnetically shielded room. 
For example, the BMSR-2 at the PTB in Berlin reaches a magnetic-field noise on the fT/$\sqrt{\mathrm{Hz}}$ level \cite{Liu2021}. 
Such environment conditions would eliminate the external magnetic-field noise limitations in antiproton precision experiments. Therefore, a transportable antiproton trap capable of relocating such measurements into magnetically-calm precision laboratories will form one of the cornerstones to enable more sensitive antiproton CPT invariance tests in the future.\\

\section{The experimental apparatus}
\subsection{Overview}

The concept of the BASE-STEP apparatus is shown in Fig.~\ref{fig:cutview}. It consists of a Penning-trap system with two electrode stacks inside the horizontal bore of a transportable \SI{1}{\tesla} superconducting magnet.
We have designed the apparatus to have only a single cryogenic setup for both the Penning-trap assembly and the superconducting magnet by using a cold-bore magnet cryostat. 
This eliminates the need for a second cooling system for the trap, and reduces the size and maintenance of the apparatus. 
The primary cooling power comes from a cryocooler used during stationary operation or while connected to a mobile power generator. 
A liquid-helium buffer tank is then used as a secondary cooling mechanism to bridge gaps in power; for example while lifting the apparatus with a crane into or out of the experiment zone or during power failures. \\

As the vacuum conditions are crucial for the antiproton storage, the design of the vacuum system has been an important aspect.
We expect that we need to store antiprotons for about three months to conduct antiproton precision measurements in an offline apparatus.
Therefore, the cryogenic trap system, which consists of a catching trap (CT) and a storage trap (ST), is contained in a dedicated cryogenic vacuum chamber designed to reach below $10^{-16}\,$mbar at the center of the ST trapping region. 
For the injection and ejection of particles, the trap chamber is connected via an extensive differential pumping system  to the inlet chamber at room temperature with a target residual gas pressure of at most $10^{-10}$ mbar. 
To prevent hydrogen diffusion from the inlet chamber into the trap chamber, the differential pumping section also includes an inlet valve to close the differential pumping channel for storage operation, and two rotatable trap electrodes are placed in the trap stack to block the direct flow of residual gas to the position of the trapped antiprotons. 
The successful implementation of the vacuum system will be tested by monitoring and counting the number of trapped antiprotons by non-destructive image-current detection \cite{Nag16}. \\ 

The main frame of the transportable apparatus is designed to be as compact as possible and is installed on an aluminum transport frame that is 2.00 m long, 0.87 m wide, and 1.85 m high. 
Weighing less than \SI{1000}{\kilo\gram}, the system can be transported by a forklift and an overhead crane, in particular through the exits of the AD hall and through the door frames, for example on the path to the BASE-Mainz laboratory at the University of Mainz \cite{Sch17,Boh21}, or to the BASE Hannover trap system\cite{Nie19}. 
The compressor of the cryocooler and its chiller system are transported on a separate frame.
We plan to initially demonstrate the transport of antiprotons in the trap system on the CERN site, and to demonstrate the transfer of antiprotons into a separate permanent-magnet based Penning trap that is presently under development to establish the methods necessary for offline experiment operation.

\subsection{Antiproton injection beamline}

The BASE-STEP apparatus will be commissioned at CERN and receive antiprotons from ELENA. To this end, the former ATRAP I experiment area \cite{DiS2013} has been modified into a dedicated experiment zone for BASE-STEP. 
The experiment zone has a vertical connection to ELENA, so that additional beamline elements are required to inject antiprotons into the horizontal trap system. \\

An electrostatic deflector with a bending radius of 600$\,$mm and $\pm 10\,$kV nominal voltage has been designed to redirect the beam into the horizontal plane. 
The design is adapted from the ELENA deflectors with lower bending angles used in the ELENA ejection lines \cite{Barna2015}, but modified for 90$^\circ$ deflection. 
The deflector electrodes are spherical to obtain focusing in the direction orthogonal to the bending plane and have 60$\,$mm spacing.
The stray field at the entrance and exit of the electrostatic deflector causes a further deflection and results in a deviation of the centered beam from the reference trajectory if not considered in the design.
Therefore, potential calculations and trajectory simulations using the geometry of the electrodes and the surroundings were performed in COMSOL, and the transfer matrix including the edge effects of the deflector was determined.
Subsequently, the transfer matrix was included into MADX modelling of the injection beamline to investigate the injection of antiprotons into the BASE-STEP trap system. 
These transport simulations comprise all elements from the injection into the vertical ELENA ejection beamline that delivers the beam into the STEP zone, the electrostatic deflector, the quadrupole assembly, and the center of the BASE-STEP magnet as target.  
The simulations project a root-mean-square (RMS) beamsize of \SI{0.53}{\milli\meter} at the center of the catching trap, well within the $\sim 2\,$mm radial trap acceptance \cite{Smo21}. \\

\begin{figure}
    \centering
    \includegraphics[width= \linewidth]{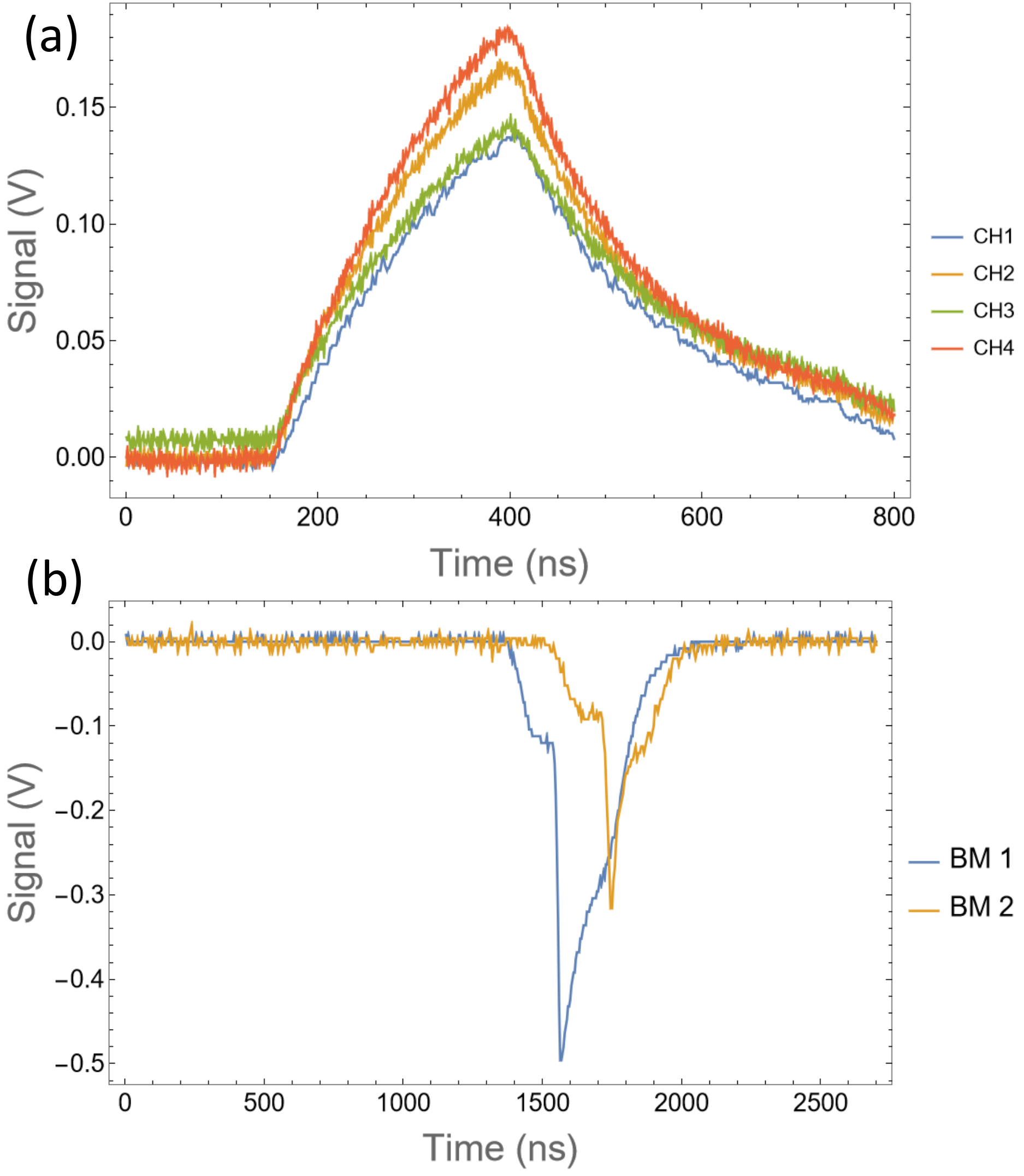}
    \caption{(a) Test signal response of the charge amplifiers for the readout of four beam monitor plates (CH1 to CH4). The variation in the pulse shape is due to the difference in the parasitic feedback capacitance of the individual channels. (b) Detection of a defocused antiproton beam in the BASE-STEP beamline that is partly deposited on the beam monitor in the vertical line (BM 1) and partly on the first beam monitor in the horizontal line (BM 2).}
    \label{fig:BMfigure}
\end{figure}

The quadupole assembly following the electrostatic deflector is crucial to achieve the steering of the antiprotons to the trap center and to set the focal length correctly, and must be configured properly for each injection. 
To center the antiproton pulse on the trap system for injection, the injection beamline is equipped with three beam monitors that are indicated in Fig.~\ref{fig:cutview}. 
The first beam monitor (BM 1) is directly above the handover point, the second one (BM 2) follows the electrostatic deflector, and the third one (BM 3) is placed around the entrance of the differential pumping tube. 
The beam monitors consist of either plates or wires which are each connected to a charge amplifier outside the vacuum system. 
The charge amplifiers consist of an inverting integrator using the LMV793 operational amplifier with 0.3$\,$pF feedback capacitance to gain a high charge sensitivity. 
Fig.~\ref{fig:BMfigure}(a) shows the response of the charge amplifiers to a 20$\,$mV test signal that places about 40$\,$000 elementary charges on the feedback capacitor. 
The beam monitors were also tested during online operation at the end of the CERN physics run in 2022 with antiprotons in the BASE-STEP beamline. 
Fig.~\ref{fig:BMfigure}(b) shows the response of two beam monitor plates following the ejection of a defocused antiproton pulse with about 5 mm RMS diameter. 
A fraction of the beam was detected on BM 1 in the vertical line and a smaller fraction was deflected by the electrostatic deflector and observed on BM 2 in the horizontal transport section (see Fig.~\ref{fig:cutview}). This demonstrates the functionality of the beam monitors and the electrostatic deflector. \\ 

\subsection{Transportable superconducting magnet}
The transportable superconducting magnet, designed in collaboration with \textit{Bilfinger Noell GmbH}, is a key component in the BASE-STEP apparatus, requiring both high magnetic-field stability and continuous operation during transport. 
Compared to stationary NMR-type magnets, the BASE-STEP magnet uses a hybrid cryocooler-backed liquid-helium bath cryostat for cooling, and a reinforced mechanical support structure to withstand the accelerating forces during the transport. \\

\begin{figure}
    \centering
    \includegraphics[width=\linewidth]{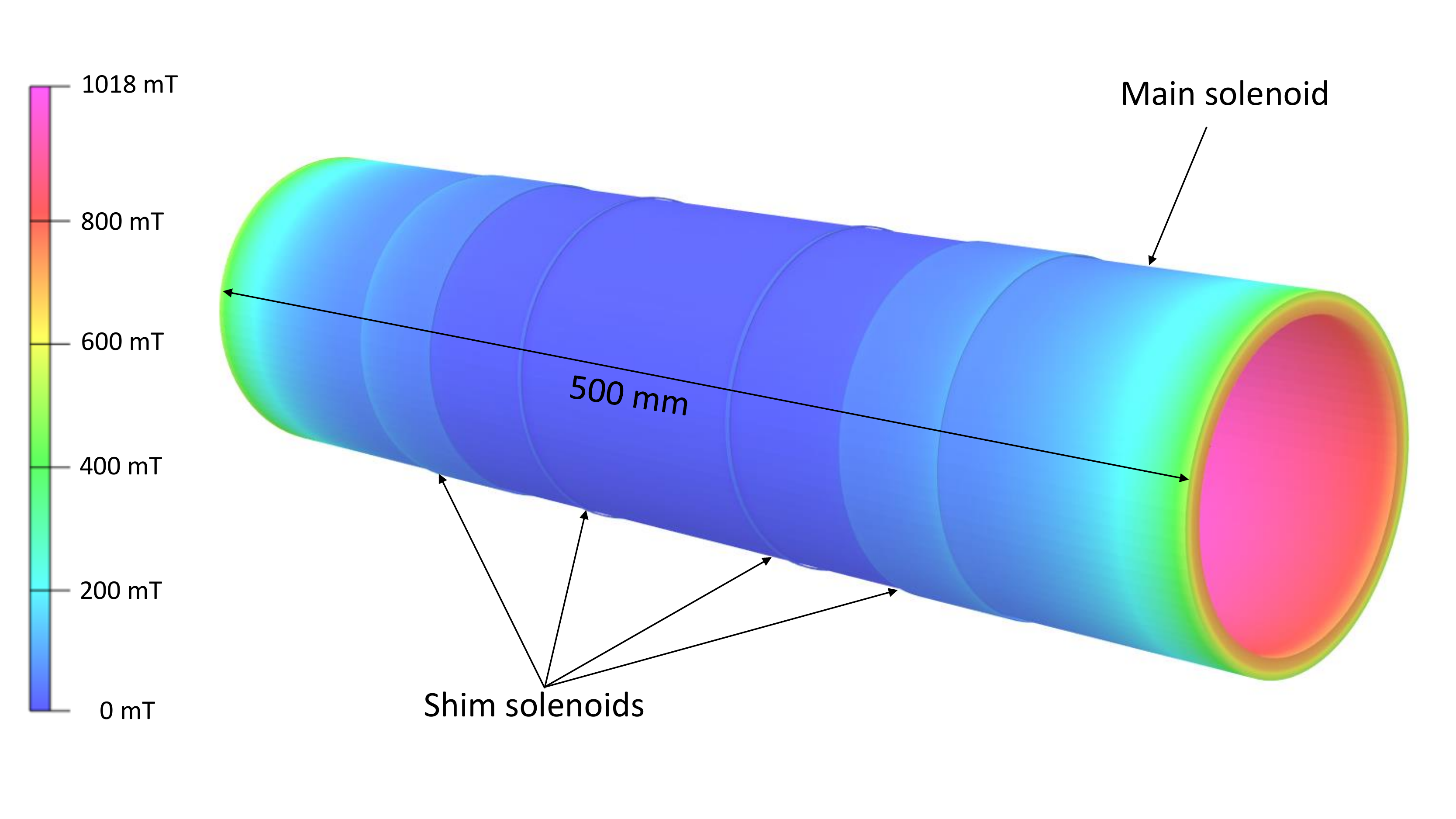}
    \caption{The surfaces of the main solenoid and of the two sets of shim coils are shown. The surface contours show the magnetic field strength on the surface of the superconductors.  }
    \label{fig:BoverSC}
\end{figure}

As the primary cooling source, we use a two-stage pulse-tube cooler (\textit{Sumitomo RP-082B2S}) with 35$\,$W cooling power at the first 45$\,$K stage, and 900$\,$mW cooling power at the 4$\,$K stage. 
The first stage is connected to the outer heatshield that is cooled to about 50$\,$K, and the second stage is connected to the inner heat shield, the liquid helium tank, the superconducting magnet coil and the Penning-trap system. 
The cool down of the entire cold mass to 4$\,$K without precooling requires 3 days.
With a power consumption of $\sim 8 \,$kW, the cryocooler can be connected to the power grid in stationary operation and to a mobile power generator during long-distance transport. 
As additional features, it is also possible to operate the magnet in dry mode without any liquid helium, or use the recondenser to build up liquid helium in the helium tank by introducing helium gas.
This ensures the possibility of long-term operation without access to cryogenic liquids. \\

The liquid helium tank has a volume of 29$\,\ell$ and can keep the system cold while transporting short distances or if external power is unavailable. 
The heat load on the second stage of the cryocooler is estimated to be around 350 mW in thermal equilibrium, and it has an additional $\sim1.2\,$ W heat load through the cold head when the cryocooler is switched off. 
As a result, we anticipate a liquid helium holding time of eight hours after the cryocooler is switched off. 
This is sufficient to transport the apparatus from the AD hall to a transport vehicle or even another laboratory at CERN. 
Continuous operation of other essential equipment during the transport, such as the trap voltage power supply (15 W) and temperature, pressure, and liquid helium sensors (200 W) is provided by batteries and uninterruptible power sources. \\

A highly homogeneous magnetic field is essential for the image-current detection of the trapped antiprotons, since frequency shifts that scale with the magnetic field gradients \cite{Ket14} would compromise the detection and particle cooling in the trap. 
Consequently, the magnet is designed to have a homogeneous center that is 150 mm long and 5 mm in diameter with low magnetic field gradients. 
To this end, the main coil and two sets of shim coils are connected in a single current loop as shown in Fig.~\ref{fig:BoverSC}. 
This geometry of the coils minimizes the first and second magnetic field gradients along the main axis of the coils. The magnetic flux density on the surface of the superconducting coil is only 2$\,\%$ higher than the field in the coil center. \\ 

\begin{figure}
    \centering
    \includegraphics[width=\linewidth]{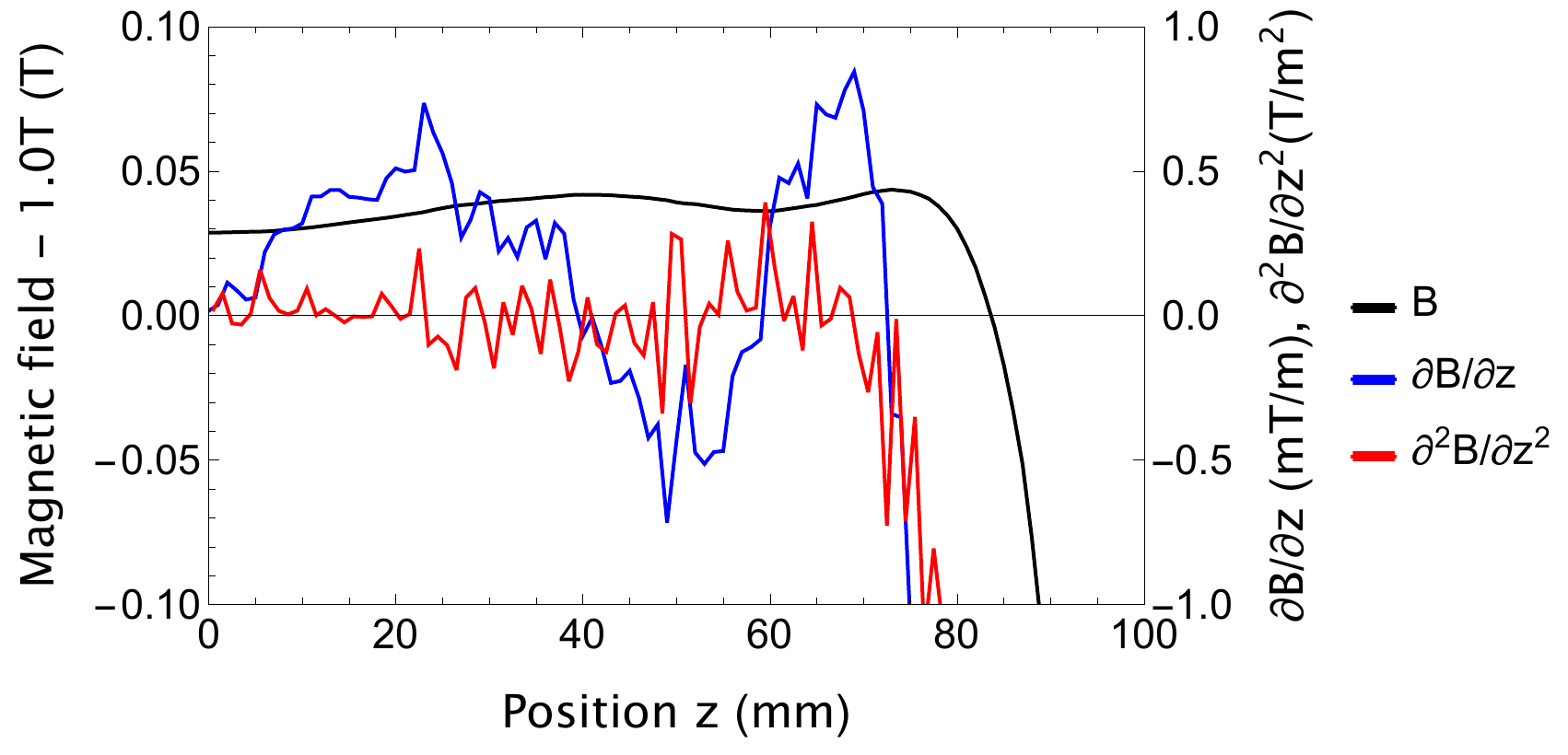}
    \caption{Calculation of $B_z$ (black, left axis), $dB_z/dz$ (blue, right axis) and $d^2B_z/dz^2$ (red, right axis) along the center axis of the coil in the homogeneous part. $z=0$ is the center plane of the coil. The maximum values of the $dB_z/dz$ and $d^2B_z/dz^2$ within the specified homogeneous region $\left|z\right|<75\,$mm are 0.85$\,$mT/m and 0.64$\,$T/m$^2$. The numeric fluctuations of the gradients are higher than for the magnetic field due to the numeric methods involved.}
    \label{fig:Bfield}
\end{figure}

Further, the calculation of the magnetic field gradients shown in Fig.~\ref{fig:Bfield} estimates that the linear and quadratic componentents of this coil configuration are expected to be below \SI{1}{\milli\tesla/\meter} and \SI{1}{\tesla/\meter^2}, respectively, and comparable with those of the first-generation BASE antiproton precision measurements \cite{Ulm15,Smo17}.
As a result, image-current detection and particle cooling can be performed in the entire homogeneous volume without relevant line-width contributions from the residual magnetic field inhomogeneity. 
Similarly, the magnet has a sufficiently strong field of \SI{1}{T} to enable efficient ($\tau \sim \SI{2.6}{\second}$) cooling of electrons via cyclotron radiation - a crucial ingredient for trapping antiprotons with $\sim\,\SI{100}{keV}$ initial energy \cite{Bar18}. 
To ease the transportation, the magnet will be operated in persistent mode so that a transportable current source is not required. 
To this end, a persistent current switch is installed on the 4 K stage, and the current leads are connected with permanently-installed high-temperature superconductors up to the \SI{50}{\kelvin} heat shield. 
The operating current to charge the magnet to 1$\,$T magnetic field is 33.27$\,$A, which is less than 25$\%$ of the critical current. \\

The superconducting magnet system is specifically designed to withstand mechanical stress during transport, including while being cold and in operation. 
Conventional NMR-magnets are shipped warm and uncharged, usually with additional removable mechanical support parts to prevent damage to the fragile cold stage from acceleration and shocks. 
The mechanical construction of our transportable trap cryostat employs a support structure for the cold stages that can withstand acceleration up to 1$\,g$ in all directions as required by CTU road shipping standards (packing of Cargo Transport Units). 
As the magnet is in operation during transport, the stray field needs to be suppressed below 0.5\,mT for safety reasons. 
This is achieved by making the vacuum vessel of the magnet out of carbon steel. \\

\subsection{Trap system}
\begin{figure*}
    \centering
    \includegraphics[width=\linewidth]{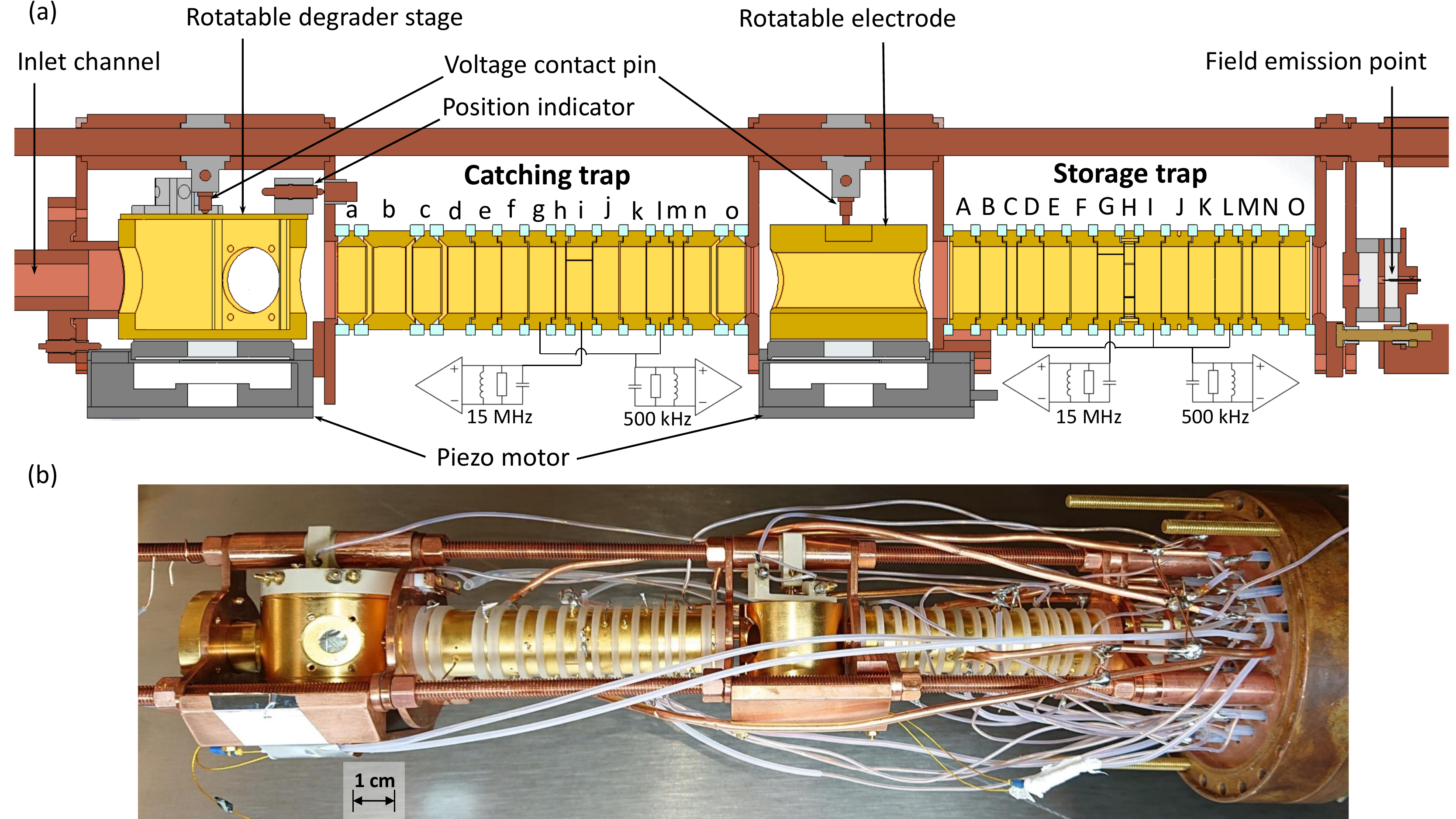}
    \caption{(a) Technical illustration of the trap system. The catching trap and the storage trap electrode stacks are shown and the individual electrodes are labeled in lower case (a-o) and upper case (A-O) letters, respectively. The rotatable degrader stage and the rotatable electrode are shown including the piezo motors driving the rotation stages, the voltage contact pins and one of the position indicators. The inlet channel for the antiprotons is shown on the left end and the field emission point for electrons on the right. The electric circuit diagrams indicate the connections of the axial (500$\,$kHz) and cyclotron (15$\,$MHz) image-current detectors to the traps. Color representation: Gold-plated OFE copper (dark yellow) with cutouts (yellow), uncoated OFE copper (brown) with cutouts (light brown), sapphire (light blue), other insulators (light gray), and the piezo motors (dark gray). (b) Photo of the trap stack. On the right side, the trap stack is mounted on the pin base flange with the feedthroughs of the cryogenic trap chamber to the electronics section. On the left side, the inlet channel that extends up to the trap chamber flange is not shown. For details see text.}
    \label{fig:STEPTRAP}
\end{figure*}

The trap system is composed of two independent, coaxial, electrode stacks shown in Fig.~\ref{fig:STEPTRAP}, and is placed in the cryogenic vacuum chamber in the magnet bore. 
One electrode stack forms the catching trap (CT), the second the storage trap (ST). 
The electrode stacks are made based on an established procedure from gold-plated oxygen-free electrolytic (OFE) copper and are separated by sapphire rings with machining tolerances $\leq$ 10 $\rm{\mu}$m \cite{Smo15}. This ensures good thermal conductance and low dissipation losses in the image-current detection circuits.
Compared to the BASE trap system with 9 mm diameter \cite{Smo15} traps, the trap diameter was increased to 12 mm to provide a larger harmonic trapping region and a larger catching volume for the antiproton injection from ELENA. \\
%A photo of the assembled trap electrodes is shown in in Fig.~\ref{fig:STEPTRAP} b). \\

The purpose of the CT is to form an interface between ELENA, the ST, and other external trap systems. 
Therefore, the CT has to manage the initial antiproton catching and cooling \cite{Smo15} of the ELENA beam. 
This requires the transmission of the 100 keV antiprotons through a degrader foil to reduce their kinetic energy so that they can be captured in the trap system by high-voltage pulses ($\left|\Delta U\right|\leq 2\,$kV). 
Moreover, the CT is also used to prepare cold antiproton clouds and separate fractions of the stored antiprotons down to single particles.
These can subsequently be transferred into the ST for long-term storage.
Further, the CT works as an ``airlock" during the antiproton ejection into external trap systems to reduce the risk of antiproton loss in the ST due to residual gas exposure. 
It is designed to operate for a short time at a higher local pressure when the window to the inlet channel is opened while closing the connection to the ST.\\

To achieve all these functions, we placed two rotatable elements around the CT.
On the side of the inlet channel, a rotatable degrader stage (see Fig.~\ref{fig:STEPTRAP}) can switch between several elements on the trap-axis: a degrader foil for injection, a thick copper surface to block residual gas during storage, and an open channel for the ejection of low-energy antiprotons into another trap system. 
Between CT and ST, a rotatable electrode with a channel for particle transmission is placed.
Both are operated indiviually by cryogenic piezo motors (\textit{Smaract SR-2812}).
They are also equipped with spring-loaded pins as a position indicators, which close an electrical contact at each designated position. The contacts have each a characteristic electrical resistance so that we can determine the position of both rotation stages by resistance measurements. 
Both rotatable electrodes are insulated via a 0.3 mm PTFE sheet between the electrode and the motor and biased via a spring-loaded pin, so that voltages can be applied during catching or transport. \\

During the antiproton injection, the degrader foil is placed on the trap axis. 
It is an aluminum-coated Mylar\textregistered~foil and its thickness is matched to the required stopping power of the antiprotons, so that the low-energy tail of the transmitted antiprotons with a kinetic energy below 2 keV can be stopped by high-voltage pulses in the CT. 
The two fast-switching, high-voltage electrodes for the catching pulses are located at both ends of the CT, electrodes \textit{a} and \textit{o} in Fig.~\ref{fig:STEPTRAP}(a), and a fast low-noise high-voltage switch can apply a pulse to -2 kV. 
Subsequently, sympathetic electron cooling \cite{Gab89EC,Smo15} can be applied to accumulate antiprotons in the central trap electrode \textit{h}, which forms together with the surrounding electrodes \textit{f} to \textit{j} a cylindrical five-electrode Penning trap in a compensated and orthogonal design \cite{Gab89}. 
Trapped particles in the resulting harmonic potential move with the well-defined motional frequencies of the Penning trap, can be manipulated by radiofrequency drives\cite{Bla06}, and detected and resistively cooled via image-current detectors \cite{Ulm13,Nag16}. 
Axial excitation and axial-radial sideband coupling signals can be either applied to dedicated electrodes, e.g.~to electrode \textit{f} and to the radially-segmented electrode \textit{i}, respectively, or through the slits between the electrodes by wires placed in the vicinity of the electrodes. 
An image-current detection system picks up signals of the axial mode on electrode \textit{g}, whereas a cyclotron detection system is connected to the second segment of electrode \textit{i}. 
We will use this trap configuration in the CT to apply electron and contaminant cleaning procedures, and to prepare a clean antiproton cloud that is cooled to the temperature of the image-current detectors, typically close to 4.2$\,$K. 
More details on this procedure were reported earlier \cite{Smo15}. \\

\begin{figure*}
    \centering
    \includegraphics[width=\linewidth]{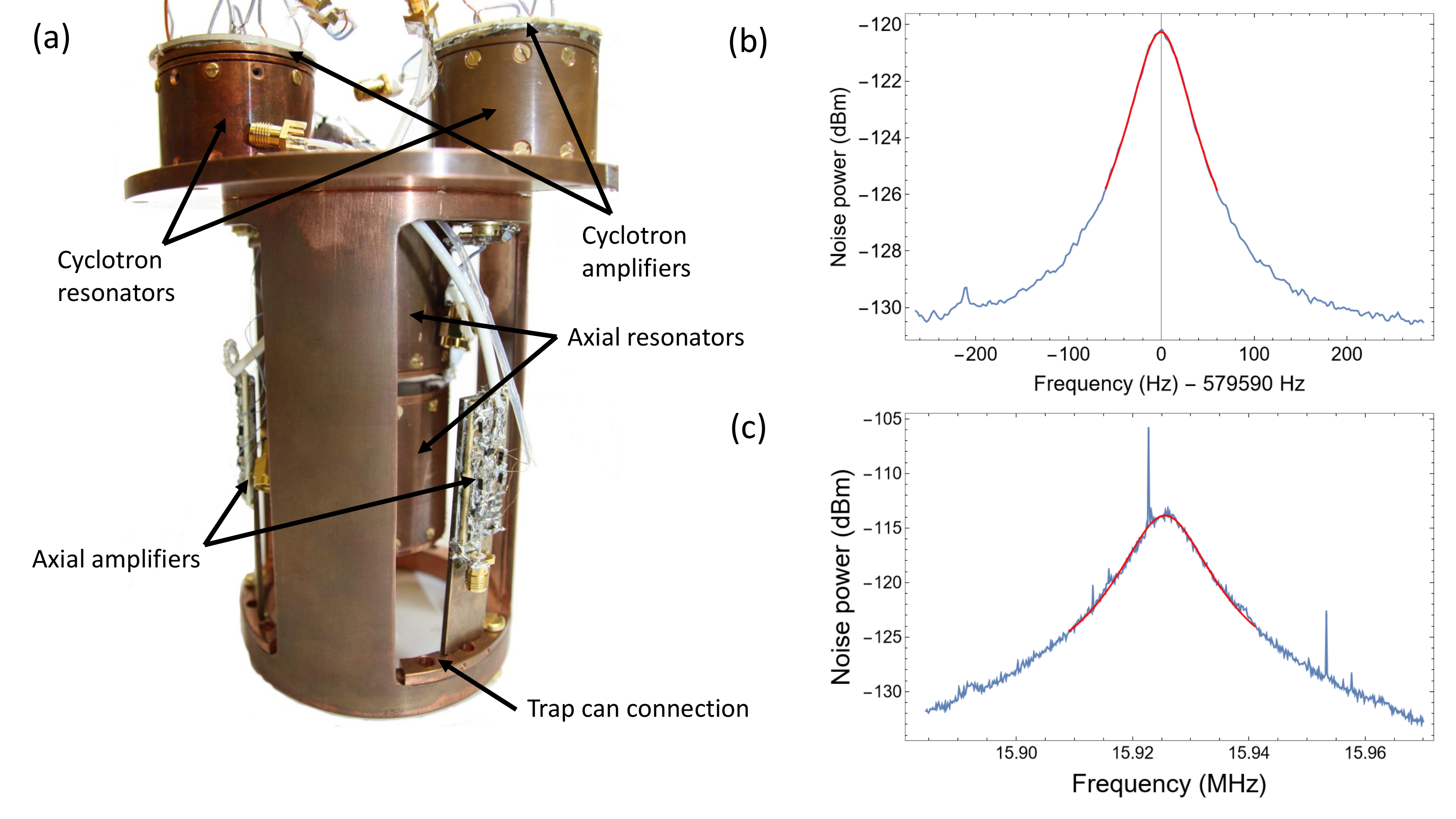}
    \caption{(a) Photograph of the detection segment of the BASE-STEP trap system. (b) Noise spectrum of one axial detector with a $Q$-value of 9200. (c) Noise spectrum of a cyclotron detector with a toroidal coil at 15.9 MHz, close to the antiproton modified cyclotron frequency. The $Q$ value of the detector is 1700. In (b) and (c) the blue line shows the measured noise spectrum, and the red curve a Lorentzian curve fitted to the data to extract the $Q$ value. The spectrum of the cyclotron detector shows several peaks due to noise signals that were pickup up in the cold-head cryostat.}
    \label{fig:STEPDET}
\end{figure*}

While particles are being manipulated in the CT, the inlet channel is closed to suppress the residual gas flow into the trap chamber and to increase the monolayer formation time (see details below). 
To this end, the wall of the degrader stage housing is rotated in front of the inlet channel. 
Note that the closed position of the degrader stage is not completely sealed, but a small channel of 0.1$\,$mm is left between the rotating part and the end of the inlet channel, so that the rotating stage can operate with minimal torque. 
Among the most essential manipulation steps is the application of separation and merging schemes that non-destructively separate and combine small clouds of antiprotons with the main trap content, respectively \cite{Smo15b}. 
Consequently, we can transfer and eject even single antiprotons, which decreases the risk of losing the entire trap population in a single operation.
To implement the separation and merging sequences in the CT, we require several locations where particles can be trapped. Since the potential ramps of these sequences typically form anharmonic potential wells so that the image-current signals cannot be observed during the procedure, we have included a second harmonic trapping region that enables detection at the end or start of the separation or merging procedure, respectively. 
This second harmonic trap is formed by the electrodes \textit{k} to \textit {o}, and we connect the axial detection system for the CT in addition also to electrode \textit{l}. 
Thereby, we can count the particles after separation in both potential wells simultaneously, which eliminates additional transport ramps and a sequential counting measurement from our previous procedure \cite{Smo15b}.
The transfer of particles into the ST is managed by the electrodes \textit{m}, \textit{n} and \textit{o} in the CT and the electrodes \textit{A}, \textit{B} and \textit{C} in the ST. 
The rotatable electrode that is placed between the CT and ST is opened for this procedure. 
As in our previous measurements, we will first move the antiprotons via adiabatic ramps of the axial trapping potential along the magnetic field lines using slow voltage ramps with the velocity set by the time constant of our low-pass filters, $\tau \approx $\SI{100}{\milli\second}, and store the particles in electrode \textit{n}. 
The electrodes \textit{o} and \textit{A} are biased via low-pass filters with a diode-bridged resistor so that they can be pulsed on a timescale of \SI{100}{\nano\second} to transfer the antiprotons with well-timed pulses as a bunch. 
This method is routinely used in other Penning-trap experiments, for example, SHIPTRAP has two traps that are similarly separated by a differential pumping barrier in the same superconducting magnet \cite{Blo05} and the transfer of ions between the traps uses low acceleration voltages, typically less than $100\,$V \cite{Blo15}. 
Our electrodes used to transfer ions are equipped with a biasing network for \SI{200}{\volt} except for electrode \textit{o} that is also intended for operation at \SI{-2}{\kilo\volt}  for the injection procedure. \\

A similar transfer procedure is implemented for the ejection out of the magnetic field into an external trap system except that the transport is conducted at higher kinetic energies. 
This ejection procedure, in which the antiprotons exit the magnetic field, is similarly applied to ions of radionuclides in the ISOLTRAP experiment at the ISOLDE facility, in which electrostatic transport between two traps in separate superconducting magnets is performed with around 3 $\,$keV kinetic energy \cite{Muk08}.
For the ejection, a small fraction of the stored antiprotons is cooled in the CT and transported into 
the electrodes \textit{a}, \textit{b} and \textit{c}. 
To accelerate the antiprotons to the transport kinetic energy of $\sim 3\,$keV, these electrodes are designed as high-voltage electrodes for up to $\pm$3 kV potential limited by the cryogenic high-voltage feedthroughs, and electrode \textit{a} is pulsed open for the ejection.
Before the ejection pulse, the other remaining antiprotons are stored in the ST, and the rotatable electrode between CT and ST is closed before the open channel of the rotatable degrader stage is aligned with the trap axis. 
This avoids annihilation in the ST with incoming residual gas. 
The ejected antiprotons need to pass the differential pumping section at much lower kinetic energy than during the ejection, and there is a lack of magnetic-field free regions for placing electrostatic steering elements that are capable of adjusting the trajectory and refocus the ejected pulse. 
Such elements can be only placed outside of the transportable BASE-STEP setup, and the only parameter that significantly impacts the collimation of the ejected pulse is the choice of the potentials of electrodes \textit{a} and \textit{b} that define the radial electric field during the acceleration. From simulations we estimate an ejection efficiency of $30\,\%$ up to the valve at the entrance of the the room temperatue vacuum chamber without additional measures.
We are presently evaluating several possibilities to increase the ejection efficiency, but the details are beyond the scope of this paper. One option is to add a magnetic solenoid field to limit the expansion of the antiproton pulse in the differential pumping section. Another possibility would be to look into an ion carpet design \cite{Ham16} that interfaces well with the Penning trap and use the ion surfing technique \cite{Bol11} to transmit the antiprotons through the differential pumping channel. \\

The ST is the trap that is best protected from the incoming residual gas and is used as long-term storage trap region for the antiprotons. 
The ST electrode stack features three harmonic trapping regions with the electrodes \textit{C}, \textit{H}, and \textit{M} as trap centers, and the primary storage region centered around electrode \textit{H} is similar to the one of the CT except for the segmented ring electrode \textit{H} that can apply a rotating wall potential to radially compress larger clouds of antiprotons \cite{Surko2006}. 
The separation and merge schemes can also be applied in the ST to control the number of antiprotons that are transferred into the CT.
Since only two trap regions are required for this procedure, the third harmonic region can implement additional functions, such as a backup antiproton reservoir or, in the scope of an antiproton lifetime measurement, a highly-charged ions trap for an independent vacuum measurement \cite{Sel17,Mic19}. 
To this end, an electron gun is mounted on the far end of the ST. It primarily produces cold electrons for the initial catching procedure \cite{Smo15}, however it can also be used for electron impact ionization or charge-breeding of trapped ions.\\

\begin{figure*}[!t]
    \centering
    \includegraphics[width=\linewidth]{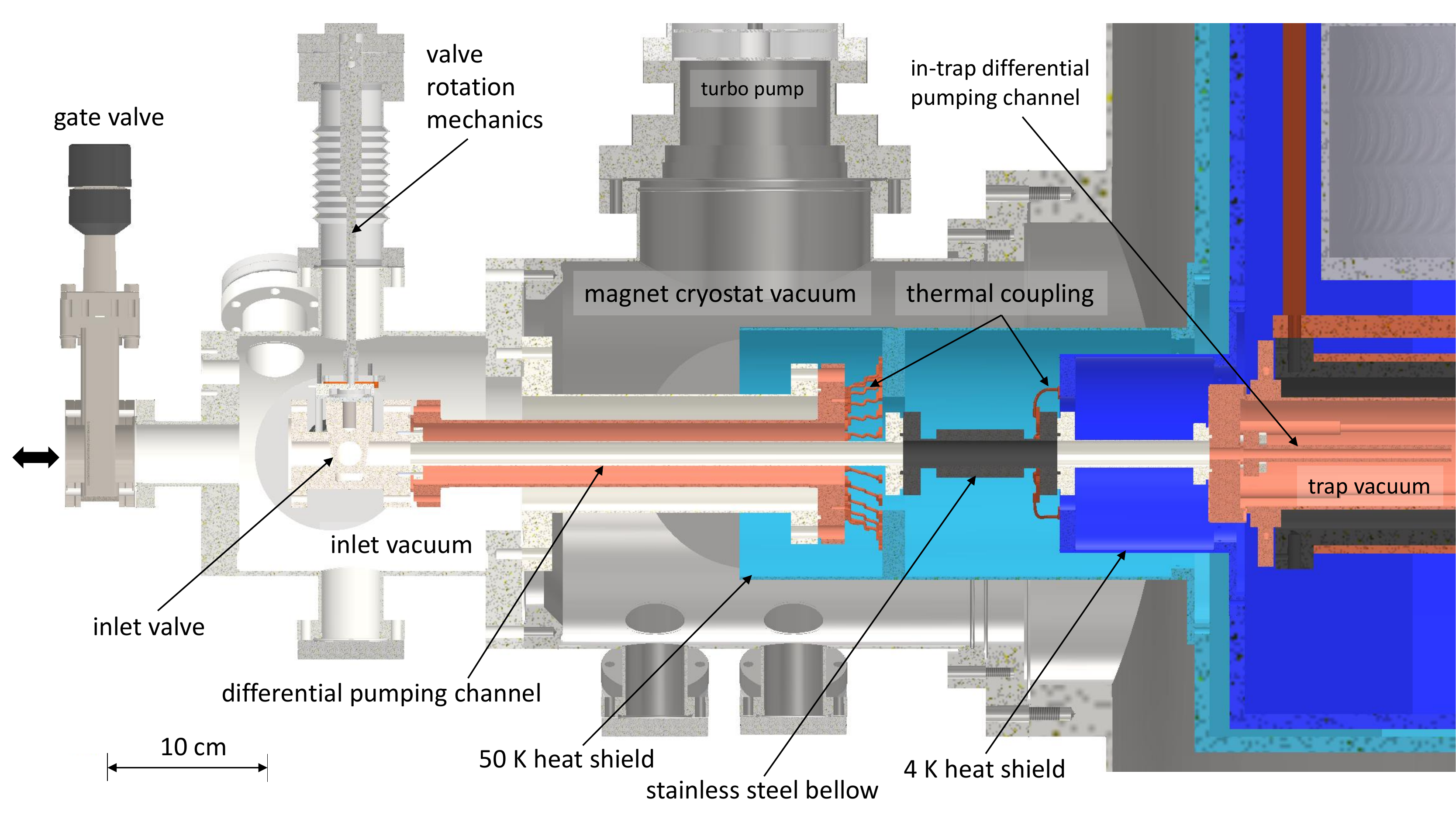}
    \caption{The differential pumping segment is shown with connections to the trap chamber on the right and to the injection/ejection port on the left. See text for details.
    }
    \label{fig:diffpump}
\end{figure*}

The non-destructive detection systems form also an essential part of the trap system.
These are employed for frequency measurements, cooling, and to count and monitor the number of trapped particles \cite{Sel17}, which is essential to characterize the particle separation procedures \cite{Smo15b}. 
For the BASE-STEP trap system, we have developed the detection segment shown in Fig.~\ref{fig:STEPDET}, that contains two axial detectors near 500 kHz \cite{Nag16} and two cyclotron detectors \cite{Ulm13} near 15 MHz that will provide the image-current detection of these two modes in each trap stack. 
The detectors consist of a superconducting toroidal coil inside a cylindrical copper housing and a low-noise cryogenic amplifier using a dual-gate GaAs FET with high input impedance as input stage \cite{Ulm13}. 
Up to now, BASE has used only the axial detection systems for catching and counting the number of trapped particles, and the BASE-STEP trap system has the cyclotron detectors for additional diagnostics. 
A novel aspect of our detection unit is that we use toroidal superconducting coils for the cyclotron detectors, which were previously implemented using copper and NbTi solenoids \cite{Ulm13, Smo15}. We show as example of the performance a resonance with $Q \sim 1700$ during tests in a cold-head cryostat, see Fig.~\ref{fig:STEPDET}(c). 
The cyclotron detectors will also provide additional information for antiproton lifetime measurements, which requires to resolve the large numbers of trapped particles as function of time.
The required averaging time of the axial detectors increases with $N^2$ which limits single-particle resolution for large clouds \cite{Sel17}, and peak counting or integral power measurements with the cycltron detector provide an alternative way to determine the number of trapped antiprotons. \\

\subsection{Differential pumping section}
An excellent vacuum in the cryogenic trap chamber is critical while transporting antiprotons. 
In concrete terms, a residual gas pressure of $10^{-10}$\,mbar results in a trap lifetime of only $\sim \SI{5}{\second}$ and we require a vacuum of at least $10^{-16}$\,mbar for a trap lifetime of about three months \cite{Fei90}. 
So far, BASE has injected antiprotons through thin vacuum windows into a hermetically-sealed cryogenic vacuum chamber \cite{Smo15}. Under these conditions residual gas pressure can only be measured via the antiproton trap lifetime and we determined an upper limit of less than $10^{-18}\,$mbar after storing antiproton clouds up to 405 days \cite{Sel17}. 

As a transportable trap requires an open trap system to transfer low-energy antiprotons into another trap system, the trap vacuum will be compromised by residual gas entering through the antiproton transfer channel. 
Initially, the trap chamber conditions will be similar to the closed cryogenic setup, since it will be cooled down after lowering the pressure below $10^{-6}$\,mbar at room temperature, which results in a low coverage ($\lesssim 1\%$) of the trap surfaces from the freeze out and efficient cryopumping to ultra-low pressure. 
However, the residual gas entering the trap chamber through the differential pumping section forms a low density atomic beam and can therefore cause annihilation by directly passing through the trap center before attaching to a cold surface. 
Further, the pressure in the trap chamber will rise after a monolayer of residual gas has formed on its surfaces. 
This causes the cryopumping to be no longer effective for hydrogen molecules and helium at 4$\,$K. Based on the molecular flow through a channel with conductance $C$, we obtain the monolayer formation time as \cite{Nai18}: 
\begin{eqnarray}
\tau_M = \frac{A_{\mathrm{trap}}}{A_{\mathrm{mol}}}\frac{k_B T}{\Delta P\,C},
\end{eqnarray}
where $A_{\mathrm{trap}}$ is the surface area of the cryogenic trap chamber, $A_{\mathrm{mol}} \sim (\pi/4) (300\,$pm$)^2$ is the area covered by a residual gas particle based on its kinetic diameter \cite{Fre06,Ism2015}, $k_B$ the Boltzmann constant, $\Delta P$ the pressure difference along the channel, and $T$ the temperature of the residual gas particles. \\

To maintain cryopumping, $\tau_M$ needs to be at least about the desired antiproton storage time. 
Therefore, we must minimize $\Delta P$ by making excellent vacuum conditions at the inlet to the differential pumping tube. 
To this end, we place a dedicated inlet chamber as last element of the room temperature vacuum chambers, see Fig.~\ref{fig:diffpump}. 
The inlet chamber is evacuated using two turbo molecular pumps in series and in parallel a non-evaporative getter (NEG) pump (pumps not shown in Fig.~\ref{fig:diffpump}). In addition, the inlet chamber has been vacuum-fired to reduce hydrogen outgassing. We aim for an operating pressure of $10^{-10}\,$mbar or better.
We also need to separate the inlet vacuum chamber from the insulation vacuum of the magnet cryostat to reach this condition, since the insulation vacuum typically has even under cryogenic conditions a higher operating pressure of $\sim \SI{1e-8}{mbar}$ due to the presence of multi-layer insulation foil and outgassing from electronics. \\

The trap chamber is connected to the inlet chamber via a series of differential pumping channels to minimize the conductance $C$. 
Starting from the inlet chamber, the first element of the differential pumping section is the inlet valve which is opened for the injection and ejection of antiprotons, and closed otherwise to increase the monolayer formation time in the cryogenic vacuum chambers. 
The inlet valve is made of a low heat-conductance titanium alloy, grade 5 titanium, and consists of a rotatable, conical stem in a titanium valve body with a free aperture of \SI{16}{mm} diameter and \SI{77.5}{mm} length in the open state. 
Following this, the residual gas must pass through a section with \SI{16}{\milli\meter} inner diameter and \SI{500}{mm} length, which is made from two grade 5 titanium tubes separated by a stainless steel bellow which compensates the mechanical tension due to the thermal contraction between the magnet assembly and differential pumping tubes. %, as shown in Fig.~\ref{fig:diffpump}.
The titanium tubes also serve as a resistance for the thermal conduction between the inlet valve, which is connected to the first stage of the cryocooler, and the trap vacuum chamber at 4 K.
This section is followed by the trap chamber flange with \SI{20}{mm} thickness and an \SI{8}{mm} diameter hole, and by a trap-chamber internal differential pumping section with \SI{6}{mm} inner diameter and \SI{135}{mm} length as shown in Fig.~\ref{fig:diffpump}. 
This channel is already in the magnetic field of the magnet where the antiproton beam diameter is constrained by the cyclotron motion, so that the channel diameter can be smaller. 
At 4.2 K, the conductance through the trap-chamber internal channel is about 0.09$\,\ell$/s and 0.07$\,\ell$/s for hydrogen molecules and helium, respectively, whereas the combined preceding elements have about 20-fold higher conductance.
%and the inlet valve at temperature $T_{\mathrm{inlet}}$, $\SI{4}{K} < T_{inlet} < \SI{300}{\kelvin}$.\\
Based on the operating pressure difference $\Delta P$ and the combined conductance $C = 0.16\,\ell$/s, we estimate the monolayer formation time to be $\tau_{M,0} \approx 430\,$days. 
For further protection of the antiproton reservoir, we have added the inlet valve at the entrance of the differential pumping tube. 
To close the inlet valve, a bellow with a rotary feedthrough is used to form a retractable key that turns the conical stem in the valve body between two defined positions and is withdrawn afterwards. Otherwise, its conductive thermal load of about 350 mW would increase the heat load of second stage of the pulse-tube cooler. 
In the closed state, the residual gas leaks between the two polished titanium surfaces into the differential pumping tube, since we have chosen a simple design without additional sealing elements. 
We have considered solutions with PTFE seals to obtain lower leak rates, however, the seal requires pressure on the sealing surface to function, requiring in turn a higher torque to turn the stem. 
We decided to operate with a low-friction design to ensure reliable operation. \\

\begin{figure}
    \centering
    \includegraphics[width= \linewidth]{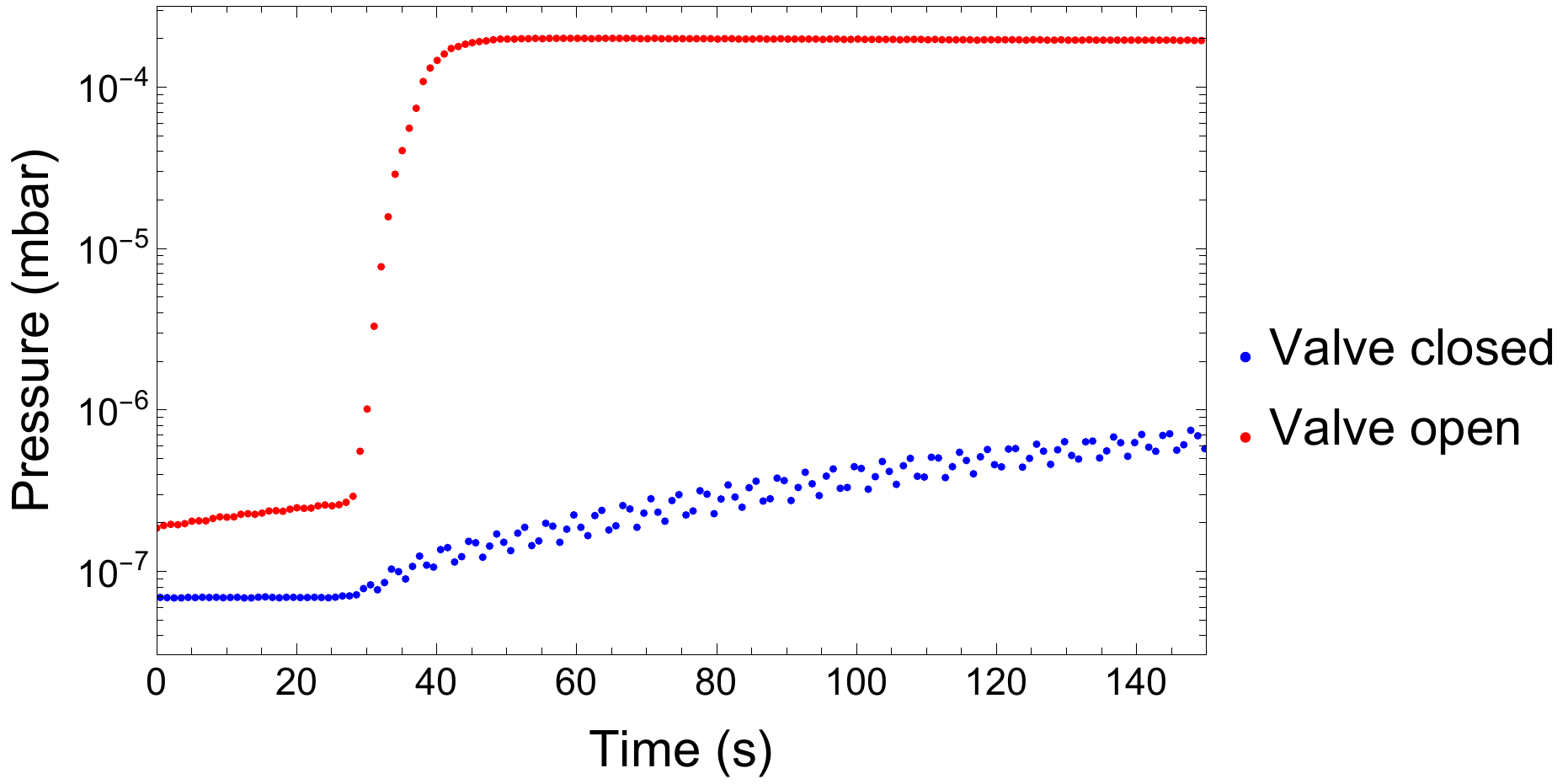}
    \caption{Pressure increase in the coldhead chamber behind the inlet valve after increasing the pressure in the inlet chamber to $6\times10^{-3}\,$mbar. In the open state, pressure equilibrium is reached after 15$\,$s, whereas the time constant for the pressure increase is estimated to be a factor 4000 lower in the closed state. 
    }
    \label{fig:valvetest}
\end{figure}

To characterize the impact of the inlet valve on the monolayer formation time, we determined the ratio of the valve conductance in the opened state to the closed state from measurements as shown in Fig.~\ref{fig:valvetest}. 
To this end, the inlet valve with its vacuum chamber and the differential pumping barrier was installed in a cold-head cryostat so that the vacuum of the inlet chamber and the cold head were separated by the inlet valve. 
The thermal couplings in Fig.~\ref{fig:diffpump} were connected by copper rods to the first and second stages of the cryocooler. 
The pressure of the inlet chamber and the cold head chamber was pumped for each test to below $5\times10^{-5}\,$mbar and $5 \times 10^{-7}\,$mbar, respectively. 
Subsequently, the valves to the pumps were closed so that only cryopumping remained active in the cold head chamber, but helium was no longer pumped out from the setup. 
We then injected a burst of helium gas into the inlet chamber that raised the pressure in the inlet chamber increasing the pressure in the inlet chamber to values in between $10^{-4}\,$mbar to 10$^{-1}\,$mbar. 
 In Fig.~\ref{fig:valvetest}, the helium injection occurs at around $t=30\,$s, and the rate of the following exponential pressure increase in the cold head chamber is proportional to the conductance of the valve.
 We compared the conductance ratio in the opened and closed state for several different conditions. 
We estimate the conductance in the closed state to be at least a factor 1000 lower than in the open state. 
However, we note that this test was conducted at a pressure above the molecular flow regime, and several other effects limit the accuracy of the measurement, e.g.~the finite volume of the inlet chamber, outgassing rates, tempeature changes, etc.~so that this can be regarded only as an estimation for the operation under realistic conditions. 
Nevertheless, our estimation projects an increase of the monolayer formation time by an additional factor of three if we also consider the low conductance of the other differential pumping tubes. \\

The direct exposure of the antiproton trapping regions to incoming residual gas particles is most of the time suppressed in comparison to the open state by the inlet valve and the two other rotatable electrodes in the trap stack. During the injection of antiprotons from ELENA ($E_{\rm{kin}}\sim 100\,$keV), the antiprotons pass through a degrader foil that blocks the incoming gas from entering the CT.
However, when antiprotons need to be ejected from the trap system at low energy ($E_{\rm{kin}}\sim 2\,$keV) a direct opening to the CT must be realized, and in this case we expect the highest annihilation rate of the trapped antiprotons in the CT.
To estimate the loss rate in this configuration, we conducted a MOLFLOW+ simulation to estimate an upper limit for the residual gas flow into the CT center. 
We simulated the pressure conditions under the assumption that the monolayer formation in the differential pumping channel is already completed, but the trap chamber surfaces are still perfectly cryopumping. 
We find the local pressure of residual gas particles at the position of the CT center to be 10$^{-13}\,$mbar, a factor 1000 lower than the inlet pressure of 10$^{-10}\,$mbar, which results in an effective storage time of about 1.5 hours. 
Since the time window required to eject the antiprotons can be less than one minute, we conclude that it is possible to perform the transfer without significant annihilation losses in the CT.
 
The differential pumping section also serves as vacuum separation of the cryostat isolation vacuum and the inlet and trap chamber vacuum. 
To this end, the differential pumping channel is mechanically connected to the room temperature vacuum chamber as part of a series of concentric tubes, see Fig.~\ref{fig:diffpump}. 
To manage the thermal load, the outer and the inner tubes are made of low heat-conductance grade 5 titanium, and the end of the outer tube is thermally connected to the first-stage heat shield ($\sim $\SI{50}{\kelvin}) of the magnet. 
An OFE copper tube in between both tubes connects both and sinks the radiative heat load on the inlet valve to the 50 K heat shield. Finally, a thermal anchor to the second stage heat shield of the magnet intercepts the remaining heat load and prevents the heating of the trap chamber and the coil body. The estimated heat loads on the first stage and second stage of the cryocooler are 6 W and 80 mW, respectively. 

\section{Outlook}

We have reported the design of the BASE-STEP transportable antiproton trapping apparatus that will be installed in the AD/ELENA facility at CERN.
Transportable traps are a promising development in precision measurements on antiprotons, and potentially also for measurements on other accelerator-produced ions with a long lifetime.
For precision measurements and fundamental physics studies, the transportable trap will be operated in a similar way to other external ion sources that are presently in use \cite{Mic18,Stu19}. \\

In particular, we anticipate that frequency ratio measurements on antiprotons stored in precision Penning traps will greatly benefit from reduced magnetic field noise, and that relocating the antiproton measurements away from the AD/ELENA facility to low-noise laboratories will improve tests of CPT symmetry in the baryonic sector.
Although transporting antimatter to distant laboratories has long been desired, it is a significant technical challenge that benefits from recent improvements in ion trap technologies. 
In particular, we conceived a careful design of the cryogenic vacuum system that will enable long antiproton storage times. 
Pending an initial demonstration of transporting particles using the BASE-STEP apparatus, we aim to improve direct limits on the antiproton lifetime that constrain dark decay channels of antiprotons \cite{Sel17}.  
Further, transportable traps would also enable a dedicated search for millicharged dark matter particles in underground laboratories. Particles with less than a $10^{-4}$ fraction of the electron charge require a large overburden ($\sim$ 1 km) to stop, thermalize, and ultimately scatter with a trapped antiproton \cite{Bud22}. Placing the transportable trap inside a high-voltage platform to accumulate or expel millicharged particles in an underground laboratory would form a novel dedicated type of dark matter search.
Further, it becomes possible to perform the first distant, simultaneous frequency measurements with trapped antiprotons using the transportable STEP apparatus and the current stationary BASE apparatus in the AD hall, as direct antiparticle test of Lorentz invariance\cite{Din19}, and to implement domain-wall searches of dark matter, such as implemented with magnetometers \cite{Afa21} or optical clocks \cite{Der14}, but with an antiparticle probe instead. \\

As a long-term benefit of developing transportable traps, it becomes possible to supply multiple low-energy antiproton experiments beyond the space available in the AD/ELENA facility, with the constraint that these experiments have to be conducted non-destructively, with a low antiproton consumption rate, or a low reloading frequency.

\begin{acknowledgments}
We are in particular thankful for the support by CERN for supporting the implementation of BASE-STEP in the AD/ELENA facility and the support by all CERN teams contributing to antiproton experiments. We acknowledge financial support by the ERC (starting grant no. 852818), RIKEN, the Max-Planck Society, the Max Planck, RIKEN, PTB Center for Time, Constants, and Fundamental Symmetries (C-TCFS), and the QUANTUM group at the Institute of Physics in Mainz. 
\end{acknowledgments}

\bibliographystyle{apsrev4-1mod.bst}
\bibliography{aipsamp}% Produces the bibliography via BibTeX.

\begin{thebibliography}{92}%
\makeatletter
\providecommand \@ifxundefined [1]{%
 \@ifx{#1\undefined}
}%
\providecommand \@ifnum [1]{%
 \ifnum #1\expandafter \@firstoftwo
 \else \expandafter \@secondoftwo
 \fi
}%
\providecommand \@ifx [1]{%
 \ifx #1\expandafter \@firstoftwo
 \else \expandafter \@secondoftwo
 \fi
}%
\providecommand \natexlab [1]{#1}%
\providecommand \enquote  [1]{``#1''}%
\providecommand \bibnamefont  [1]{#1}%
\providecommand \bibfnamefont [1]{#1}%
\providecommand \citenamefont [1]{#1}%
\providecommand \href@noop [0]{\@secondoftwo}%
\providecommand \href [0]{\begingroup \@sanitize@url \@href}%
\providecommand \@href[1]{\@@startlink{#1}\@@href}%
\providecommand \@@href[1]{\endgroup#1\@@endlink}%
\providecommand \@sanitize@url [0]{\catcode `\\12\catcode `\$12\catcode
  `\&12\catcode `\#12\catcode `\^12\catcode `\_12\catcode `\%12\relax}%
\providecommand \@@startlink[1]{}%
\providecommand \@@endlink[0]{}%
\providecommand \url  [0]{\begingroup\@sanitize@url \@url }%
\providecommand \@url [1]{\endgroup\@href {#1}{\urlprefix }}%
\providecommand \urlprefix  [0]{URL }%
\providecommand \Eprint [0]{\href }%
\providecommand \doibase [0]{http://dx.doi.org/}%
\providecommand \selectlanguage [0]{\@gobble}%
\providecommand \bibinfo  [0]{\@secondoftwo}%
\providecommand \bibfield  [0]{\@secondoftwo}%
\providecommand \translation [1]{[#1]}%
\providecommand \BibitemOpen [0]{}%
\providecommand \bibitemStop [0]{}%
\providecommand \bibitemNoStop [0]{.\EOS\space}%
\providecommand \EOS [0]{\spacefactor3000\relax}%
\providecommand \BibitemShut  [1]{\csname bibitem#1\endcsname}%
\let\auto@bib@innerbib\@empty
%</preamble>
\bibitem [{\citenamefont {Dine}\ and\ \citenamefont {Kusenko}(2003)}]{Din03}%
  \BibitemOpen
  \bibfield  {author} {\bibinfo {author} {\bibfnamefont {M.}~\bibnamefont
  {Dine}}\ and\ \bibinfo {author} {\bibfnamefont {A.}~\bibnamefont {Kusenko}},\
  }\href {\doibase 10.1103/RevModPhys.76.1} {\bibfield  {journal} {\bibinfo
  {journal} {Rev. Mod. Phys.}\ }\textbf {\bibinfo {volume} {76}},\ \bibinfo
  {pages} {1} (\bibinfo {year} {2003})}\BibitemShut {NoStop}%
\bibitem [{\citenamefont {Bertone}\ \emph {et~al.}(2005)\citenamefont
  {Bertone}, \citenamefont {Hooper},\ and\ \citenamefont {Silk}}]{Ber05}%
  \BibitemOpen
  \bibfield  {author} {\bibinfo {author} {\bibfnamefont {G.}~\bibnamefont
  {Bertone}}, \bibinfo {author} {\bibfnamefont {D.}~\bibnamefont {Hooper}}, \
  and\ \bibinfo {author} {\bibfnamefont {J.}~\bibnamefont {Silk}},\ }\href
  {\doibase https://doi.org/10.1016/j.physrep.2004.08.031} {\bibfield
  {journal} {\bibinfo  {journal} {Physics Reports}\ }\textbf {\bibinfo {volume}
  {405}},\ \bibinfo {pages} {279} (\bibinfo {year} {2005})}\BibitemShut
  {NoStop}%
\bibitem [{\citenamefont {Safronova}\ \emph {et~al.}(2018)\citenamefont
  {Safronova}, \citenamefont {Budker}, \citenamefont {DeMille}, \citenamefont
  {Kimball}, \citenamefont {Derevianko},\ and\ \citenamefont {Clark}}]{Saf18}%
  \BibitemOpen
  \bibfield  {author} {\bibinfo {author} {\bibfnamefont {M.~S.}\ \bibnamefont
  {Safronova}}, \bibinfo {author} {\bibfnamefont {D.}~\bibnamefont {Budker}},
  \bibinfo {author} {\bibfnamefont {D.}~\bibnamefont {DeMille}},  \emph
  {et~al.},\ }\href {\doibase 10.1103/RevModPhys.90.025008} {\bibfield
  {journal} {\bibinfo  {journal} {Rev. Mod. Phys.}\ }\textbf {\bibinfo {volume}
  {90}},\ \bibinfo {pages} {025008} (\bibinfo {year} {2018})}\BibitemShut
  {NoStop}%
\bibitem [{\citenamefont {Hanneke}\ \emph {et~al.}(2008)\citenamefont
  {Hanneke}, \citenamefont {Fogwell},\ and\ \citenamefont {Gabrielse}}]{Han08}%
  \BibitemOpen
  \bibfield  {author} {\bibinfo {author} {\bibfnamefont {D.}~\bibnamefont
  {Hanneke}}, \bibinfo {author} {\bibfnamefont {S.}~\bibnamefont {Fogwell}}, \
  and\ \bibinfo {author} {\bibfnamefont {G.}~\bibnamefont {Gabrielse}},\ }\href
  {\doibase 10.1103/PhysRevLett.100.120801} {\bibfield  {journal} {\bibinfo
  {journal} {Phys. Rev. Lett.}\ }\textbf {\bibinfo {volume} {100}},\ \bibinfo
  {pages} {120801} (\bibinfo {year} {2008})}\BibitemShut {NoStop}%
\bibitem [{\citenamefont {Fan}\ \emph {et~al.}(2023)\citenamefont {Fan},
  \citenamefont {Myers}, \citenamefont {Sukra},\ and\ \citenamefont
  {Gabrielse}}]{Fan23}%
  \BibitemOpen
  \bibfield  {author} {\bibinfo {author} {\bibfnamefont {X.}~\bibnamefont
  {Fan}}, \bibinfo {author} {\bibfnamefont {T.~G.}\ \bibnamefont {Myers}},
  \bibinfo {author} {\bibfnamefont {B.~A.~D.}\ \bibnamefont {Sukra}}, \ and\
  \bibinfo {author} {\bibfnamefont {G.}~\bibnamefont {Gabrielse}},\ }\href
  {\doibase 10.1103/PhysRevLett.130.071801} {\bibfield  {journal} {\bibinfo
  {journal} {Phys. Rev. Lett.}\ }\textbf {\bibinfo {volume} {130}},\ \bibinfo
  {pages} {071801} (\bibinfo {year} {2023})}\BibitemShut {NoStop}%
\bibitem [{\citenamefont {Parker}\ \emph {et~al.}(2018)\citenamefont {Parker},
  \citenamefont {Yu}, \citenamefont {Zhong}, \citenamefont {Estey},\ and\
  \citenamefont {M{\"u}ller}}]{Par18}%
  \BibitemOpen
  \bibfield  {author} {\bibinfo {author} {\bibfnamefont {R.~H.}\ \bibnamefont
  {Parker}}, \bibinfo {author} {\bibfnamefont {C.}~\bibnamefont {Yu}}, \bibinfo
  {author} {\bibfnamefont {W.}~\bibnamefont {Zhong}},  \emph {et~al.},\ }\href
  {\doibase 10.1126/science.aap7706} {\bibfield  {journal} {\bibinfo  {journal}
  {Science}\ }\textbf {\bibinfo {volume} {360}},\ \bibinfo {pages} {191}
  (\bibinfo {year} {2018})}\BibitemShut {NoStop}%
\bibitem [{\citenamefont {Morel}\ \emph {et~al.}(2020)\citenamefont {Morel},
  \citenamefont {Yao}, \citenamefont {Cladé},\ and\ \citenamefont
  {Guellati-Khélifa}}]{Mor20}%
  \BibitemOpen
  \bibfield  {author} {\bibinfo {author} {\bibfnamefont {L.}~\bibnamefont
  {Morel}}, \bibinfo {author} {\bibfnamefont {Z.}~\bibnamefont {Yao}}, \bibinfo
  {author} {\bibfnamefont {P.}~\bibnamefont {Cladé}}, \ and\ \bibinfo {author}
  {\bibfnamefont {S.}~\bibnamefont {Guellati-Khélifa}},\ }\href {\doibase
  10.1038/s41586-020-2964-7} {\bibfield  {journal} {\bibinfo  {journal}
  {Nature}\ }\textbf {\bibinfo {volume} {588}},\ \bibinfo {pages} {61}
  (\bibinfo {year} {2020})}\BibitemShut {NoStop}%
\bibitem [{\citenamefont {Abi}\ \emph {et~al.}(2021)\citenamefont {Abi},
  \citenamefont {Albahri}, \citenamefont {Al-Kilani}, \citenamefont {Allspach},
  \citenamefont {Alonzi}, \citenamefont {Anastasi}, \citenamefont {Anisenkov},
  \citenamefont {Azfar}, \citenamefont {Badgley}, \citenamefont {Bae\ss{}ler},
  \citenamefont {Bailey}, \citenamefont {Baranov}, \citenamefont
  {Barlas-Yucel}, \citenamefont {Barrett}, \citenamefont {Barzi}, \citenamefont
  {Basti}, \citenamefont {Bedeschi}, \citenamefont {Behnke}, \citenamefont
  {Berz}, \citenamefont {Bhattacharya}, \citenamefont {Binney}, \citenamefont
  {Bjorkquist}, \citenamefont {Bloom}, \citenamefont {Bono}, \citenamefont
  {Bottalico}, \citenamefont {Bowcock}, \citenamefont {Boyden}, \citenamefont
  {Cantatore}, \citenamefont {Carey}, \citenamefont {Carroll}, \citenamefont
  {Casey}, \citenamefont {Cauz}, \citenamefont {Ceravolo}, \citenamefont
  {Chakraborty}, \citenamefont {Chang}, \citenamefont {Chapelain},
  \citenamefont {Chappa}, \citenamefont {Charity}, \citenamefont {Chislett},
  \citenamefont {Choi}, \citenamefont {Chu}, \citenamefont {Chupp},
  \citenamefont {Convery}, \citenamefont {Conway}, \citenamefont {Corradi},
  \citenamefont {Corrodi}, \citenamefont {Cotrozzi}, \citenamefont {Crnkovic},
  \citenamefont {Dabagov}, \citenamefont {De~Lurgio}, \citenamefont {Debevec},
  \citenamefont {Di~Falco}, \citenamefont {Di~Meo}, \citenamefont
  {Di~Sciascio}, \citenamefont {Di~Stefano}, \citenamefont {Drendel},
  \citenamefont {Driutti}, \citenamefont {Duginov}, \citenamefont {Eads},
  \citenamefont {Eggert}, \citenamefont {Epps}, \citenamefont {Esquivel},
  \citenamefont {Farooq}, \citenamefont {Fatemi}, \citenamefont {Ferrari},
  \citenamefont {Fertl}, \citenamefont {Fiedler}, \citenamefont {Fienberg},
  \citenamefont {Fioretti}, \citenamefont {Flay}, \citenamefont {Foster},
  \citenamefont {Friedsam}, \citenamefont {Frle\ifmmode~\check{z}\else
  \v{z}\fi{}}, \citenamefont {Froemming}, \citenamefont {Fry}, \citenamefont
  {Fu}, \citenamefont {Gabbanini}, \citenamefont {Galati}, \citenamefont
  {Ganguly}, \citenamefont {Garcia}, \citenamefont {Gastler}, \citenamefont
  {George}, \citenamefont {Gibbons}, \citenamefont {Gioiosa}, \citenamefont
  {Giovanetti}, \citenamefont {Girotti}, \citenamefont {Gohn}, \citenamefont
  {Gorringe}, \citenamefont {Grange}, \citenamefont {Grant}, \citenamefont
  {Gray}, \citenamefont {Haciomeroglu}, \citenamefont {Hahn}, \citenamefont
  {Halewood-Leagas}, \citenamefont {Hampai}, \citenamefont {Han}, \citenamefont
  {Hazen}, \citenamefont {Hempstead}, \citenamefont {Henry}, \citenamefont
  {Herrod}, \citenamefont {Hertzog}, \citenamefont {Hesketh}, \citenamefont
  {Hibbert}, \citenamefont {Hodge}, \citenamefont {Holzbauer}, \citenamefont
  {Hong}, \citenamefont {Hong}, \citenamefont {Iacovacci}, \citenamefont
  {Incagli}, \citenamefont {Johnstone}, \citenamefont {Johnstone},
  \citenamefont {Kammel}, \citenamefont {Kargiantoulakis}, \citenamefont
  {Karuza}, \citenamefont {Kaspar}, \citenamefont {Kawall}, \citenamefont
  {Kelton}, \citenamefont {Keshavarzi}, \citenamefont {Kessler}, \citenamefont
  {Khaw}, \citenamefont {Khechadoorian}, \citenamefont {Khomutov},
  \citenamefont {Kiburg}, \citenamefont {Kiburg}, \citenamefont {Kim},
  \citenamefont {Kim}, \citenamefont {Kim}, \citenamefont {King}, \citenamefont
  {Kinnaird}, \citenamefont {Korostelev}, \citenamefont {Kourbanis},
  \citenamefont {Kraegeloh}, \citenamefont {Krylov}, \citenamefont
  {Kuchibhotla}, \citenamefont {Kuchinskiy}, \citenamefont {Labe},
  \citenamefont {LaBounty}, \citenamefont {Lancaster}, \citenamefont {Lee},
  \citenamefont {Lee}, \citenamefont {Leo}, \citenamefont {Li}, \citenamefont
  {Li}, \citenamefont {Li}, \citenamefont {Logashenko}, \citenamefont
  {Lorente~Campos}, \citenamefont {Luc\`a}, \citenamefont {Lukicov},
  \citenamefont {Luo}, \citenamefont {Lusiani}, \citenamefont {Lyon},
  \citenamefont {MacCoy}, \citenamefont {Madrak}, \citenamefont {Makino},
  \citenamefont {Marignetti}, \citenamefont {Mastroianni}, \citenamefont
  {Maxfield}, \citenamefont {McEvoy}, \citenamefont {Merritt}, \citenamefont
  {Mikhailichenko}, \citenamefont {Miller}, \citenamefont {Miozzi},
  \citenamefont {Morgan}, \citenamefont {Morse}, \citenamefont {Mott},
  \citenamefont {Motuk}, \citenamefont {Nath}, \citenamefont {Newton},
  \citenamefont {Nguyen}, \citenamefont {Oberling}, \citenamefont {Osofsky},
  \citenamefont {Ostiguy}, \citenamefont {Park}, \citenamefont {Pauletta},
  \citenamefont {Piacentino}, \citenamefont {Pilato}, \citenamefont {Pitts},
  \citenamefont {Plaster}, \citenamefont {Po\ifmmode \check{c}\else
  \v{c}\fi{}ani\ifmmode~\acute{c}\else \'{c}\fi{}}, \citenamefont {Pohlman},
  \citenamefont {Polly}, \citenamefont {Popovic}, \citenamefont {Price},
  \citenamefont {Quinn}, \citenamefont {Raha}, \citenamefont {Ramachandran},
  \citenamefont {Ramberg}, \citenamefont {Rider}, \citenamefont {Ritchie},
  \citenamefont {Roberts}, \citenamefont {Rubin}, \citenamefont {Santi},
  \citenamefont {Sathyan}, \citenamefont {Schellman}, \citenamefont
  {Schlesier}, \citenamefont {Schreckenberger}, \citenamefont {Semertzidis},
  \citenamefont {Shatunov}, \citenamefont {Shemyakin}, \citenamefont {Shenk},
  \citenamefont {Sim}, \citenamefont {Smith}, \citenamefont {Smith},
  \citenamefont {Soha}, \citenamefont {Sorbara}, \citenamefont {St\"ockinger},
  \citenamefont {Stapleton}, \citenamefont {Still}, \citenamefont {Stoughton},
  \citenamefont {Stratakis}, \citenamefont {Strohman}, \citenamefont
  {Stuttard}, \citenamefont {Swanson}, \citenamefont {Sweetmore}, \citenamefont
  {Sweigart}, \citenamefont {Syphers}, \citenamefont {Tarazona}, \citenamefont
  {Teubner}, \citenamefont {Tewsley-Booth}, \citenamefont {Thomson},
  \citenamefont {Tishchenko}, \citenamefont {Tran}, \citenamefont {Turner},
  \citenamefont {Valetov}, \citenamefont {Vasilkova}, \citenamefont
  {Venanzoni}, \citenamefont {Volnykh}, \citenamefont {Walton}, \citenamefont
  {Warren}, \citenamefont {Weisskopf}, \citenamefont {Welty-Rieger},
  \citenamefont {Whitley}, \citenamefont {Winter}, \citenamefont {Wolski},
  \citenamefont {Wormald}, \citenamefont {Wu},\ and\ \citenamefont
  {Yoshikawa}}]{Abi21}%
  \BibitemOpen
  \bibfield  {author} {\bibinfo {author} {\bibfnamefont {B.}~\bibnamefont
  {Abi}}, \bibinfo {author} {\bibfnamefont {T.}~\bibnamefont {Albahri}},
  \bibinfo {author} {\bibfnamefont {S.}~\bibnamefont {Al-Kilani}},  \emph
  {et~al.} (\bibinfo {collaboration} {Muon $g\ensuremath{-}2$ Collaboration}),\
  }\href {\doibase 10.1103/PhysRevLett.126.141801} {\bibfield  {journal}
  {\bibinfo  {journal} {Phys. Rev. Lett.}\ }\textbf {\bibinfo {volume} {126}},\
  \bibinfo {pages} {141801} (\bibinfo {year} {2021})}\BibitemShut {NoStop}%
\bibitem [{\citenamefont {Cairncross}\ \emph {et~al.}(2017)\citenamefont
  {Cairncross}, \citenamefont {Gresh}, \citenamefont {Grau}, \citenamefont
  {Cossel}, \citenamefont {Roussy}, \citenamefont {Ni}, \citenamefont {Zhou},
  \citenamefont {Ye},\ and\ \citenamefont {Cornell}}]{Cai17}%
  \BibitemOpen
  \bibfield  {author} {\bibinfo {author} {\bibfnamefont {W.~B.}\ \bibnamefont
  {Cairncross}}, \bibinfo {author} {\bibfnamefont {D.~N.}\ \bibnamefont
  {Gresh}}, \bibinfo {author} {\bibfnamefont {M.}~\bibnamefont {Grau}},  \emph
  {et~al.},\ }\href {\doibase 10.1103/PhysRevLett.119.153001} {\bibfield
  {journal} {\bibinfo  {journal} {Phys. Rev. Lett.}\ }\textbf {\bibinfo
  {volume} {119}},\ \bibinfo {pages} {153001} (\bibinfo {year}
  {2017})}\BibitemShut {NoStop}%
\bibitem [{\citenamefont {Andreev}\ \emph {et~al.}(2018)\citenamefont
  {Andreev}, \citenamefont {Ang}, \citenamefont {DeMille}, \citenamefont
  {Doyle}, \citenamefont {Gabrielse}, \citenamefont {Haefner}, \citenamefont
  {Hutzler}, \citenamefont {Lasner}, \citenamefont {Meisenhelder},
  \citenamefont {O’Leary}, \citenamefont {Panda}, \citenamefont {West},
  \citenamefont {West}, \citenamefont {Wu},\ and\ \citenamefont {{ACME
  Collaboration}}}]{And18}%
  \BibitemOpen
  \bibfield  {author} {\bibinfo {author} {\bibfnamefont {V.}~\bibnamefont
  {Andreev}}, \bibinfo {author} {\bibfnamefont {D.~G.}\ \bibnamefont {Ang}},
  \bibinfo {author} {\bibfnamefont {D.}~\bibnamefont {DeMille}},  \emph
  {et~al.},\ }\href {\doibase 10.1038/s41586-018-0599-8} {\bibfield  {journal}
  {\bibinfo  {journal} {Nature}\ }\textbf {\bibinfo {volume} {562}},\ \bibinfo
  {pages} {355} (\bibinfo {year} {2018})}\BibitemShut {NoStop}%
\bibitem [{\citenamefont {Abel}\ \emph {et~al.}(2020)\citenamefont {Abel},
  \citenamefont {Afach}, \citenamefont {Ayres}, \citenamefont {Baker},
  \citenamefont {Ban}, \citenamefont {Bison}, \citenamefont {Bodek},
  \citenamefont {Bondar}, \citenamefont {Burghoff}, \citenamefont {Chanel},
  \citenamefont {Chowdhuri}, \citenamefont {Chiu}, \citenamefont {Clement},
  \citenamefont {Crawford}, \citenamefont {Daum}, \citenamefont {Emmenegger},
  \citenamefont {Ferraris-Bouchez}, \citenamefont {Fertl}, \citenamefont
  {Flaux}, \citenamefont {Franke}, \citenamefont {Fratangelo}, \citenamefont
  {Geltenbort}, \citenamefont {Green}, \citenamefont {Griffith}, \citenamefont
  {van~der Grinten}, \citenamefont {Gruji\ifmmode~\acute{c}\else \'{c}\fi{}},
  \citenamefont {Harris}, \citenamefont {Hayen}, \citenamefont {Heil},
  \citenamefont {Henneck}, \citenamefont {H\'elaine}, \citenamefont {Hild},
  \citenamefont {Hodge}, \citenamefont {Horras}, \citenamefont {Iaydjiev},
  \citenamefont {Ivanov}, \citenamefont {Kasprzak}, \citenamefont {Kermaidic},
  \citenamefont {Kirch}, \citenamefont {Knecht}, \citenamefont {Knowles},
  \citenamefont {Koch}, \citenamefont {Koss}, \citenamefont {Komposch},
  \citenamefont {Kozela}, \citenamefont {Kraft}, \citenamefont {Krempel},
  \citenamefont {Ku\ifmmode~\acute{z}\else \'{z}\fi{}niak}, \citenamefont
  {Lauss}, \citenamefont {Lefort}, \citenamefont {Lemi\`ere}, \citenamefont
  {Leredde}, \citenamefont {Mohanmurthy}, \citenamefont {Mtchedlishvili},
  \citenamefont {Musgrave}, \citenamefont {Naviliat-Cuncic}, \citenamefont
  {Pais}, \citenamefont {Piegsa}, \citenamefont {Pierre}, \citenamefont
  {Pignol}, \citenamefont {Plonka-Spehr}, \citenamefont {Prashanth},
  \citenamefont {Qu\'em\'ener}, \citenamefont {Rawlik}, \citenamefont
  {Rebreyend}, \citenamefont {Rien\"acker}, \citenamefont {Ries}, \citenamefont
  {Roccia}, \citenamefont {Rogel}, \citenamefont {Rozpedzik}, \citenamefont
  {Schnabel}, \citenamefont {Schmidt-Wellenburg}, \citenamefont {Severijns},
  \citenamefont {Shiers}, \citenamefont {Tavakoli~Dinani}, \citenamefont
  {Thorne}, \citenamefont {Virot}, \citenamefont {Voigt}, \citenamefont {Weis},
  \citenamefont {Wursten}, \citenamefont {Wyszynski}, \citenamefont {Zejma},
  \citenamefont {Zenner},\ and\ \citenamefont {Zsigmond}}]{Abe20}%
  \BibitemOpen
  \bibfield  {author} {\bibinfo {author} {\bibfnamefont {C.}~\bibnamefont
  {Abel}}, \bibinfo {author} {\bibfnamefont {S.}~\bibnamefont {Afach}},
  \bibinfo {author} {\bibfnamefont {N.~J.}\ \bibnamefont {Ayres}},  \emph
  {et~al.},\ }\href {\doibase 10.1103/PhysRevLett.124.081803} {\bibfield
  {journal} {\bibinfo  {journal} {Phys. Rev. Lett.}\ }\textbf {\bibinfo
  {volume} {124}},\ \bibinfo {pages} {081803} (\bibinfo {year}
  {2020})}\BibitemShut {NoStop}%
\bibitem [{\citenamefont {Hori}\ \emph {et~al.}(2016)\citenamefont {Hori},
  \citenamefont {Aghai-Khozani}, \citenamefont {S{\'o}t{\'e}r}, \citenamefont
  {Barna}, \citenamefont {Dax}, \citenamefont {Hayano}, \citenamefont
  {Kobayashi}, \citenamefont {Murakami}, \citenamefont {Todoroki},
  \citenamefont {Yamada}, \citenamefont {Horv{\'a}th},\ and\ \citenamefont
  {Venturelli}}]{Hor16}%
  \BibitemOpen
  \bibfield  {author} {\bibinfo {author} {\bibfnamefont {M.}~\bibnamefont
  {Hori}}, \bibinfo {author} {\bibfnamefont {H.}~\bibnamefont {Aghai-Khozani}},
  \bibinfo {author} {\bibfnamefont {A.}~\bibnamefont {S{\'o}t{\'e}r}},  \emph
  {et~al.},\ }\href {\doibase 10.1126/science.aaf6702} {\bibfield  {journal}
  {\bibinfo  {journal} {Science}\ }\textbf {\bibinfo {volume} {354}},\ \bibinfo
  {pages} {610} (\bibinfo {year} {2016})}\BibitemShut {NoStop}%
\bibitem [{\citenamefont {Ahmadi}\ \emph {et~al.}(2018)\citenamefont {Ahmadi},
  \citenamefont {Alves}, \citenamefont {Baker}, \citenamefont {Bertsche},
  \citenamefont {Capra}, \citenamefont {Carruth}, \citenamefont {Cesar},
  \citenamefont {Charlton}, \citenamefont {Cohen}, \citenamefont {Collister},
  \citenamefont {Eriksson}, \citenamefont {Evans}, \citenamefont {Evetts},
  \citenamefont {Fajans}, \citenamefont {Friesen}, \citenamefont {Fujiwara},
  \citenamefont {Gill}, \citenamefont {Hangst}, \citenamefont {Hardy},
  \citenamefont {Hayden}, \citenamefont {Isaac}, \citenamefont {Johnson},
  \citenamefont {Jones}, \citenamefont {Jones}, \citenamefont {Jonsell},
  \citenamefont {Khramov}, \citenamefont {Knapp}, \citenamefont {Kurchaninov},
  \citenamefont {Madsen}, \citenamefont {Maxwell}, \citenamefont {McKenna},
  \citenamefont {Menary}, \citenamefont {Momose}, \citenamefont {Munich},
  \citenamefont {OlcH.ki}, \citenamefont {Olin}, \citenamefont {Pusa},
  \citenamefont {Rasmussen}, \citenamefont {Robicheaux}, \citenamefont
  {Sacramento}, \citenamefont {Sameed}, \citenamefont {Sarid}, \citenamefont
  {Silveira}, \citenamefont {Stutter}, \citenamefont {So}, \citenamefont
  {Tharp}, \citenamefont {Thompson}, \citenamefont {van~der Werf},\ and\
  \citenamefont {Wurtele}}]{Ahm18}%
  \BibitemOpen
  \bibfield  {author} {\bibinfo {author} {\bibfnamefont {M.}~\bibnamefont
  {Ahmadi}}, \bibinfo {author} {\bibfnamefont {B.~X.~R.}\ \bibnamefont
  {Alves}}, \bibinfo {author} {\bibfnamefont {C.~J.}\ \bibnamefont {Baker}},
  \emph {et~al.},\ }\href {\doibase 10.1038/s41586-018-0017-2} {\bibfield
  {journal} {\bibinfo  {journal} {Nature}\ }\textbf {\bibinfo {volume} {557}},\
  \bibinfo {pages} {71} (\bibinfo {year} {2018})}\BibitemShut {NoStop}%
\bibitem [{\citenamefont {Ahmadi}\ \emph {et~al.}(2020)\citenamefont {Ahmadi},
  \citenamefont {Alves}, \citenamefont {Baker}, \citenamefont {Bertsche},
  \citenamefont {Capra}, \citenamefont {Carruth}, \citenamefont {Cesar},
  \citenamefont {Charlton}, \citenamefont {Cohen}, \citenamefont {Collister},
  \citenamefont {Eriksson}, \citenamefont {Evans}, \citenamefont {Evetts},
  \citenamefont {Fajans}, \citenamefont {Friesen}, \citenamefont {Fujiwara},
  \citenamefont {Gill}, \citenamefont {Granum}, \citenamefont {Hangst},
  \citenamefont {Hardy}, \citenamefont {Hayden}, \citenamefont {Hunter},
  \citenamefont {Isaac}, \citenamefont {Johnson}, \citenamefont {Jones},
  \citenamefont {Jones}, \citenamefont {Jonsell}, \citenamefont {Khramov},
  \citenamefont {Knapp}, \citenamefont {Kurchaninov}, \citenamefont {Madsen},
  \citenamefont {Maxwell}, \citenamefont {McKenna}, \citenamefont {Menary},
  \citenamefont {Michan}, \citenamefont {Momose}, \citenamefont {Munich},
  \citenamefont {OlcH.ki}, \citenamefont {Olin}, \citenamefont {Pusa},
  \citenamefont {Rasmussen}, \citenamefont {Robicheaux}, \citenamefont
  {Sacramento}, \citenamefont {Sameed}, \citenamefont {Sarid}, \citenamefont
  {Silveira}, \citenamefont {So}, \citenamefont {Starko}, \citenamefont
  {Stutter}, \citenamefont {Tharp}, \citenamefont {Thompson}, \citenamefont
  {van~der Werf}, \citenamefont {Wurtele},\ and\ \citenamefont
  {Collaboration}}]{Ahm20}%
  \BibitemOpen
  \bibfield  {author} {\bibinfo {author} {\bibfnamefont {M.}~\bibnamefont
  {Ahmadi}}, \bibinfo {author} {\bibfnamefont {B.~X.~R.}\ \bibnamefont
  {Alves}}, \bibinfo {author} {\bibfnamefont {C.~J.}\ \bibnamefont {Baker}},
  \emph {et~al.},\ }\href {\doibase 10.1038/s41586-020-2006-5} {\bibfield
  {journal} {\bibinfo  {journal} {Nature}\ }\textbf {\bibinfo {volume} {578}},\
  \bibinfo {pages} {375} (\bibinfo {year} {2020})}\BibitemShut {NoStop}%
\bibitem [{\citenamefont {Charlton}\ \emph {et~al.}(2020)\citenamefont
  {Charlton}, \citenamefont {Eriksson},\ and\ \citenamefont {Shore}}]{Char20}%
  \BibitemOpen
  \bibfield  {author} {\bibinfo {author} {\bibfnamefont {M.}~\bibnamefont
  {Charlton}}, \bibinfo {author} {\bibfnamefont {S.}~\bibnamefont {Eriksson}},
  \ and\ \bibinfo {author} {\bibfnamefont {G.~M.}\ \bibnamefont {Shore}},\
  }\href@noop {} {\enquote {\bibinfo {title} {Testing fundamental physics in
  antihydrogen experiments},}\ } (\bibinfo {year} {2020}),\ \Eprint
  {http://arxiv.org/abs/2002.09348} {arXiv:2002.09348 [hep-ph]} \BibitemShut
  {NoStop}%
\bibitem [{\citenamefont {Ulmer}\ \emph {et~al.}(2015)\citenamefont {Ulmer},
  \citenamefont {Smorra}, \citenamefont {Mooser}, \citenamefont {Franke},
  \citenamefont {Nagahama}, \citenamefont {Schneider}, \citenamefont {Higuchi},
  \citenamefont {Van~Gorp}, \citenamefont {Blaum}, \citenamefont {Matsuda},
  \citenamefont {Quint}, \citenamefont {Walz},\ and\ \citenamefont
  {Yamazaki}}]{Ulm15}%
  \BibitemOpen
  \bibfield  {author} {\bibinfo {author} {\bibfnamefont {S.}~\bibnamefont
  {Ulmer}}, \bibinfo {author} {\bibfnamefont {C.}~\bibnamefont {Smorra}},
  \bibinfo {author} {\bibfnamefont {A.}~\bibnamefont {Mooser}},  \emph
  {et~al.},\ }\href {\doibase 10.1038/nature14861} {\bibfield  {journal}
  {\bibinfo  {journal} {Nature}\ }\textbf {\bibinfo {volume} {524}},\ \bibinfo
  {pages} {196} (\bibinfo {year} {2015})}\BibitemShut {NoStop}%
\bibitem [{\citenamefont {Smorra}\ \emph {et~al.}(2017)\citenamefont {Smorra},
  \citenamefont {Sellner}, \citenamefont {Borchert}, \citenamefont
  {Harrington}, \citenamefont {Higuchi}, \citenamefont {Nagahama},
  \citenamefont {Tanaka}, \citenamefont {Mooser}, \citenamefont {Schneider},
  \citenamefont {Bohman}, \citenamefont {Blaum}, \citenamefont {Matsuda},
  \citenamefont {Ospelkaus}, \citenamefont {Quint}, \citenamefont {Walz},
  \citenamefont {Yamazaki},\ and\ \citenamefont {Ulmer}}]{Smo17}%
  \BibitemOpen
  \bibfield  {author} {\bibinfo {author} {\bibfnamefont {C.}~\bibnamefont
  {Smorra}}, \bibinfo {author} {\bibfnamefont {S.}~\bibnamefont {Sellner}},
  \bibinfo {author} {\bibfnamefont {M.~J.}\ \bibnamefont {Borchert}},  \emph
  {et~al.},\ }\href {\doibase 10.1038/nature24048} {\bibfield  {journal}
  {\bibinfo  {journal} {Nature}\ }\textbf {\bibinfo {volume} {550}},\ \bibinfo
  {pages} {371} (\bibinfo {year} {2017})}\BibitemShut {NoStop}%
\bibitem [{\citenamefont {Borchert}\ \emph {et~al.}(2022)\citenamefont
  {Borchert}, \citenamefont {Devlin}, \citenamefont {Erlewein}, \citenamefont
  {Fleck}, \citenamefont {Harrington}, \citenamefont {Higuchi}, \citenamefont
  {Latacz}, \citenamefont {Voelksen}, \citenamefont {Wursten}, \citenamefont
  {Abbass}, \citenamefont {Bohman}, \citenamefont {Mooser}, \citenamefont
  {Popper}, \citenamefont {Wiesinger}, \citenamefont {Will}, \citenamefont
  {Blaum}, \citenamefont {Matsuda}, \citenamefont {Ospelkaus}, \citenamefont
  {Quint}, \citenamefont {Walz}, \citenamefont {Yamazaki}, \citenamefont
  {Smorra},\ and\ \citenamefont {Ulmer}}]{Bor22}%
  \BibitemOpen
  \bibfield  {author} {\bibinfo {author} {\bibfnamefont {M.}~\bibnamefont
  {Borchert}}, \bibinfo {author} {\bibfnamefont {J.}~\bibnamefont {Devlin}},
  \bibinfo {author} {\bibfnamefont {S.}~\bibnamefont {Erlewein}},  \emph
  {et~al.},\ }\href {\doibase 10.1038/s41586-021-04203-w} {\bibfield  {journal}
  {\bibinfo  {journal} {Nature}\ }\textbf {\bibinfo {volume} {601}},\ \bibinfo
  {pages} {53} (\bibinfo {year} {2022})}\BibitemShut {NoStop}%
\bibitem [{\citenamefont {Hori}\ and\ \citenamefont {Walz}(2013)}]{Hor13}%
  \BibitemOpen
  \bibfield  {author} {\bibinfo {author} {\bibfnamefont {M.}~\bibnamefont
  {Hori}}\ and\ \bibinfo {author} {\bibfnamefont {J.}~\bibnamefont {Walz}},\
  }\href {\doibase https://doi.org/10.1016/j.ppnp.2013.02.004} {\bibfield
  {journal} {\bibinfo  {journal} {Progress in Particle and Nuclear Physics}\
  }\textbf {\bibinfo {volume} {72}},\ \bibinfo {pages} {206} (\bibinfo {year}
  {2013})}\BibitemShut {NoStop}%
\bibitem [{\citenamefont {Bartmann}\ \emph {et~al.}(2018)\citenamefont
  {Bartmann}, \citenamefont {Belochitskii}, \citenamefont {Breuker},
  \citenamefont {Butin}, \citenamefont {Carli}, \citenamefont {Eriksson},
  \citenamefont {Oelert}, \citenamefont {Ostojic}, \citenamefont {Pasinelli},\
  and\ \citenamefont {Tranquille}}]{Bar18}%
  \BibitemOpen
  \bibfield  {author} {\bibinfo {author} {\bibfnamefont {W.}~\bibnamefont
  {Bartmann}}, \bibinfo {author} {\bibfnamefont {P.}~\bibnamefont
  {Belochitskii}}, \bibinfo {author} {\bibfnamefont {H.}~\bibnamefont
  {Breuker}},  \emph {et~al.},\ }\href {\doibase 10.1098/rsta.2017.0266}
  {\bibfield  {journal} {\bibinfo  {journal} {Philosophical Transactions of the
  Royal Society A: Mathematical, Physical and Engineering Sciences}\ }\textbf
  {\bibinfo {volume} {376}},\ \bibinfo {pages} {20170266} (\bibinfo {year}
  {2018})}\BibitemShut {NoStop}%
\bibitem [{\citenamefont {Schneider}\ \emph {et~al.}(2017)\citenamefont
  {Schneider}, \citenamefont {Mooser}, \citenamefont {Bohman}, \citenamefont
  {Sch{\"o}n}, \citenamefont {Harrington}, \citenamefont {Higuchi},
  \citenamefont {Nagahama}, \citenamefont {Sellner}, \citenamefont {Smorra},
  \citenamefont {Blaum}, \citenamefont {Matsuda}, \citenamefont {Quint},
  \citenamefont {Walz},\ and\ \citenamefont {Ulmer}}]{Sch17}%
  \BibitemOpen
  \bibfield  {author} {\bibinfo {author} {\bibfnamefont {G.}~\bibnamefont
  {Schneider}}, \bibinfo {author} {\bibfnamefont {A.}~\bibnamefont {Mooser}},
  \bibinfo {author} {\bibfnamefont {M.}~\bibnamefont {Bohman}},  \emph
  {et~al.},\ }\href {\doibase 10.1126/science.aan0207} {\bibfield  {journal}
  {\bibinfo  {journal} {Science}\ }\textbf {\bibinfo {volume} {358}},\ \bibinfo
  {pages} {1081} (\bibinfo {year} {2017})}\BibitemShut {NoStop}%
\bibitem [{\citenamefont {Hughes}\ and\ \citenamefont
  {Holzscheiter}(1991)}]{Hug91}%
  \BibitemOpen
  \bibfield  {author} {\bibinfo {author} {\bibfnamefont {R.~J.}\ \bibnamefont
  {Hughes}}\ and\ \bibinfo {author} {\bibfnamefont {M.~H.}\ \bibnamefont
  {Holzscheiter}},\ }\href {\doibase 10.1103/PhysRevLett.66.854} {\bibfield
  {journal} {\bibinfo  {journal} {Phys. Rev. Lett.}\ }\textbf {\bibinfo
  {volume} {66}},\ \bibinfo {pages} {854} (\bibinfo {year} {1991})}\BibitemShut
  {NoStop}%
\bibitem [{\citenamefont {Bluhm}\ \emph {et~al.}(1998)\citenamefont {Bluhm},
  \citenamefont {Kosteleck\'y},\ and\ \citenamefont {Russell}}]{Blu98}%
  \BibitemOpen
  \bibfield  {author} {\bibinfo {author} {\bibfnamefont {R.}~\bibnamefont
  {Bluhm}}, \bibinfo {author} {\bibfnamefont {V.~A.}\ \bibnamefont
  {Kosteleck\'y}}, \ and\ \bibinfo {author} {\bibfnamefont {N.}~\bibnamefont
  {Russell}},\ }\href {\doibase 10.1103/PhysRevD.57.3932} {\bibfield  {journal}
  {\bibinfo  {journal} {Phys. Rev. D}\ }\textbf {\bibinfo {volume} {57}},\
  \bibinfo {pages} {3932} (\bibinfo {year} {1998})}\BibitemShut {NoStop}%
\bibitem [{\citenamefont {Ding}(2019)}]{Din19}%
  \BibitemOpen
  \bibfield  {author} {\bibinfo {author} {\bibfnamefont {Y.}~\bibnamefont
  {Ding}},\ }\href@noop {} {\bibfield  {journal} {\bibinfo  {journal}
  {Symmetry}\ }\textbf {\bibinfo {volume} {11}},\ \bibinfo {pages} {1220}
  (\bibinfo {year} {2019})}\BibitemShut {NoStop}%
\bibitem [{\citenamefont {Smorra}\ \emph {et~al.}(2019)\citenamefont {Smorra},
  \citenamefont {Stadnik}, \citenamefont {Blessing}, \citenamefont {Bohman},
  \citenamefont {Borchert}, \citenamefont {Devlin}, \citenamefont {Erlewein},
  \citenamefont {Harrington}, \citenamefont {Higuchi}, \citenamefont {Mooser},
  \citenamefont {Schneider}, \citenamefont {Wiesinger}, \citenamefont
  {Wursten}, \citenamefont {Blaum}, \citenamefont {Matsuda}, \citenamefont
  {Ospelkaus}, \citenamefont {Quint}, \citenamefont {Walz}, \citenamefont
  {Yamazaki}, \citenamefont {Budker},\ and\ \citenamefont {Ulmer}}]{Smo19}%
  \BibitemOpen
  \bibfield  {author} {\bibinfo {author} {\bibfnamefont {C.}~\bibnamefont
  {Smorra}}, \bibinfo {author} {\bibfnamefont {Y.~V.}\ \bibnamefont {Stadnik}},
  \bibinfo {author} {\bibfnamefont {P.~E.}\ \bibnamefont {Blessing}},  \emph
  {et~al.},\ }\href {\doibase 10.1038/s41586-019-1727-9} {\bibfield  {journal}
  {\bibinfo  {journal} {Nature}\ }\textbf {\bibinfo {volume} {575}},\ \bibinfo
  {pages} {310} (\bibinfo {year} {2019})}\BibitemShut {NoStop}%
\bibitem [{\citenamefont {Bediaga}\ and\ \citenamefont {Göbel}(2020)}]{Bed20}%
  \BibitemOpen
  \bibfield  {author} {\bibinfo {author} {\bibfnamefont {I.}~\bibnamefont
  {Bediaga}}\ and\ \bibinfo {author} {\bibfnamefont {C.}~\bibnamefont
  {Göbel}},\ }\href {\doibase https://doi.org/10.1016/j.ppnp.2020.103808}
  {\bibfield  {journal} {\bibinfo  {journal} {Progress in Particle and Nuclear
  Physics}\ }\textbf {\bibinfo {volume} {114}},\ \bibinfo {pages} {103808}
  (\bibinfo {year} {2020})}\BibitemShut {NoStop}%
\bibitem [{\citenamefont {Dove}\ \emph {et~al.}(2021)\citenamefont {Dove},
  \citenamefont {Kerns}, \citenamefont {McClellan}, \citenamefont {Miyasaka},
  \citenamefont {Morton}, \citenamefont {Nagai}, \citenamefont {Prasad},
  \citenamefont {Sanftl}, \citenamefont {Scott}, \citenamefont {Tadepalli}
  \emph {et~al.}}]{Dov21}%
  \BibitemOpen
  \bibfield  {author} {\bibinfo {author} {\bibfnamefont {J.}~\bibnamefont
  {Dove}}, \bibinfo {author} {\bibfnamefont {B.}~\bibnamefont {Kerns}},
  \bibinfo {author} {\bibfnamefont {R.}~\bibnamefont {McClellan}},  \emph
  {et~al.},\ }\href@noop {} {\bibfield  {journal} {\bibinfo  {journal}
  {Nature}\ }\textbf {\bibinfo {volume} {590}},\ \bibinfo {pages} {561}
  (\bibinfo {year} {2021})}\BibitemShut {NoStop}%
\bibitem [{\citenamefont {Ulmer}(2019)}]{Ulmer2019a}%
  \BibitemOpen
  \bibfield  {author} {\bibinfo {author} {\bibfnamefont {S.}~\bibnamefont
  {Ulmer}},\ }\href@noop {} {\emph {\bibinfo {title} {BASE Annual Report
  2019}}},\ \bibinfo {type} {Tech. Rep.}\ (\bibinfo  {institution} {CERN,
  Geneva},\ \bibinfo {year} {2019})\BibitemShut {NoStop}%
\bibitem [{\citenamefont {Tseng}\ and\ \citenamefont
  {Gabrielse}(1993)}]{Tse93}%
  \BibitemOpen
  \bibfield  {author} {\bibinfo {author} {\bibfnamefont {C.~H.}\ \bibnamefont
  {Tseng}}\ and\ \bibinfo {author} {\bibfnamefont {G.}~\bibnamefont
  {Gabrielse}},\ }\href {\doibase 10.1007/BF02316739} {\bibfield  {journal}
  {\bibinfo  {journal} {Hyperfine Interactions}\ }\textbf {\bibinfo {volume}
  {76}},\ \bibinfo {pages} {381} (\bibinfo {year} {1993})}\BibitemShut
  {NoStop}%
\bibitem [{\citenamefont {Dehmelt}(1995)}]{Deh95}%
  \BibitemOpen
  \bibfield  {author} {\bibinfo {author} {\bibfnamefont {H.}~\bibnamefont
  {Dehmelt}},\ }\href {\doibase 10.1088/0031-8949/1995/t59/060} {\bibfield
  {journal} {\bibinfo  {journal} {Physica Scripta}\ }\textbf {\bibinfo {volume}
  {T59}},\ \bibinfo {pages} {423} (\bibinfo {year} {1995})}\BibitemShut
  {NoStop}%
\bibitem [{\citenamefont {Huntemann}\ \emph {et~al.}(2016)\citenamefont
  {Huntemann}, \citenamefont {Sanner}, \citenamefont {Lipphardt}, \citenamefont
  {Tamm},\ and\ \citenamefont {Peik}}]{Hun16}%
  \BibitemOpen
  \bibfield  {author} {\bibinfo {author} {\bibfnamefont {N.}~\bibnamefont
  {Huntemann}}, \bibinfo {author} {\bibfnamefont {C.}~\bibnamefont {Sanner}},
  \bibinfo {author} {\bibfnamefont {B.}~\bibnamefont {Lipphardt}},  \emph
  {et~al.},\ }\href {\doibase 10.1103/PhysRevLett.116.063001} {\bibfield
  {journal} {\bibinfo  {journal} {Phys. Rev. Lett.}\ }\textbf {\bibinfo
  {volume} {116}},\ \bibinfo {pages} {063001} (\bibinfo {year}
  {2016})}\BibitemShut {NoStop}%
\bibitem [{\citenamefont {Bothwell}\ \emph {et~al.}(2019)\citenamefont
  {Bothwell}, \citenamefont {Kedar}, \citenamefont {Oelker}, \citenamefont
  {Robinson}, \citenamefont {Bromley}, \citenamefont {Tew}, \citenamefont
  {Ye},\ and\ \citenamefont {Kennedy}}]{Bot19}%
  \BibitemOpen
  \bibfield  {author} {\bibinfo {author} {\bibfnamefont {T.}~\bibnamefont
  {Bothwell}}, \bibinfo {author} {\bibfnamefont {D.}~\bibnamefont {Kedar}},
  \bibinfo {author} {\bibfnamefont {E.}~\bibnamefont {Oelker}},  \emph
  {et~al.},\ }\href {\doibase 10.1088/1681-7575/ab4089} {\bibfield  {journal}
  {\bibinfo  {journal} {Metrologia}\ }\textbf {\bibinfo {volume} {56}},\
  \bibinfo {pages} {065004} (\bibinfo {year} {2019})}\BibitemShut {NoStop}%
\bibitem [{\citenamefont {Brewer}\ \emph {et~al.}(2019)\citenamefont {Brewer},
  \citenamefont {Chen}, \citenamefont {Hankin}, \citenamefont {Clements},
  \citenamefont {Chou}, \citenamefont {Wineland}, \citenamefont {Hume},\ and\
  \citenamefont {Leibrandt}}]{Bre18}%
  \BibitemOpen
  \bibfield  {author} {\bibinfo {author} {\bibfnamefont {S.~M.}\ \bibnamefont
  {Brewer}}, \bibinfo {author} {\bibfnamefont {J.-S.}\ \bibnamefont {Chen}},
  \bibinfo {author} {\bibfnamefont {A.~M.}\ \bibnamefont {Hankin}},  \emph
  {et~al.},\ }\href {\doibase 10.1103/PhysRevLett.123.033201} {\bibfield
  {journal} {\bibinfo  {journal} {Phys. Rev. Lett.}\ }\textbf {\bibinfo
  {volume} {123}},\ \bibinfo {pages} {033201} (\bibinfo {year}
  {2019})}\BibitemShut {NoStop}%
\bibitem [{\citenamefont {Huang}\ \emph {et~al.}(2020)\citenamefont {Huang},
  \citenamefont {Zhang}, \citenamefont {Zhang}, \citenamefont {Hao},
  \citenamefont {Guan}, \citenamefont {Zeng}, \citenamefont {Chen},
  \citenamefont {Lin}, \citenamefont {Wang}, \citenamefont {Cao}, \citenamefont
  {Liang}, \citenamefont {Fang}, \citenamefont {Fang}, \citenamefont {Li},\
  and\ \citenamefont {Gao}}]{Hua20}%
  \BibitemOpen
  \bibfield  {author} {\bibinfo {author} {\bibfnamefont {Y.}~\bibnamefont
  {Huang}}, \bibinfo {author} {\bibfnamefont {H.}~\bibnamefont {Zhang}},
  \bibinfo {author} {\bibfnamefont {B.}~\bibnamefont {Zhang}},  \emph
  {et~al.},\ }\href {\doibase 10.1103/PhysRevA.102.050802} {\bibfield
  {journal} {\bibinfo  {journal} {Phys. Rev. A}\ }\textbf {\bibinfo {volume}
  {102}},\ \bibinfo {pages} {050802} (\bibinfo {year} {2020})}\BibitemShut
  {NoStop}%
\bibitem [{\citenamefont {Grotti}\ \emph {et~al.}(2018)\citenamefont {Grotti},
  \citenamefont {Koller}, \citenamefont {Vogt}, \citenamefont {Häfner},
  \citenamefont {Sterr}, \citenamefont {Lisdat}, \citenamefont {Denker},
  \citenamefont {Voigt}, \citenamefont {Timmen}, \citenamefont {Rolland},
  \citenamefont {Baynes}, \citenamefont {Margolis}, \citenamefont {Zampaolo},
  \citenamefont {Thoumany}, \citenamefont {Pizzocaro}, \citenamefont {Rauf},
  \citenamefont {Bregolin}, \citenamefont {Tampellini}, \citenamefont
  {Barbieri}, \citenamefont {Zucco}, \citenamefont {Costanzo}, \citenamefont
  {Clivati}, \citenamefont {Levi},\ and\ \citenamefont {Calonico}}]{Gro18}%
  \BibitemOpen
  \bibfield  {author} {\bibinfo {author} {\bibfnamefont {J.}~\bibnamefont
  {Grotti}}, \bibinfo {author} {\bibfnamefont {S.}~\bibnamefont {Koller}},
  \bibinfo {author} {\bibfnamefont {S.}~\bibnamefont {Vogt}},  \emph {et~al.},\
  }\href@noop {} {\bibfield  {journal} {\bibinfo  {journal} {Nature Physics}\
  }\textbf {\bibinfo {volume} {14}},\ \bibinfo {pages} {437} (\bibinfo {year}
  {2018})}\BibitemShut {NoStop}%
\bibitem [{\citenamefont {Takamoto}\ \emph {et~al.}(2020)\citenamefont
  {Takamoto}, \citenamefont {Ushijima}, \citenamefont {Ohmae}, \citenamefont
  {Yahagi}, \citenamefont {Kokado}, \citenamefont {Shinkai},\ and\
  \citenamefont {Katori}}]{Tak20}%
  \BibitemOpen
  \bibfield  {author} {\bibinfo {author} {\bibfnamefont {M.}~\bibnamefont
  {Takamoto}}, \bibinfo {author} {\bibfnamefont {I.}~\bibnamefont {Ushijima}},
  \bibinfo {author} {\bibfnamefont {N.}~\bibnamefont {Ohmae}},  \emph
  {et~al.},\ }\href {\doibase 10.1038/s41566-020-0619-8} {\bibfield  {journal}
  {\bibinfo  {journal} {Nature Photonics}\ }\textbf {\bibinfo {volume} {14}},\
  \bibinfo {pages} {411} (\bibinfo {year} {2020})}\BibitemShut {NoStop}%
\bibitem [{\citenamefont {Wada}\ and\ \citenamefont {Yamazaki}(2004)}]{Wad04}%
  \BibitemOpen
  \bibfield  {author} {\bibinfo {author} {\bibfnamefont {M.}~\bibnamefont
  {Wada}}\ and\ \bibinfo {author} {\bibfnamefont {Y.}~\bibnamefont
  {Yamazaki}},\ }\href@noop {} {\bibfield  {journal} {\bibinfo  {journal}
  {Nuclear Instruments and Methods in Physics Research Section B: Beam
  Interactions with Materials and Atoms}\ }\textbf {\bibinfo {volume} {214}},\
  \bibinfo {pages} {196} (\bibinfo {year} {2004})}\BibitemShut {NoStop}%
\bibitem [{\citenamefont {Nakatsuka}\ \emph {et~al.}(2019)\citenamefont
  {Nakatsuka}, \citenamefont {Obertelli}, \citenamefont {de~Gersem},
  \citenamefont {Kubota}, \citenamefont {Marsic}, \citenamefont {M{\"u}ller},
  \citenamefont {Fischer}, \citenamefont {Schmidt}, \citenamefont {Carbonell},
  \citenamefont {Corsi} \emph {et~al.}}]{Nak19}%
  \BibitemOpen
  \bibfield  {author} {\bibinfo {author} {\bibfnamefont {N.}~\bibnamefont
  {Nakatsuka}}, \bibinfo {author} {\bibfnamefont {A.}~\bibnamefont
  {Obertelli}}, \bibinfo {author} {\bibfnamefont {H.}~\bibnamefont
  {de~Gersem}},  \emph {et~al.},\ }\href@noop {} {\bibfield  {journal}
  {\bibinfo  {journal} {Verhandlungen der Deutschen Physikalischen
  Gesellschaft}\ } (\bibinfo {year} {2019})}\BibitemShut {NoStop}%
\bibitem [{\citenamefont {{Aumann, T.}}\ \emph {et~al.}(2022)\citenamefont
  {{Aumann, T.}}, \citenamefont {{Bartmann, W.}}, \citenamefont
  {{Boine-Frankenheim, O.}}, \citenamefont {{Bouvard, A.}}, \citenamefont
  {{Broche, A.}}, \citenamefont {{Butin, F.}}, \citenamefont {{Calvet, D.}},
  \citenamefont {{Carbonell, J.}}, \citenamefont {{Chiggiato, P.}},
  \citenamefont {{De Gersem, H.}}, \citenamefont {{De Oliveira, R.}},
  \citenamefont {{Dobers, T.}}, \citenamefont {{Ehm, F.}}, \citenamefont
  {{Somoza, J. Ferreira}}, \citenamefont {{Fischer, J.}}, \citenamefont
  {{Fraser, M.}}, \citenamefont {{Friedrich, E.}}, \citenamefont {{Frotscher,
  A.}}, \citenamefont {{Gomez-Ramos, M.}}, \citenamefont {{Grenard, J.-L.}},
  \citenamefont {{Hobl, A.}}, \citenamefont {{Hupin, G.}}, \citenamefont
  {{Husson, A.}}, \citenamefont {{Indelicato, P.}}, \citenamefont {{Johnston,
  K.}}, \citenamefont {{Klink, C.}}, \citenamefont {{Kubota, Y.}},
  \citenamefont {{Lazauskas, R.}}, \citenamefont {{Malbrunot-Ettenauer, S.}},
  \citenamefont {{Marsic, N.}}, \citenamefont {{O M\"uller, W. F.}},
  \citenamefont {{Naimi, S.}}, \citenamefont {{Nakatsuka, N.}}, \citenamefont
  {{Necca, R.}}, \citenamefont {{Neidherr, D.}}, \citenamefont {{Neyens, G.}},
  \citenamefont {{Obertelli, A.}}, \citenamefont {{Ono, Y.}}, \citenamefont
  {{Pasinelli, S.}}, \citenamefont {{Paul, N.}}, \citenamefont {{Pollacco, E.
  C.}}, \citenamefont {{Rossi, D.}}, \citenamefont {{Scheit, H.}},
  \citenamefont {{Schlaich, M.}}, \citenamefont {{Schmidt, A.}}, \citenamefont
  {{Schweikhard, L.}}, \citenamefont {{Seki, R.}}, \citenamefont {{Sels, S.}},
  \citenamefont {{Siesling, E.}}, \citenamefont {{Uesaka, T.}}, \citenamefont
  {{Vil\'en, M.}}, \citenamefont {{Wada, M.}}, \citenamefont {{Wienholtz, F.}},
  \citenamefont {{Wycech, S.}},\ and\ \citenamefont {{Zacarias,
  S.}}}]{PUMA2022}%
  \BibitemOpen
  \bibfield  {author} {\bibinfo {author} {\bibnamefont {{Aumann, T.}}},
  \bibinfo {author} {\bibnamefont {{Bartmann, W.}}}, \bibinfo {author}
  {\bibnamefont {{Boine-Frankenheim, O.}}},  \emph {et~al.},\ }\href {\doibase
  10.1140/epja/s10050-022-00713-x} {\bibfield  {journal} {\bibinfo  {journal}
  {Eur. Phys. J. A}\ }\textbf {\bibinfo {volume} {58}},\ \bibinfo {pages} {88}
  (\bibinfo {year} {2022})}\BibitemShut {NoStop}%
\bibitem [{\citenamefont {Bohman}\ \emph {et~al.}(2018)\citenamefont {Bohman},
  \citenamefont {Mooser}, \citenamefont {Schneider}, \citenamefont {Schön},
  \citenamefont {Wiesinger}, \citenamefont {Harrington}, \citenamefont
  {Higuchi}, \citenamefont {Nagahama}, \citenamefont {Smorra}, \citenamefont
  {Sellner}, \citenamefont {Blaum}, \citenamefont {Matsuda}, \citenamefont
  {Quint}, \citenamefont {Walz},\ and\ \citenamefont {Ulmer}}]{Boh17}%
  \BibitemOpen
  \bibfield  {author} {\bibinfo {author} {\bibfnamefont {M.}~\bibnamefont
  {Bohman}}, \bibinfo {author} {\bibfnamefont {A.}~\bibnamefont {Mooser}},
  \bibinfo {author} {\bibfnamefont {G.}~\bibnamefont {Schneider}},  \emph
  {et~al.},\ }\href {\doibase 10.1080/09500340.2017.1404656} {\bibfield
  {journal} {\bibinfo  {journal} {Journal of Modern Optics}\ }\textbf {\bibinfo
  {volume} {65}},\ \bibinfo {pages} {568} (\bibinfo {year} {2018})}\BibitemShut
  {NoStop}%
\bibitem [{\citenamefont {Bohman}\ \emph {et~al.}(2021)\citenamefont {Bohman},
  \citenamefont {Grunhofer}, \citenamefont {Smorra}, \citenamefont {Wiesinger},
  \citenamefont {Will}, \citenamefont {Borchert}, \citenamefont {Devlin},
  \citenamefont {Erlewein}, \citenamefont {Fleck}, \citenamefont {Gavranovic},
  \citenamefont {Harrington}, \citenamefont {Latacz}, \citenamefont {Mooser},
  \citenamefont {Popper}, \citenamefont {Wursten}, \citenamefont {Blaum},
  \citenamefont {Matsuda}, \citenamefont {Ospelkaus}, \citenamefont {Quint},
  \citenamefont {Walz},\ and\ \citenamefont {Ulmer}}]{Boh21}%
  \BibitemOpen
  \bibfield  {author} {\bibinfo {author} {\bibfnamefont {M.}~\bibnamefont
  {Bohman}}, \bibinfo {author} {\bibfnamefont {V.}~\bibnamefont {Grunhofer}},
  \bibinfo {author} {\bibfnamefont {C.}~\bibnamefont {Smorra}},  \emph
  {et~al.},\ }\href {\doibase 10.1038/s41586-021-03784-w} {\bibfield  {journal}
  {\bibinfo  {journal} {Nature}\ }\textbf {\bibinfo {volume} {596}},\ \bibinfo
  {pages} {514} (\bibinfo {year} {2021})}\BibitemShut {NoStop}%
\bibitem [{\citenamefont {Will}\ \emph {et~al.}(2022)\citenamefont {Will},
  \citenamefont {Bohman}, \citenamefont {Driscoll}, \citenamefont {Wiesinger},
  \citenamefont {Abbass}, \citenamefont {Borchert}, \citenamefont {Devlin},
  \citenamefont {Erlewein}, \citenamefont {Fleck}, \citenamefont {Latacz},
  \citenamefont {Moller}, \citenamefont {Mooser}, \citenamefont {Popper},
  \citenamefont {Wursten}, \citenamefont {Blaum}, \citenamefont {Matsuda},
  \citenamefont {Ospelkaus}, \citenamefont {Quint}, \citenamefont {Walz},
  \citenamefont {Smorra},\ and\ \citenamefont {Ulmer}}]{Wil22}%
  \BibitemOpen
  \bibfield  {author} {\bibinfo {author} {\bibfnamefont {C.}~\bibnamefont
  {Will}}, \bibinfo {author} {\bibfnamefont {M.}~\bibnamefont {Bohman}},
  \bibinfo {author} {\bibfnamefont {T.}~\bibnamefont {Driscoll}},  \emph
  {et~al.},\ }\href {\doibase 10.1088/1367-2630/ac55b3} {\bibfield  {journal}
  {\bibinfo  {journal} {New Journal of Physics}\ }\textbf {\bibinfo {volume}
  {24}},\ \bibinfo {pages} {033021} (\bibinfo {year} {2022})}\BibitemShut
  {NoStop}%
\bibitem [{\citenamefont {Niemann}\ \emph {et~al.}(2019)\citenamefont
  {Niemann}, \citenamefont {Meiners}, \citenamefont {Mielke}, \citenamefont
  {Borchert}, \citenamefont {Cornejo}, \citenamefont {Ulmer},\ and\
  \citenamefont {Ospelkaus}}]{Nie19}%
  \BibitemOpen
  \bibfield  {author} {\bibinfo {author} {\bibfnamefont {M.}~\bibnamefont
  {Niemann}}, \bibinfo {author} {\bibfnamefont {T.}~\bibnamefont {Meiners}},
  \bibinfo {author} {\bibfnamefont {J.}~\bibnamefont {Mielke}},  \emph
  {et~al.},\ }\href {\doibase 10.1088/1361-6501/ab5722} {\bibfield  {journal}
  {\bibinfo  {journal} {Measurement Science and Technology}\ }\textbf {\bibinfo
  {volume} {31}},\ \bibinfo {pages} {035003} (\bibinfo {year}
  {2019})}\BibitemShut {NoStop}%
\bibitem [{\citenamefont {Mielke}\ \emph {et~al.}(2021)\citenamefont {Mielke},
  \citenamefont {Pick}, \citenamefont {Coenders}, \citenamefont {Meiners},
  \citenamefont {Niemann}, \citenamefont {Cornejo}, \citenamefont {Ulmer},\
  and\ \citenamefont {Ospelkaus}}]{Mie21}%
  \BibitemOpen
  \bibfield  {author} {\bibinfo {author} {\bibfnamefont {J.}~\bibnamefont
  {Mielke}}, \bibinfo {author} {\bibfnamefont {J.}~\bibnamefont {Pick}},
  \bibinfo {author} {\bibfnamefont {J.~A.}\ \bibnamefont {Coenders}},  \emph
  {et~al.},\ }\href@noop {} {\bibfield  {journal} {\bibinfo  {journal} {Journal
  of Physics B-Atomic, Molecular, and Optical Physics}\ }\textbf {\bibinfo
  {volume} {54}} (\bibinfo {year} {2021})}\BibitemShut {NoStop}%
\bibitem [{\citenamefont {Rau}\ \emph {et~al.}(2020)\citenamefont {Rau},
  \citenamefont {Hei{\ss}e}, \citenamefont {K{\"o}hler-Langes}, \citenamefont
  {Sasidharan}, \citenamefont {Haas}, \citenamefont {Renisch}, \citenamefont
  {D{\"u}llmann}, \citenamefont {Quint}, \citenamefont {Sturm},\ and\
  \citenamefont {Blaum}}]{Rau20}%
  \BibitemOpen
  \bibfield  {author} {\bibinfo {author} {\bibfnamefont {S.}~\bibnamefont
  {Rau}}, \bibinfo {author} {\bibfnamefont {F.}~\bibnamefont {Hei{\ss}e}},
  \bibinfo {author} {\bibfnamefont {F.}~\bibnamefont {K{\"o}hler-Langes}},
  \emph {et~al.},\ }\href {\doibase 10.1038/s41586-020-2628-7} {\bibfield
  {journal} {\bibinfo  {journal} {Nature}\ }\textbf {\bibinfo {volume} {585}},\
  \bibinfo {pages} {43} (\bibinfo {year} {2020})}\BibitemShut {NoStop}%
\bibitem [{\citenamefont {Myers}(2018)}]{Mey18}%
  \BibitemOpen
  \bibfield  {author} {\bibinfo {author} {\bibfnamefont {E.~G.}\ \bibnamefont
  {Myers}},\ }\href {\doibase 10.1103/PhysRevA.98.010101} {\bibfield  {journal}
  {\bibinfo  {journal} {Phys. Rev. A}\ }\textbf {\bibinfo {volume} {98}},\
  \bibinfo {pages} {010101} (\bibinfo {year} {2018})}\BibitemShut {NoStop}%
\bibitem [{\citenamefont {Afach}\ \emph {et~al.}(2021)\citenamefont {Afach},
  \citenamefont {Buchler}, \citenamefont {Budker}, \citenamefont {Dailey},
  \citenamefont {Derevianko}, \citenamefont {Dumont}, \citenamefont {Figueroa},
  \citenamefont {Gerhardt}, \citenamefont {Grujić}, \citenamefont {Guo},
  \citenamefont {Hao}, \citenamefont {Hamilton}, \citenamefont {Hedges},
  \citenamefont {Kimball}, \citenamefont {Kim}, \citenamefont {Khamis},
  \citenamefont {Kornack}, \citenamefont {Lebedev}, \citenamefont {Lu},
  \citenamefont {Masia-Roig}, \citenamefont {Monroy}, \citenamefont {Padniuk},
  \citenamefont {Palm}, \citenamefont {Park}, \citenamefont {Paul},
  \citenamefont {Penaflor}, \citenamefont {Peng}, \citenamefont {Pospelov},
  \citenamefont {Preston}, \citenamefont {Pustelny}, \citenamefont {Scholtes},
  \citenamefont {Segura}, \citenamefont {Y.~K.~Semertzidis}, \citenamefont
  {Shin}, \citenamefont {Smiga}, \citenamefont {Stalnaker}, \citenamefont
  {Sulai}, \citenamefont {Tandon}, \citenamefont {Wang}, \citenamefont {Weis},
  \citenamefont {Wickenbrock}, \citenamefont {Wilson}, \citenamefont {Wu},
  \citenamefont {Wurm}, \citenamefont {Xiao}, \citenamefont {Yang},
  \citenamefont {Yu},\ and\ \citenamefont {Zhang}}]{Afa21}%
  \BibitemOpen
  \bibfield  {author} {\bibinfo {author} {\bibfnamefont {S.}~\bibnamefont
  {Afach}}, \bibinfo {author} {\bibfnamefont {B.}~\bibnamefont {Buchler}},
  \bibinfo {author} {\bibfnamefont {D.}~\bibnamefont {Budker}},  \emph
  {et~al.},\ }\href {\doibase 10.1038/s41567-021-01393-y} {\bibfield  {journal}
  {\bibinfo  {journal} {Nature Physics}\ }\textbf {\bibinfo {volume} {17}},\
  \bibinfo {pages} {1396} (\bibinfo {year} {2021})}\BibitemShut {NoStop}%
\bibitem [{\citenamefont {Wcis{\l}o}\ \emph {et~al.}(2018)\citenamefont
  {Wcis{\l}o}, \citenamefont {Ablewski}, \citenamefont {Beloy}, \citenamefont
  {Bilicki}, \citenamefont {Bober}, \citenamefont {Brown}, \citenamefont
  {Fasano}, \citenamefont {Ciury{\l}o}, \citenamefont {Hachisu}, \citenamefont
  {Ido}, \citenamefont {Lodewyck}, \citenamefont {Ludlow}, \citenamefont
  {McGrew}, \citenamefont {Morzy{\'n}ski}, \citenamefont {Nicolodi},
  \citenamefont {Schioppo}, \citenamefont {Sekido}, \citenamefont {Le~Targat},
  \citenamefont {Wolf}, \citenamefont {Zhang}, \citenamefont {Zjawin},\ and\
  \citenamefont {Zawada}}]{Wis18}%
  \BibitemOpen
  \bibfield  {author} {\bibinfo {author} {\bibfnamefont {P.}~\bibnamefont
  {Wcis{\l}o}}, \bibinfo {author} {\bibfnamefont {P.}~\bibnamefont {Ablewski}},
  \bibinfo {author} {\bibfnamefont {K.}~\bibnamefont {Beloy}},  \emph
  {et~al.},\ }\href {https://advances.sciencemag.org/content/4/12/eaau4869}
  {\bibfield  {journal} {\bibinfo  {journal} {Science Advances}\ }\textbf
  {\bibinfo {volume} {4}},\ \bibinfo {pages} {12} (\bibinfo {year}
  {2018})}\BibitemShut {NoStop}%
\bibitem [{\citenamefont {Budker}\ \emph {et~al.}(2022)\citenamefont {Budker},
  \citenamefont {Graham}, \citenamefont {Ramani}, \citenamefont
  {Schmidt-Kaler}, \citenamefont {Smorra},\ and\ \citenamefont
  {Ulmer}}]{Bud22}%
  \BibitemOpen
  \bibfield  {author} {\bibinfo {author} {\bibfnamefont {D.}~\bibnamefont
  {Budker}}, \bibinfo {author} {\bibfnamefont {P.~W.}\ \bibnamefont {Graham}},
  \bibinfo {author} {\bibfnamefont {H.}~\bibnamefont {Ramani}},  \emph
  {et~al.},\ }\href {\doibase 10.1103/PRXQuantum.3.010330} {\bibfield
  {journal} {\bibinfo  {journal} {PRX Quantum}\ }\textbf {\bibinfo {volume}
  {3}},\ \bibinfo {pages} {010330} (\bibinfo {year} {2022})}\BibitemShut
  {NoStop}%
\bibitem [{\citenamefont {Smorra}\ \emph
  {et~al.}(2015{\natexlab{a}})\citenamefont {Smorra}, \citenamefont {Blaum},
  \citenamefont {Bojtar}, \citenamefont {Borchert}, \citenamefont {Franke},
  \citenamefont {Higuchi}, \citenamefont {Leefer}, \citenamefont {Nagahama},
  \citenamefont {Matsuda}, \citenamefont {Mooser} \emph {et~al.}}]{Smo15}%
  \BibitemOpen
  \bibfield  {author} {\bibinfo {author} {\bibfnamefont {C.}~\bibnamefont
  {Smorra}}, \bibinfo {author} {\bibfnamefont {K.}~\bibnamefont {Blaum}},
  \bibinfo {author} {\bibfnamefont {L.}~\bibnamefont {Bojtar}},  \emph
  {et~al.},\ }\href@noop {} {\bibfield  {journal} {\bibinfo  {journal} {The
  European Physical Journal Special Topics}\ }\textbf {\bibinfo {volume}
  {224}},\ \bibinfo {pages} {3055} (\bibinfo {year}
  {2015}{\natexlab{a}})}\BibitemShut {NoStop}%
\bibitem [{\citenamefont {Smorra}\ and\ \citenamefont {Mooser}(2020)}]{Smo20}%
  \BibitemOpen
  \bibfield  {author} {\bibinfo {author} {\bibfnamefont {C.}~\bibnamefont
  {Smorra}}\ and\ \bibinfo {author} {\bibfnamefont {A.}~\bibnamefont
  {Mooser}},\ }in\ \href@noop {} {\emph {\bibinfo {booktitle} {Journal of
  Physics: Conference Series}}},\ Vol.\ \bibinfo {volume} {1412}\ (\bibinfo
  {organization} {IOP Publishing},\ \bibinfo {year} {2020})\ p.\ \bibinfo
  {pages} {032001}\BibitemShut {NoStop}%
\bibitem [{\citenamefont {Brown}\ and\ \citenamefont
  {Gabrielse}(1982)}]{Bro82}%
  \BibitemOpen
  \bibfield  {author} {\bibinfo {author} {\bibfnamefont {L.~S.}\ \bibnamefont
  {Brown}}\ and\ \bibinfo {author} {\bibfnamefont {G.}~\bibnamefont
  {Gabrielse}},\ }\href {\doibase 10.1103/PhysRevA.25.2423} {\bibfield
  {journal} {\bibinfo  {journal} {Phys. Rev. A}\ }\textbf {\bibinfo {volume}
  {25}},\ \bibinfo {pages} {2423} (\bibinfo {year} {1982})}\BibitemShut
  {NoStop}%
\bibitem [{\citenamefont {Dehmelt}(1986)}]{Deh86}%
  \BibitemOpen
  \bibfield  {author} {\bibinfo {author} {\bibfnamefont {H.}~\bibnamefont
  {Dehmelt}},\ }\href@noop {} {\bibfield  {journal} {\bibinfo  {journal}
  {Proceedings of the National Academy of Sciences}\ }\textbf {\bibinfo
  {volume} {83}},\ \bibinfo {pages} {2291} (\bibinfo {year}
  {1986})}\BibitemShut {NoStop}%
\bibitem [{\citenamefont {Nagahama}\ \emph {et~al.}(2016)\citenamefont
  {Nagahama}, \citenamefont {Schneider}, \citenamefont {Mooser}, \citenamefont
  {Smorra}, \citenamefont {Sellner}, \citenamefont {Harrington}, \citenamefont
  {Higuchi}, \citenamefont {Borchert}, \citenamefont {Tanaka}, \citenamefont
  {Besirli}, \citenamefont {Blaum}, \citenamefont {Matsuda}, \citenamefont
  {Ospelkaus}, \citenamefont {Quint}, \citenamefont {Walz}, \citenamefont
  {Yamazaki},\ and\ \citenamefont {Ulmer}}]{Nag16}%
  \BibitemOpen
  \bibfield  {author} {\bibinfo {author} {\bibfnamefont {H.}~\bibnamefont
  {Nagahama}}, \bibinfo {author} {\bibfnamefont {G.}~\bibnamefont {Schneider}},
  \bibinfo {author} {\bibfnamefont {A.}~\bibnamefont {Mooser}},  \emph
  {et~al.},\ }\href {\doibase 10.1063/1.4967493} {\bibfield  {journal}
  {\bibinfo  {journal} {Review of Scientific Instruments}\ }\textbf {\bibinfo
  {volume} {87}},\ \bibinfo {pages} {113305} (\bibinfo {year}
  {2016})}\BibitemShut {NoStop}%
\bibitem [{\citenamefont {Thompson}\ \emph {et~al.}(2004)\citenamefont
  {Thompson}, \citenamefont {Rainville},\ and\ \citenamefont
  {Pritchard}}]{Tho04}%
  \BibitemOpen
  \bibfield  {author} {\bibinfo {author} {\bibfnamefont {J.~K.}\ \bibnamefont
  {Thompson}}, \bibinfo {author} {\bibfnamefont {S.}~\bibnamefont {Rainville}},
  \ and\ \bibinfo {author} {\bibfnamefont {D.~E.}\ \bibnamefont {Pritchard}},\
  }\href {\doibase 10.1038/nature02682} {\bibfield  {journal} {\bibinfo
  {journal} {Nature}\ }\textbf {\bibinfo {volume} {430}},\ \bibinfo {pages}
  {58} (\bibinfo {year} {2004})}\BibitemShut {NoStop}%
\bibitem [{\citenamefont {Van Dyck~Jr}\ \emph {et~al.}(1999)\citenamefont {Van
  Dyck~Jr}, \citenamefont {Farnham}, \citenamefont {Zafonte},\ and\
  \citenamefont {Schwinberg}}]{Van99}%
  \BibitemOpen
  \bibfield  {author} {\bibinfo {author} {\bibfnamefont {R.~S.}\ \bibnamefont
  {Van Dyck~Jr}}, \bibinfo {author} {\bibfnamefont {D.}~\bibnamefont
  {Farnham}}, \bibinfo {author} {\bibfnamefont {S.}~\bibnamefont {Zafonte}}, \
  and\ \bibinfo {author} {\bibfnamefont {P.}~\bibnamefont {Schwinberg}},\
  }\href@noop {} {\bibfield  {journal} {\bibinfo  {journal} {Review of
  scientific instruments}\ }\textbf {\bibinfo {volume} {70}},\ \bibinfo {pages}
  {1665} (\bibinfo {year} {1999})}\BibitemShut {NoStop}%
\bibitem [{\citenamefont {Kromer}\ \emph {et~al.}(2022)\citenamefont {Kromer},
  \citenamefont {Lyu}, \citenamefont {Door}, \citenamefont {Filianin},
  \citenamefont {Harman}, \citenamefont {Herkenhoff}, \citenamefont {Huang},
  \citenamefont {Keitel}, \citenamefont {Lange}, \citenamefont {Novikov},
  \citenamefont {Schweiger}, \citenamefont {Eliseev},\ and\ \citenamefont
  {Blaum}}]{Kro22}%
  \BibitemOpen
  \bibfield  {author} {\bibinfo {author} {\bibfnamefont {K.}~\bibnamefont
  {Kromer}}, \bibinfo {author} {\bibfnamefont {C.}~\bibnamefont {Lyu}},
  \bibinfo {author} {\bibfnamefont {M.}~\bibnamefont {Door}},  \emph {et~al.},\
  }\href@noop {} {\bibfield  {journal} {\bibinfo  {journal} {The European
  Physical Journal A}\ }\textbf {\bibinfo {volume} {58}},\ \bibinfo {pages}
  {202} (\bibinfo {year} {2022})}\BibitemShut {NoStop}%
\bibitem [{\citenamefont {Gabrielse}\ \emph
  {et~al.}(1989{\natexlab{a}})\citenamefont {Gabrielse}, \citenamefont
  {Haarsma},\ and\ \citenamefont {Rolston}}]{Gab89}%
  \BibitemOpen
  \bibfield  {author} {\bibinfo {author} {\bibfnamefont {G.}~\bibnamefont
  {Gabrielse}}, \bibinfo {author} {\bibfnamefont {L.}~\bibnamefont {Haarsma}},
  \ and\ \bibinfo {author} {\bibfnamefont {S.~L.}\ \bibnamefont {Rolston}},\
  }\href {\doibase https://doi.org/10.1016/0168-1176(89)85027-X} {\bibfield
  {journal} {\bibinfo  {journal} {International Journal of Mass Spectrometry
  and Ion Processes}\ }\textbf {\bibinfo {volume} {88}},\ \bibinfo {pages} {319
  } (\bibinfo {year} {1989}{\natexlab{a}})}\BibitemShut {NoStop}%
\bibitem [{\citenamefont {Devlin}\ \emph {et~al.}(2019)\citenamefont {Devlin},
  \citenamefont {Wursten}, \citenamefont {Harrington}, \citenamefont {Higuchi},
  \citenamefont {Blessing}, \citenamefont {Borchert}, \citenamefont {Erlewein},
  \citenamefont {Hansen}, \citenamefont {Morgner}, \citenamefont {Bohman},
  \citenamefont {Mooser}, \citenamefont {Smorra}, \citenamefont {Wiesinger},
  \citenamefont {Blaum}, \citenamefont {Matsuda}, \citenamefont {Ospelkaus},
  \citenamefont {Quint}, \citenamefont {Walz}, \citenamefont {Yamazaki},\ and\
  \citenamefont {Ulmer}}]{Dev19}%
  \BibitemOpen
  \bibfield  {author} {\bibinfo {author} {\bibfnamefont {J.~A.}\ \bibnamefont
  {Devlin}}, \bibinfo {author} {\bibfnamefont {E.}~\bibnamefont {Wursten}},
  \bibinfo {author} {\bibfnamefont {J.~A.}\ \bibnamefont {Harrington}},  \emph
  {et~al.},\ }\href {\doibase 10.1103/PhysRevApplied.12.044012} {\bibfield
  {journal} {\bibinfo  {journal} {Phys. Rev. Applied}\ }\textbf {\bibinfo
  {volume} {12}},\ \bibinfo {pages} {044012} (\bibinfo {year}
  {2019})}\BibitemShut {NoStop}%
\bibitem [{\citenamefont {Cornell}\ \emph {et~al.}(1990)\citenamefont
  {Cornell}, \citenamefont {Weisskoff}, \citenamefont {Boyce},\ and\
  \citenamefont {Pritchard}}]{Cor90}%
  \BibitemOpen
  \bibfield  {author} {\bibinfo {author} {\bibfnamefont {E.~A.}\ \bibnamefont
  {Cornell}}, \bibinfo {author} {\bibfnamefont {R.~M.}\ \bibnamefont
  {Weisskoff}}, \bibinfo {author} {\bibfnamefont {K.~R.}\ \bibnamefont
  {Boyce}}, \ and\ \bibinfo {author} {\bibfnamefont {D.~E.}\ \bibnamefont
  {Pritchard}},\ }\href {\doibase 10.1103/PhysRevA.41.312} {\bibfield
  {journal} {\bibinfo  {journal} {Phys. Rev. A}\ }\textbf {\bibinfo {volume}
  {41}},\ \bibinfo {pages} {312} (\bibinfo {year} {1990})}\BibitemShut
  {NoStop}%
\bibitem [{\citenamefont {Sturm}\ \emph {et~al.}(2011)\citenamefont {Sturm},
  \citenamefont {Wagner}, \citenamefont {Schabinger},\ and\ \citenamefont
  {Blaum}}]{Stu11}%
  \BibitemOpen
  \bibfield  {author} {\bibinfo {author} {\bibfnamefont {S.}~\bibnamefont
  {Sturm}}, \bibinfo {author} {\bibfnamefont {A.}~\bibnamefont {Wagner}},
  \bibinfo {author} {\bibfnamefont {B.}~\bibnamefont {Schabinger}}, \ and\
  \bibinfo {author} {\bibfnamefont {K.}~\bibnamefont {Blaum}},\ }\href
  {\doibase 10.1103/PhysRevLett.107.143003} {\bibfield  {journal} {\bibinfo
  {journal} {Phys. Rev. Lett.}\ }\textbf {\bibinfo {volume} {107}},\ \bibinfo
  {pages} {143003} (\bibinfo {year} {2011})}\BibitemShut {NoStop}%
\bibitem [{\citenamefont {Gabrielse}\ \emph {et~al.}(1999)\citenamefont
  {Gabrielse}, \citenamefont {Khabbaz}, \citenamefont {Hall}, \citenamefont
  {Heimann}, \citenamefont {Kalinowsky},\ and\ \citenamefont {Jhe}}]{Gab99}%
  \BibitemOpen
  \bibfield  {author} {\bibinfo {author} {\bibfnamefont {G.}~\bibnamefont
  {Gabrielse}}, \bibinfo {author} {\bibfnamefont {A.}~\bibnamefont {Khabbaz}},
  \bibinfo {author} {\bibfnamefont {D.~S.}\ \bibnamefont {Hall}},  \emph
  {et~al.},\ }\href {\doibase 10.1103/PhysRevLett.82.3198} {\bibfield
  {journal} {\bibinfo  {journal} {Phys. Rev. Lett.}\ }\textbf {\bibinfo
  {volume} {82}},\ \bibinfo {pages} {3198} (\bibinfo {year}
  {1999})}\BibitemShut {NoStop}%
\bibitem [{\citenamefont {Cornell}\ \emph {et~al.}(1989)\citenamefont
  {Cornell}, \citenamefont {Weisskoff}, \citenamefont {Boyce}, \citenamefont
  {Flanagan}, \citenamefont {Lafyatis},\ and\ \citenamefont
  {Pritchard}}]{Cor89}%
  \BibitemOpen
  \bibfield  {author} {\bibinfo {author} {\bibfnamefont {E.~A.}\ \bibnamefont
  {Cornell}}, \bibinfo {author} {\bibfnamefont {R.~M.}\ \bibnamefont
  {Weisskoff}}, \bibinfo {author} {\bibfnamefont {K.~R.}\ \bibnamefont
  {Boyce}},  \emph {et~al.},\ }\href {\doibase 10.1103/PhysRevLett.63.1674}
  {\bibfield  {journal} {\bibinfo  {journal} {Phys. Rev. Lett.}\ }\textbf
  {\bibinfo {volume} {63}},\ \bibinfo {pages} {1674} (\bibinfo {year}
  {1989})}\BibitemShut {NoStop}%
\bibitem [{\citenamefont {Borchert}(2021)}]{BorTh}%
  \BibitemOpen
  \bibfield  {author} {\bibinfo {author} {\bibfnamefont {M.~J.}\ \bibnamefont
  {Borchert}},\ }\emph {\bibinfo {title} {Challenging the Standard Model by
  high precision comparisons of the fundamental properties of antiprotons and
  protons}},\ \href@noop {} {Ph.D. thesis},\ \bibinfo  {school} {Gottfried
  Wilhelm Leibniz Universität Hannover} (\bibinfo {year} {2021})\BibitemShut
  {NoStop}%
\bibitem [{\citenamefont {Ulmer}\ \emph {et~al.}(2013)\citenamefont {Ulmer},
  \citenamefont {Blaum}, \citenamefont {Kracke}, \citenamefont {Mooser},
  \citenamefont {Quint}, \citenamefont {Rodegheri},\ and\ \citenamefont
  {Walz}}]{Ulm13}%
  \BibitemOpen
  \bibfield  {author} {\bibinfo {author} {\bibfnamefont {S.}~\bibnamefont
  {Ulmer}}, \bibinfo {author} {\bibfnamefont {K.}~\bibnamefont {Blaum}},
  \bibinfo {author} {\bibfnamefont {H.}~\bibnamefont {Kracke}},  \emph
  {et~al.},\ }\href {\doibase https://doi.org/10.1016/j.nima.2012.12.071}
  {\bibfield  {journal} {\bibinfo  {journal} {Nuclear Instruments and Methods
  in Physics Research Section A: Accelerators, Spectrometers, Detectors and
  Associated Equipment}\ }\textbf {\bibinfo {volume} {705}},\ \bibinfo {pages}
  {55} (\bibinfo {year} {2013})}\BibitemShut {NoStop}%
\bibitem [{\citenamefont {Ketter}\ \emph {et~al.}(2014)\citenamefont {Ketter},
  \citenamefont {Eronen}, \citenamefont {H{\"o}cker}, \citenamefont
  {Streubel},\ and\ \citenamefont {Blaum}}]{Ket14}%
  \BibitemOpen
  \bibfield  {author} {\bibinfo {author} {\bibfnamefont {J.}~\bibnamefont
  {Ketter}}, \bibinfo {author} {\bibfnamefont {T.}~\bibnamefont {Eronen}},
  \bibinfo {author} {\bibfnamefont {M.}~\bibnamefont {H{\"o}cker}},  \emph
  {et~al.},\ }\href {\doibase https://doi.org/10.1016/j.ijms.2013.10.005}
  {\bibfield  {journal} {\bibinfo  {journal} {International Journal of Mass
  Spectrometry}\ }\textbf {\bibinfo {volume} {358}},\ \bibinfo {pages} {1 }
  (\bibinfo {year} {2014})}\BibitemShut {NoStop}%
\bibitem [{\citenamefont {Sch{\"u}ssler}\ \emph {et~al.}(2020)\citenamefont
  {Sch{\"u}ssler}, \citenamefont {Bekker}, \citenamefont {Bra{\ss}},
  \citenamefont {Cakir}, \citenamefont {Crespo L{\'o}pez-Urrutia},
  \citenamefont {Door}, \citenamefont {Filianin}, \citenamefont {Harman},
  \citenamefont {Haverkort}, \citenamefont {Huang}, \citenamefont {Indelicato},
  \citenamefont {Keitel}, \citenamefont {K{\"o}nig}, \citenamefont {Kromer},
  \citenamefont {M{\"u}ller}, \citenamefont {Novikov}, \citenamefont {Rischka},
  \citenamefont {Schweiger}, \citenamefont {Sturm}, \citenamefont {Ulmer},
  \citenamefont {Eliseev},\ and\ \citenamefont {Blaum}}]{Sch20}%
  \BibitemOpen
  \bibfield  {author} {\bibinfo {author} {\bibfnamefont {R.~X.}\ \bibnamefont
  {Sch{\"u}ssler}}, \bibinfo {author} {\bibfnamefont {H.}~\bibnamefont
  {Bekker}}, \bibinfo {author} {\bibfnamefont {M.}~\bibnamefont {Bra{\ss}}},
  \emph {et~al.},\ }\href {\doibase 10.1038/s41586-020-2221-0} {\bibfield
  {journal} {\bibinfo  {journal} {Nature}\ }\textbf {\bibinfo {volume} {581}},\
  \bibinfo {pages} {42} (\bibinfo {year} {2020})}\BibitemShut {NoStop}%
\bibitem [{\citenamefont {Heiße}\ \emph {et~al.}(2017)\citenamefont {Heiße},
  \citenamefont {Köhler-Langes}, \citenamefont {Rau}, \citenamefont {Hou},
  \citenamefont {Junck}, \citenamefont {Kracke}, \citenamefont {Mooser},
  \citenamefont {Quint}, \citenamefont {Ulmer}, \citenamefont {Werth},
  \citenamefont {Blaum},\ and\ \citenamefont {Sturm}}]{Hei17}%
  \BibitemOpen
  \bibfield  {author} {\bibinfo {author} {\bibfnamefont {F.}~\bibnamefont
  {Heiße}}, \bibinfo {author} {\bibfnamefont {F.}~\bibnamefont
  {Köhler-Langes}}, \bibinfo {author} {\bibfnamefont {S.}~\bibnamefont {Rau}},
   \emph {et~al.},\ }\href {\doibase 10.1103/PhysRevLett.119.033001} {\bibfield
   {journal} {\bibinfo  {journal} {Phys. Rev. Lett.}\ }\textbf {\bibinfo
  {volume} {119}},\ \bibinfo {pages} {033001} (\bibinfo {year}
  {2017})}\BibitemShut {NoStop}%
\bibitem [{\citenamefont {Myers}(2013)}]{Mye13}%
  \BibitemOpen
  \bibfield  {author} {\bibinfo {author} {\bibfnamefont {E.~G.}\ \bibnamefont
  {Myers}},\ }\href
  {https://www.sciencedirect.com/science/article/pii/S1387380613001097}
  {\bibfield  {journal} {\bibinfo  {journal} {International Journal of Mass
  Spectrometry}\ }\textbf {\bibinfo {volume} {349-350}},\ \bibinfo {pages}
  {107} (\bibinfo {year} {2013})},\ \bibinfo {note} {100 years of Mass
  Spectrometry}\BibitemShut {NoStop}%
\bibitem [{\citenamefont {Riley}(1995)}]{Ril95}%
  \BibitemOpen
  \bibfield  {author} {\bibinfo {author} {\bibfnamefont {W.}~\bibnamefont
  {Riley}},\ }\href@noop {} {\emph {\bibinfo {title} {Handbook of Frequency
  Stability Analysis}}}\ (\bibinfo  {publisher} {National Institute of
  Standards and Technology},\ \bibinfo {address} {Boulder, CO},\ \bibinfo
  {year} {1995})\BibitemShut {NoStop}%
\bibitem [{\citenamefont {Liu}\ \emph {et~al.}(2021)\citenamefont {Liu},
  \citenamefont {Schnabel}, \citenamefont {Voigt}, \citenamefont {Kilian},
  \citenamefont {Sun}, \citenamefont {Li},\ and\ \citenamefont
  {Trahms}}]{Liu2021}%
  \BibitemOpen
  \bibfield  {author} {\bibinfo {author} {\bibfnamefont {T.}~\bibnamefont
  {Liu}}, \bibinfo {author} {\bibfnamefont {A.}~\bibnamefont {Schnabel}},
  \bibinfo {author} {\bibfnamefont {J.}~\bibnamefont {Voigt}},  \emph
  {et~al.},\ }\href {\doibase 10.1063/5.0027848} {\bibfield  {journal}
  {\bibinfo  {journal} {Review of Scientific Instruments}\ }\textbf {\bibinfo
  {volume} {92}},\ \bibinfo {pages} {024709} (\bibinfo {year}
  {2021})}\BibitemShut {NoStop}%
\bibitem [{\citenamefont {DiSciacca}\ \emph {et~al.}(2013)\citenamefont
  {DiSciacca}, \citenamefont {Marshall}, \citenamefont {Marable}, \citenamefont
  {Gabrielse}, \citenamefont {Ettenauer}, \citenamefont {Tardiff},
  \citenamefont {Kalra}, \citenamefont {Fitzakerley}, \citenamefont {George},
  \citenamefont {Hessels}, \citenamefont {Storry}, \citenamefont {Weel},
  \citenamefont {Grzonka}, \citenamefont {Oelert},\ and\ \citenamefont
  {Sefzick}}]{DiS2013}%
  \BibitemOpen
  \bibfield  {author} {\bibinfo {author} {\bibfnamefont {J.}~\bibnamefont
  {DiSciacca}}, \bibinfo {author} {\bibfnamefont {M.}~\bibnamefont {Marshall}},
  \bibinfo {author} {\bibfnamefont {K.}~\bibnamefont {Marable}},  \emph
  {et~al.} (\bibinfo {collaboration} {ATRAP Collaboration}),\ }\href {\doibase
  10.1103/PhysRevLett.110.130801} {\bibfield  {journal} {\bibinfo  {journal}
  {Phys. Rev. Lett.}\ }\textbf {\bibinfo {volume} {110}},\ \bibinfo {pages}
  {130801} (\bibinfo {year} {2013})}\BibitemShut {NoStop}%
\bibitem [{\citenamefont {Barna}\ \emph {et~al.}(2015)\citenamefont {Barna},
  \citenamefont {Bartmann}, \citenamefont {Fraser},\ and\ \citenamefont
  {Ostojić}}]{Barna2015}%
  \BibitemOpen
  \bibfield  {author} {\bibinfo {author} {\bibfnamefont {D.}~\bibnamefont
  {Barna}}, \bibinfo {author} {\bibfnamefont {W.}~\bibnamefont {Bartmann}},
  \bibinfo {author} {\bibfnamefont {M.}~\bibnamefont {Fraser}}, \ and\ \bibinfo
  {author} {\bibfnamefont {R.}~\bibnamefont {Ostojić}},\ }in\ \href
  {http://jacow.org/ipac2015/papers/mopje043.pdf} {\emph {\bibinfo {booktitle}
  {Proc. 6th International Particle Accelerator Conference (IPAC'15), Richmond,
  VA, USA, May 3-8, 2015}}},\ \bibinfo {series and number} {\bibinfo {series}
  {International Particle Accelerator Conference}\ No.~\bibinfo {number} {6}}\
  (\bibinfo  {publisher} {JACoW},\ \bibinfo {address} {Geneva, Switzerland},\
  \bibinfo {year} {2015})\ pp.\ \bibinfo {pages} {382--384},\ \bibinfo {note}
  {https://doi.org/10.18429/JACoW-IPAC2015-MOPJE043}\BibitemShut {NoStop}%
\bibitem [{\citenamefont {Smorra}\ \emph {et~al.}(2021)\citenamefont {Smorra},
  \citenamefont {Ulmer}, \citenamefont {Walz}, \citenamefont {Latacz},
  \citenamefont {Popper}, \citenamefont {Wiesinger}, \citenamefont {Mooser},
  \citenamefont {Borchert}, \citenamefont {Dutheil}, \citenamefont {Matsuda}
  \emph {et~al.}}]{Smo21}%
  \BibitemOpen
  \bibfield  {author} {\bibinfo {author} {\bibfnamefont {C.}~\bibnamefont
  {Smorra}}, \bibinfo {author} {\bibfnamefont {S.}~\bibnamefont {Ulmer}},
  \bibinfo {author} {\bibfnamefont {J.}~\bibnamefont {Walz}},  \emph {et~al.},\
  }\href@noop {} {\emph {\bibinfo {title} {Technical Design Report of
  BASE-STEP}}},\ \bibinfo {type} {Tech. Rep.}\ (\bibinfo  {institution}
  {CERN},\ \bibinfo {year} {2021})\BibitemShut {NoStop}%
\bibitem [{\citenamefont {Gabrielse}\ \emph
  {et~al.}(1989{\natexlab{b}})\citenamefont {Gabrielse}, \citenamefont {Fei},
  \citenamefont {Orozco}, \citenamefont {Tjoelker}, \citenamefont {Haas},
  \citenamefont {Kalinowsky}, \citenamefont {Trainor},\ and\ \citenamefont
  {Kells}}]{Gab89EC}%
  \BibitemOpen
  \bibfield  {author} {\bibinfo {author} {\bibfnamefont {G.}~\bibnamefont
  {Gabrielse}}, \bibinfo {author} {\bibfnamefont {X.}~\bibnamefont {Fei}},
  \bibinfo {author} {\bibfnamefont {L.~A.}\ \bibnamefont {Orozco}},  \emph
  {et~al.},\ }\href {\doibase 10.1103/PhysRevLett.63.1360} {\bibfield
  {journal} {\bibinfo  {journal} {Phys. Rev. Lett.}\ }\textbf {\bibinfo
  {volume} {63}},\ \bibinfo {pages} {1360} (\bibinfo {year}
  {1989}{\natexlab{b}})}\BibitemShut {NoStop}%
\bibitem [{\citenamefont {Blaum}(2006)}]{Bla06}%
  \BibitemOpen
  \bibfield  {author} {\bibinfo {author} {\bibfnamefont {K.}~\bibnamefont
  {Blaum}},\ }\href {\doibase https://doi.org/10.1016/j.physrep.2005.10.011}
  {\bibfield  {journal} {\bibinfo  {journal} {Physics Reports}\ }\textbf
  {\bibinfo {volume} {425}},\ \bibinfo {pages} {1} (\bibinfo {year}
  {2006})}\BibitemShut {NoStop}%
\bibitem [{\citenamefont {Smorra}\ \emph
  {et~al.}(2015{\natexlab{b}})\citenamefont {Smorra}, \citenamefont {Mooser},
  \citenamefont {Franke}, \citenamefont {Nagahama}, \citenamefont {Schneider},
  \citenamefont {Higuchi}, \citenamefont {Gorp}, \citenamefont {Blaum},
  \citenamefont {Matsuda}, \citenamefont {Quint}, \citenamefont {Walz},
  \citenamefont {Yamazaki},\ and\ \citenamefont {Ulmer}}]{Smo15b}%
  \BibitemOpen
  \bibfield  {author} {\bibinfo {author} {\bibfnamefont {C.}~\bibnamefont
  {Smorra}}, \bibinfo {author} {\bibfnamefont {A.}~\bibnamefont {Mooser}},
  \bibinfo {author} {\bibfnamefont {K.}~\bibnamefont {Franke}},  \emph
  {et~al.},\ }\href {\doibase https://doi.org/10.1016/j.ijms.2015.08.007}
  {\bibfield  {journal} {\bibinfo  {journal} {International Journal of Mass
  Spectrometry}\ }\textbf {\bibinfo {volume} {389}},\ \bibinfo {pages} {10}
  (\bibinfo {year} {2015}{\natexlab{b}})}\BibitemShut {NoStop}%
\bibitem [{\citenamefont {Block}\ \emph {et~al.}(2005)\citenamefont {Block},
  \citenamefont {Ackermann}, \citenamefont {Beck}, \citenamefont {Blaum},
  \citenamefont {Breitenfeldt}, \citenamefont {Chauduri}, \citenamefont
  {Doemer}, \citenamefont {Eliseev}, \citenamefont {Habs}, \citenamefont
  {Heinz} \emph {et~al.}}]{Blo05}%
  \BibitemOpen
  \bibfield  {author} {\bibinfo {author} {\bibfnamefont {M.}~\bibnamefont
  {Block}}, \bibinfo {author} {\bibfnamefont {D.}~\bibnamefont {Ackermann}},
  \bibinfo {author} {\bibfnamefont {D.}~\bibnamefont {Beck}},  \emph {et~al.},\
  }\href@noop {} {\bibfield  {journal} {\bibinfo  {journal} {The European
  Physical Journal A-Hadrons and Nuclei}\ }\textbf {\bibinfo {volume} {25}},\
  \bibinfo {pages} {49} (\bibinfo {year} {2005})}\BibitemShut {NoStop}%
\bibitem [{\citenamefont {Block}(2015)}]{Blo15}%
  \BibitemOpen
  \bibfield  {author} {\bibinfo {author} {\bibfnamefont {M.}~\bibnamefont
  {Block}},\ }\href@noop {} {\bibfield  {journal} {\bibinfo  {journal} {Nuclear
  Physics A}\ }\textbf {\bibinfo {volume} {944}},\ \bibinfo {pages} {471}
  (\bibinfo {year} {2015})}\BibitemShut {NoStop}%
\bibitem [{\citenamefont {Mukherjee}\ \emph {et~al.}(2008)\citenamefont
  {Mukherjee}, \citenamefont {Beck}, \citenamefont {Blaum}, \citenamefont
  {Bollen}, \citenamefont {Dilling}, \citenamefont {George}, \citenamefont
  {Herfurth}, \citenamefont {Herlert}, \citenamefont {Kellerbauer},
  \citenamefont {Kluge} \emph {et~al.}}]{Muk08}%
  \BibitemOpen
  \bibfield  {author} {\bibinfo {author} {\bibfnamefont {M.}~\bibnamefont
  {Mukherjee}}, \bibinfo {author} {\bibfnamefont {D.}~\bibnamefont {Beck}},
  \bibinfo {author} {\bibfnamefont {K.}~\bibnamefont {Blaum}},  \emph
  {et~al.},\ }\href@noop {} {\bibfield  {journal} {\bibinfo  {journal} {The
  European Physical Journal A}\ }\textbf {\bibinfo {volume} {35}},\ \bibinfo
  {pages} {1} (\bibinfo {year} {2008})}\BibitemShut {NoStop}%
\bibitem [{\citenamefont {Hamaker}\ \emph {et~al.}(2016)\citenamefont
  {Hamaker}, \citenamefont {Brodeur}, \citenamefont {Kelly}, \citenamefont
  {Long}, \citenamefont {Nicoloff}, \citenamefont {Ryan}, \citenamefont
  {Schultz}, \citenamefont {Schury},\ and\ \citenamefont {Wada}}]{Ham16}%
  \BibitemOpen
  \bibfield  {author} {\bibinfo {author} {\bibfnamefont {A.}~\bibnamefont
  {Hamaker}}, \bibinfo {author} {\bibfnamefont {M.}~\bibnamefont {Brodeur}},
  \bibinfo {author} {\bibfnamefont {J.~M.}\ \bibnamefont {Kelly}},  \emph
  {et~al.},\ }\href {\doibase 10.1016/j.ijms.2016.04.004} {\bibfield  {journal}
  {\bibinfo  {journal} {Int. J. Mass Spectr.}\ }\textbf {\bibinfo {volume}
  {404}},\ \bibinfo {pages} {14} (\bibinfo {year} {2016})}\BibitemShut
  {NoStop}%
\bibitem [{\citenamefont {Bollen}(2011)}]{Bol11}%
  \BibitemOpen
  \bibfield  {author} {\bibinfo {author} {\bibfnamefont {G.}~\bibnamefont
  {Bollen}},\ }\href {\doibase 10.1016/j.ijms.2010.09.032} {\bibfield
  {journal} {\bibinfo  {journal} {Int. J. Mass Spectr.}\ }\textbf {\bibinfo
  {volume} {299}},\ \bibinfo {pages} {131} (\bibinfo {year}
  {2011})}\BibitemShut {NoStop}%
\bibitem [{\citenamefont {Danielson}\ and\ \citenamefont
  {Surko}(2006)}]{Surko2006}%
  \BibitemOpen
  \bibfield  {author} {\bibinfo {author} {\bibfnamefont {J.~R.}\ \bibnamefont
  {Danielson}}\ and\ \bibinfo {author} {\bibfnamefont {C.~M.}\ \bibnamefont
  {Surko}},\ }\href {\doibase 10.1063/1.2179410} {\bibfield  {journal}
  {\bibinfo  {journal} {Physics of Plasmas}\ }\textbf {\bibinfo {volume}
  {13}},\ \bibinfo {pages} {055706} (\bibinfo {year} {2006})}\BibitemShut
  {NoStop}%
\bibitem [{\citenamefont {Sellner}\ \emph {et~al.}(2017)\citenamefont
  {Sellner}, \citenamefont {Besirli}, \citenamefont {Bohman}, \citenamefont
  {Borchert}, \citenamefont {Harrington}, \citenamefont {Higuchi},
  \citenamefont {Mooser}, \citenamefont {Nagahama}, \citenamefont {Schneider},
  \citenamefont {Smorra}, \citenamefont {Tanaka}, \citenamefont {Blaum},
  \citenamefont {Matsuda}, \citenamefont {Ospelkaus}, \citenamefont {Quint},
  \citenamefont {Walz}, \citenamefont {Yamazaki},\ and\ \citenamefont
  {Ulmer}}]{Sel17}%
  \BibitemOpen
  \bibfield  {author} {\bibinfo {author} {\bibfnamefont {S.}~\bibnamefont
  {Sellner}}, \bibinfo {author} {\bibfnamefont {M.}~\bibnamefont {Besirli}},
  \bibinfo {author} {\bibfnamefont {M.}~\bibnamefont {Bohman}},  \emph
  {et~al.},\ }\href {\doibase 10.1088/1367-2630/aa7e73} {\bibfield  {journal}
  {\bibinfo  {journal} {New Journal of Physics}\ }\textbf {\bibinfo {volume}
  {19}},\ \bibinfo {pages} {083023} (\bibinfo {year} {2017})}\BibitemShut
  {NoStop}%
\bibitem [{\citenamefont {Micke}\ \emph {et~al.}(2019)\citenamefont {Micke},
  \citenamefont {Stark}, \citenamefont {King}, \citenamefont {Leopold},
  \citenamefont {Pfeifer}, \citenamefont {Schm{\"o}ger}, \citenamefont
  {Schwarz}, \citenamefont {Spie{\ss}}, \citenamefont {Schmidt},\ and\
  \citenamefont {Crespo L{\'o}pez-Urrutia}}]{Mic19}%
  \BibitemOpen
  \bibfield  {author} {\bibinfo {author} {\bibfnamefont {P.}~\bibnamefont
  {Micke}}, \bibinfo {author} {\bibfnamefont {J.}~\bibnamefont {Stark}},
  \bibinfo {author} {\bibfnamefont {S.~A.}\ \bibnamefont {King}},  \emph
  {et~al.},\ }\href {\doibase 10.1063/1.5088593} {\bibfield  {journal}
  {\bibinfo  {journal} {Review of Scientific Instruments}\ }\textbf {\bibinfo
  {volume} {90}},\ \bibinfo {pages} {065104} (\bibinfo {year}
  {2019})}\BibitemShut {NoStop}%
\bibitem [{\citenamefont {Fei}(1990)}]{Fei90}%
  \BibitemOpen
  \bibfield  {author} {\bibinfo {author} {\bibfnamefont {X.}~\bibnamefont
  {Fei}},\ }\emph {\bibinfo {title} {Trapping low energy antiprotons in an ion
  trap}},\ \href@noop {} {Ph.D. thesis},\ \bibinfo  {school} {Harvard
  University, Department of Physics} (\bibinfo {year} {1990})\BibitemShut
  {NoStop}%
\bibitem [{\citenamefont {Naik}(2018)}]{Nai18}%
  \BibitemOpen
  \bibfield  {author} {\bibinfo {author} {\bibfnamefont {P.~K.}\ \bibnamefont
  {Naik}},\ }\href@noop {} {\emph {\bibinfo {title} {Vacuum: Science,
  Technology and Applications}}}\ (\bibinfo  {publisher} {CRC Press},\ \bibinfo
  {year} {2018})\BibitemShut {NoStop}%
\bibitem [{\citenamefont {Freeman}\ \emph {et~al.}(2006)\citenamefont
  {Freeman}, \citenamefont {Yampolskii},\ and\ \citenamefont {Pinnau}}]{Fre06}%
  \BibitemOpen
  \bibfield  {author} {\bibinfo {author} {\bibfnamefont {B.}~\bibnamefont
  {Freeman}}, \bibinfo {author} {\bibfnamefont {Y.}~\bibnamefont {Yampolskii}},
  \ and\ \bibinfo {author} {\bibfnamefont {I.}~\bibnamefont {Pinnau}},\
  }\href@noop {} {\emph {\bibinfo {title} {Materials science of membranes for
  gas and vapor separation}}}\ (\bibinfo  {publisher} {John Wiley \& Sons},\
  \bibinfo {year} {2006})\BibitemShut {NoStop}%
\bibitem [{\citenamefont {Ismail}\ \emph {et~al.}(2015)\citenamefont {Ismail},
  \citenamefont {Khulbe},\ and\ \citenamefont {Matsuura}}]{Ism2015}%
  \BibitemOpen
  \bibfield  {author} {\bibinfo {author} {\bibfnamefont {A.~F.}\ \bibnamefont
  {Ismail}}, \bibinfo {author} {\bibfnamefont {K.~C.}\ \bibnamefont {Khulbe}},
  \ and\ \bibinfo {author} {\bibfnamefont {T.}~\bibnamefont {Matsuura}},\
  }\href@noop {} {\bibfield  {journal} {\bibinfo  {journal} {Switz. Springer}\
  }\textbf {\bibinfo {volume} {10}},\ \bibinfo {pages} {978} (\bibinfo {year}
  {2015})}\BibitemShut {NoStop}%
\bibitem [{\citenamefont {Micke}\ \emph {et~al.}(2018)\citenamefont {Micke},
  \citenamefont {Beiersdorfer}, \citenamefont {Brown}, \citenamefont {Gu},
  \citenamefont {Kilbourne}, \citenamefont {Porter}, \citenamefont
  {Reinhardt},\ and\ \citenamefont {Trabert}}]{Mic18}%
  \BibitemOpen
  \bibfield  {author} {\bibinfo {author} {\bibfnamefont {P.}~\bibnamefont
  {Micke}}, \bibinfo {author} {\bibfnamefont {P.}~\bibnamefont {Beiersdorfer}},
  \bibinfo {author} {\bibfnamefont {G.~V.}\ \bibnamefont {Brown}},  \emph
  {et~al.},\ }\href {\doibase 10.1063/1.5031422} {\bibfield  {journal}
  {\bibinfo  {journal} {Review of Scientific Instruments}\ }\textbf {\bibinfo
  {volume} {89}},\ \bibinfo {pages} {063109} (\bibinfo {year}
  {2018})}\BibitemShut {NoStop}%
\bibitem [{\citenamefont {Sturm}\ \emph {et~al.}(2019)\citenamefont {Sturm},
  \citenamefont {Arapoglou}, \citenamefont {Egl}, \citenamefont {H{\"o}cker},
  \citenamefont {Kraemer}, \citenamefont {Sailer}, \citenamefont {Tu},
  \citenamefont {Weigel}, \citenamefont {Wolf}, \citenamefont
  {L{\'o}pez-Urrutia},\ and\ \citenamefont {Blaum}}]{Stu19}%
  \BibitemOpen
  \bibfield  {author} {\bibinfo {author} {\bibfnamefont {S.}~\bibnamefont
  {Sturm}}, \bibinfo {author} {\bibfnamefont {I.}~\bibnamefont {Arapoglou}},
  \bibinfo {author} {\bibfnamefont {A.}~\bibnamefont {Egl}},  \emph {et~al.},\
  }\href {\doibase 10.1140/epjst/e2018-800225-2} {\bibfield  {journal}
  {\bibinfo  {journal} {The European Physical Journal Special Topics}\ }\textbf
  {\bibinfo {volume} {227}},\ \bibinfo {pages} {1425} (\bibinfo {year}
  {2019})}\BibitemShut {NoStop}%
\bibitem [{\citenamefont {Derevianko}\ and\ \citenamefont
  {Pospelov}(2014)}]{Der14}%
  \BibitemOpen
  \bibfield  {author} {\bibinfo {author} {\bibfnamefont {A.}~\bibnamefont
  {Derevianko}}\ and\ \bibinfo {author} {\bibfnamefont {M.}~\bibnamefont
  {Pospelov}},\ }\href@noop {} {\bibfield  {journal} {\bibinfo  {journal}
  {Nature Physics}\ }\textbf {\bibinfo {volume} {10}},\ \bibinfo {pages} {933}
  (\bibinfo {year} {2014})}\BibitemShut {NoStop}%
\end{thebibliography}%

\end{document}